\documentclass[reprint,amsmath,amssymb,aps,usenatbib,useAMS,nofootinbib]{revtex4-1}

\usepackage{graphicx}
\usepackage{dcolumn}
\usepackage{bm}
\usepackage{color}
\usepackage{lipsum}
\usepackage{multirow}
\usepackage{centernot}
\usepackage[dvipsnames,table]{xcolor}
\usepackage{hyperref}
\hypersetup{urlcolor=RedViolet,citecolor=ForestGreen,linkcolor=RoyalBlue,colorlinks=true}
\usepackage[normalem]{ulem}
\usepackage{aas_macros}
\usepackage{gensymb}
\usepackage{multirow}

\def\simlt{\stackrel{<}{{}_\sim}}
\def\simgt{\stackrel{>}{{}_\sim}}

\newcommand{\Mpch}{$h^{-1}\,\mbox{Mpc}$}

\newcommand{\ie}{\textit{i.e.}}
\newcommand{\eg}{\textit{e.g.}}

\newcommand{\elephant}{\textsc{ELEPHANT}}

\newcommand{\lcdm}{$\mathrm{\Lambda \text{CDM}}$}

\begin{document}

\title{Galaxy and halo angular clustering in \lcdm{} and Modified Gravity cosmologies}

\author{Pawe\l{} Drozda} \email{(pdrozda, hellwing, bilicki)@cft.edu.pl}
\author{Wojciech A. Hellwing}
\author{Maciej Bilicki}

\affiliation{Center for Theoretical Physics, Polish Academy of Sciences, Al. Lotników 32/46, 02-668 Warsaw, Poland}

\date{\today}

\begin{abstract}
Using a suite of $N$-body simulations, we study the angular clustering of galaxies, halos, and dark matter in
Lambda Cold Dark Matter and modified gravity (MG) scenarios.
We consider two general categories of such MG models, one is the $f(R)$ gravity, and the other is the normal branch
of the Dvali-Gabadadze-Porrati brane world (nDGP).
To measure angular clustering we construct a set of observer-frame light cones and resulting mock sky catalogs.
We focus on the area-averaged angular correlation functions, $W_J$, and the associated reduced cumulants $S_J\equiv W_J/W_2^{(J-1)}$, and robustly measure
them up to the ninth 
order using counts in cells. We find that $0.15 < z < 0.3$ is the optimal redshift range to maximize the MG signal in our light cones.
Analyzing various scales for the two types of statistics, we identify up to 20\% relative departures in MG measurements from general relativity (GR), 
with varying signal significance.
For the case of halos and galaxies, we find that third-order statistics offer the most sensitive probe of the different
structure formation scenarios, with both $W_3$ and the reduced skewness $S_3$, reaching from $2\sigma$ to $4\sigma$ significance
at angular scales $\theta\sim0.13\degree$. The MG clustering of the smooth dark matter field is characterized by even stronger deviations
($\simgt 5\sigma$) from GR, albeit at a bit smaller scales of $\theta\sim0.08\degree$, where baryonic physics is already important.
Finally, we stress out that our mock halo and galaxy catalogs are characterized by rather low surface number densities when compared
to existing and forthcoming state-of-the-art photometric surveys. This opens up exciting potential for testing GR and MG using angular clustering in future applications,
with even higher precision and significance than reported here.
\end{abstract}

\maketitle

\section{Introduction}\label{sec.intro}
One of the greatest accomplishments of modern cosmology is the formulation of the concordance standard cosmological model,  
the   so-called  {\it Lambda Cold Dark Matter} (\lcdm). This is a phenomenological model assuming that around $30\%$ of
the present-day Universe energy density is in the form of non-relativistic matter (baryonic and dark) and the remaining $70\%$ is
attributed to the ``dark energy'', an exotic phenomenon propelling the late-time accelerated expansion of the Universe. 
\lcdm{} is a simple model with six free parameters, that is able to pass successfully many stringent observational tests,
e.g., \citep[]{Planck2020,Alam2021,Abbott2022,Brout2022}.

One of the core predictions of \lcdm\ is that  the cosmic large-scale structure (LSS) originated from
gravitational instability acting on early matter
distribution \cite{peebles1980large}. The widely accepted 
scenario assumes adiabatic Gaussian initial conditions (as supported by the cosmic microwave background
measurements \cite{Planck2020}),  
which were then reshaped due to the nonlinear gravitational evolution.
The resulting LSS is organized into the so-called cosmic web \cite{bond1996filaments},
as traced by its main building blocks -- luminous galaxies. 
Among the most striking features of this cosmic-web arrangement are 
the high anisotropy of the underlying density distribution (\ie{} volume dominance of voids, mass dominance of filaments) and 
a scale-dependent clustering amplitude, observed also in the spatial distribution of galaxies \citep{bernardeau2002large,hawkins20032df,zehavi2011galaxy,Piscionere2015,Beutler2017}.

The scale-dependent hierarchical matter and galaxy clustering is one of the most striking manifestations of the gravitational instability paradigm
\citep{peebles1980large}.

These unique and characteristic features of the LSS have been extensively employed as powerful probes of the standard cosmological 
model and its core assumptions. In the past few decades galaxy photometric and spectroscopic catalogs have been
growing both in volume as well as in quality of the data, which allowed for more and more precise tests of the fundamental components of \lcdm.

Thanks to these growing observational data, presently 
the spatial distribution of galaxies and their time evolution can be readily used for performing stringent tests 
of the gravitational instability scenario.
The latter is rooted in two core assumptions: the adiabatic Gaussian initial conditions; and 
general relativity (GR) as an adequate and valid description of gravitational clustering on all scales and at all times. In this paper we 
explore the possibility of employing the
properties of late-time matter and galaxy angular clustering for testing and differentiating GR and beyond-GR structure formation scenarios.

One of the simplest characteristics of matter clustering  is the two-point correlation function (2PCF).
This statistics is relatively easy to measure, which makes it a fundamental object commonly used in
cosmology for quantifying matter and galaxy clustering
\citep{maddox1996apm,connolly2002angular,hawkins20032df,maller2005galaxy,crocce2011modelling,wang2013sdss}.
In this context, we can recall that
the Wick's theorem for a Gaussian random field states that the first two order statistics (\ie\ the mean and the variance)
are sufficient to provide a complete statistical description of the field and clustering. 
In other words, for normal distribution all the higher-order odd moments vanish, and all the higher-order even moments are proportional
to the variance. 

However, the distribution function of a developed late-time LSS on scales below a few hundred megaparsecs deviates from a pure Gaussian
distribution. Because of gravitational evolution, depending on the scale and the epoch involved, 
non-Gaussian features emerge, leading to the highly anisotropic and complex large-scale structure.
Therefore, the variance and related two-point statistics no longer provide a sufficient description
of the late-time LSS on scales important for galaxy formation and clustering. To infer additional information,
one can resort to higher-order or beyond-two-point clustering statistics. Analogically to 2PCF,
we can define $N$-point equivalents. However, already starting from $N=3$,
such correlation functions (CFs) become very expensive computationally and their usage in cosmology has been so far limited
to some special cases \citep{Takada2003,Nichol2006,Guo2015,Slepian2017,nunez2020fast}
-- but see Refs.\
\citep{slepian2018practical,umeh2021optimal} for some recent developments.

An approach that is complementary to $N$-point statistics involves
$N$th order central moments.
These are volume-averaged versions of their full $N$-point counterparts, 
but are significantly easier to compute and model.
At the same time, higher order central moments, and the associated cumulants, contain extra
information on the shape and asymmetry of the matter and galaxy distribution. Thus these statistics
bear information that is complementary to that carried by two-point statistics, e.g.,
\citep[]{white1979hierarchy,gaztanaga1994high,croton20042df}.

Both two-point and higher-order clustering statistics have proven to be insightful and rich cosmological probes. 
It is however worth mentioning that
the higher central moments are especially well suited for testing departures from standard structure formation scenarios.
This is thanks to their increased sensitivity to non-Gaussian features of the distribution functions \citep{peebles1980large,juszkiewicz1993skewness,lokas1995kurtosis}.
In general,
the non-standard gravitational instability models would involve some level of modified gravity (MG).
In such scenarios one usually deals with low-energy effective scalar-tensor modifications to the Einstein-Hilbert action integral \cite{Clifton2012}.
In that sense, these models are not new fundamental theories of gravity in their own right, but rather phenomenological
manifestations (and parametrizations) of deeper underlying theories. 

In this work we consider two such MG models,
which constitute a good representative sample of a whole family of effective phenomenological modifications to gravity. 
The first consists of the so-called $f(R)$ framework \cite{hu2007models,sotiriou2010f}, where in the gravity
action integral the classical Ricci's scalar, $R$,
is generalized to a functional $f(R)$ form \citep{Buchdahl1970}. The second family is the so-called normal branch of
the Dvali-Gabadadze-Porrati (nDGP) brane world model \citep[nDGP;][]{vainshtein1972problem,dvali20004d},
where gravity can propagate in the full five-dimensional space time, while
the standard elementary particle forces are confined to a four-dimensional subset space-time
of a brane \citep{Sahni2003}.
Both of these MG scenarios admit the action of so-called fifth force 
on cosmological scales.
This extra force
is a manifestation of the additional scalar degrees of freedom of these models which, when coupled to the usual
matter fields, affect the action of the gravitational instability and structure formation on galactic and intergalactic scales \citep{Li2009}.
The stringent tests of GR in the strong-field regime \cite{Liu2014,Wei2017,Abbott2020} and in the weak-field for the Solar System
and our own Galaxy \cite{Ingrid2003,DeMarchi2020} impose rigorous constraints on the scales and times on which such an MG-induced fifth force
is allowed to manifest itself.
In order to pass these fifth force tests, viable MG models need to suppress propagation of the extra degrees of freedom
in environments such as the Solar System or the Milky Way. The physical phenomena that lead to the fifth force suppression are called
the screening mechanisms. Both the $f(R)$ and the nDGP theories naturally admit for such effects. Tuning the related theoretical
parameters of these theories allows for finding solutions that simultaneously pass the local gravity tests and match
the global \lcdm\ expansion histories.

We focus on \textit{angular} correlations, \ie\ those projected along the line of sight.
While the full 3D CFs give direct access to such cosmologically important effects as redshift-space distortions,
they can only be observationally studied with sufficient accuracy using spectroscopic redshift catalogs.
These often suffer from small-area coverage and/or sparse sampling, and even in the era of the
forthcoming Dark Energy Spectroscopic Instrument data, will include only a small fraction of all observable
galaxies. Another problem connected to spectroscopic surveys is that their analysis requires
theoretical input on redshift-space to real-space mapping, which is a strongly model-dependent procedure
\citep{Bose2017,Bose2019,garcia2021probing}.
On the other hand, photometric (imaging) surveys typically offer a much better combination of
depth, sky coverage and completeness than the spectroscopic ones, and if accompanied by photometric redshifts,
give the possibility to perform tomographic analyses of the density field. In view of future
multi-billion galaxy catalogs from such campaigns as the Vera Rubin Observatory Legacy Survey of Space
and Time \cite{ivezic2009lsst} or the Euclid space mission \cite{laureijs2011euclid},
it is timely to investigate possible MG signals from higher-order angular clustering.

This paper is structured as follows:
In Section \ref{sec.models_sims} we describe gravity models and simulations used.
The broader picture of clustering statistics is contained within Section \ref{sec.clustering}.
Afterwards, in Section \ref{sec.calc} we show the method of clustering calculations 
and introduce corresponding analytical predictions. 
Later, in Section \ref{sec.results}
we motivate our choice of redshift ranges which we use in the search for modified gravity signals.
Then we present all our results which are summarized and concluded upon in Section \ref{sec.concl}.

\section{Modified Gravity models and simulations}
\label{sec.models_sims}

In our study we will examine and compare higher-order angular clustering in different growth-of-structure scenarios.
For this purpose we invoke 
the \elephant\ ({\it Extended LEnsing PHysics with ANalytical ray Tracing}) numerical
$N$-body simulations \citep{cautun2018santiago} performed using the \textsc{ECOSMOG} code \cite{li2012roy}.
They assume the evolution of $N_{part}=1024^3$ particles within $L_{box}=1024/h$ Mpc sized box. Our fiducial, 
or baseline, model of choice is the GR-based flat $\Lambda CDM$ model with \textsc{WMAP} 9-year cosmology \cite{hinshaw2013nine}, with matter and dark energy density parameters $\Omega_m=0.281$ and $\Omega_{\Lambda}=0.719$, and the Hubble constant $H_0=100 \, h\;\rm{km\;s^{-1} Mpc^{-1}}$ where $h=0.697$.

On top of this background $\Lambda CDM$ model we consider two beyond-GR scenarios. The first of them -- the
Hu-Sawicki variant of $f(R)$ MG -- introduces the 
fifth-force which is suppressed in dense environments thanks to the virtue of the chameleon mechanism \citep{khoury2004chameleon,Brax2008}.
Adopting a standard choice of the free parameters for this model
\citep[\eg][]{hu2007models,arnold2019modified,garcia2021probing}, we are left with only
one variable to be set  in order to characterize the late-time modifications to GR: the amplitude of
the background scalar field at the present times, usually denoted as $f_{R0}$.
Following the previous works which employ the \elephant simulation suite \citep{cautun2018santiago,alam2020testing},
we label the two $f(R)$ variants used as F6 and F5, which corresponds to $f_{R0}=\{-10^{-6},-10^{-5}\}$ respectively.

The second MG family -- normal branch Dvali-Gabadadze-Porrati models (nDGP) -- incorporate the Vainshtein screening mechanism
\cite{vainshtein1972problem,babichev2013introduction}
to suppress the fifth-force in the vicinity of massive bodies. 
In the parameterization adopted here, the nDGP models can be also fully characterized by a single choice of the model
physical parameter. This is the so-called crossing-over scale, $r_c$, which depicts a characteristic scale where the
gravity propagation starts to leak-out to the fifth spatial dimension.
Taking $c=1$, we can fix our nDGP variants to have $H_0 r_c=\{5,1\}{\rm Gpc}/h$, which we label as N5 and N1,
accordingly. 

We calculate angular counts for the projected dark matter (DM) density field from sub-sampled data, using only 
$0.1\%$ from the each initial $1024^3$ particle load. Such sub-sampling severely limits
the spatial and angular resolution of the density fields, but this is needed to facilitate
numerical calculations.
Dark matter halos were extracted using the \textsc{ROCKSTAR} halo finder
\cite{behroozi2012rockstar} and mock galaxy catalogs were generated with the Halo Occupation Distribution (HOD)
method in Ref. \cite{cautun2018santiago} using parameters from \cite{manera2013clustering}.
Unlike the dark matter particles, dark matter halos and mock galaxies are not sub-sampled, and their relatively low number density (see Sec.\ref{subsec:lightcone_depth}) is related to the very nature of the \elephant\ catalogs.

For our analysis we employ 5 independent random phase realizations of initial conditions,
and take snapshots saved at $z=0.0, 0.3, 0.5,$ and $1$ for further analysis.

In order to work in sky-projected observer frame, we need to construct proper observer light cones
from our snapshots.  The redshift range we consider, \ie\ $0 < z < 0.5$ 
for galaxies and DM particles, and $0 < z < 1$ 
for halos, corresponds to comoving scales that by far exceed the \elephant\ simulation box size.
To cope with that we locate the observer at the $\vec{r_0}=(0,0,0)$ corner of the box and 
 use the box replications method \citep[see][]{blaizot2005momaf,overzier2013millennium,smith2017lightcone} to build
 the light cones.
For each snapshot we copy the adequate box within ranges defined as half the comoving distance between the redshift of the current and each adjacent snapshot.

From our ligth cones we generate  two-dimensional sky catalogs consisting of a series of $\sim1567$ $\deg^2$ chunks,
which would correspond to sky patches of sizes $40\degree \times 40\degree$ if centered on the equator.
Each of our 2D sky catalogs is a sum of several separate sky chunks\footnote{Note that some of the sky chunks can partially overlap over the mock sky.
For details, see Appendix}. We found that for 15 chunks 
we already attain the maximum spatially-independent catalog information,
as measured by the catalog effective volume $V_{\rm eff}$. 

This effective volume needs to be defined because, due
to box replications, the catalogs contain many copies of the same structures.
Thus, the total amount of independent cosmological information is always smaller than it
normally would result from the actual light cone comoving volume. To get a total measure
of  unique (\ie\ not cloned) volume, we use the catalog effective volume, $V_{\rm eff}$.
We define it as a sum over all the simulation box texels that are used at least once, divided
by their total number inside the box. For more details on how $V_\mathrm{eff}$ is measured, see Appendix.

Following Ref.\ \cite{garcia2021probing}, for our halo samples we consider only objects with
$M_{vir}\geq 10^{12} M_{\odot}/h$. In Ref.\ \cite{garcia2021probing} it was found that the abundance of less massive halos
is already affected by the mass resolution limit of the simulations. After this initial mass cut,
with the same universal threshold for all the simulations runs, we employ secondary individual sample mass cuts.
These are administered in such a fashion to obtain the same object number density within a given initial condition realization
suite among all different physical models (\ie\ F5, F6, GR, N1, and N5). Here randomly selected 
least-massive halos are trailed-off until a given sample is reduced to the target number density.
The latter is set by the lowest number density sample within a given ensemble. This is done to mimic
a volume-limited sample selection effect. For the mock galaxy sample we applied an analogous operation.
However, since for galaxies we do not have masses nor luminosities, we trail-off all galaxies picked up randomly.

\section{Higher order clustering}\label{sec.clustering}
Our goal is to study clustering properties of matter, halo, and galaxy distributions over large
range of scales and epochs. Here, we will define the basic objects and methods of our clustering analysis.
We will work either with physical ($\vec{r}$) or co-moving ($\vec{x}$) distance units, where the usual relation is
 $\vec{r}=\vec{x}/(1+z)$. Adopting a standard notation we will refer to comoving units as either \Mpch{} or $h^{-1}$Gpc.

We define the standard density contrast which measures a local (\ie{} at point $\vec{x}$) fluctuation of the density
around a uniform background as:
\begin{equation}
    \delta_{3D}(\vec{x})=\frac{\rho(\vec{x})}{\bar{\rho}}-1,
\end{equation}
where $\bar{\rho}$ is the background density.
Converting the coordinates into 
sky-frame $\vec{x}\rightarrow [r,\vec{\Omega}]$, where $r$ is radial distance and $\vec{\Omega}$ represents 
a sky-pointing angular
vector, we can obtain the projected density contrast:
\begin{equation}
    \delta_{2D}(\vec{\Omega})=\int_{r_{min}}^{r_{max}} dr F(r) r^2
    \delta_{3D}([r,\vec{\Omega}]).
\end{equation}
Here, the $r_{min/max}$ values stand for the distance ranges covered by a given survey and $F(r)$ is its selection function.
From now on, for simplicity we will drop the '2D' sub-index whenever referring to $\delta_{2D}$.

The global properties of density fluctuations are encapsulated in their 1D probability distribution function (PDF), 
which can be estimated by averaging the fluctuations over many sky directions.
Gravitational clustering moves the shape of density PDF away from its initial Gaussian form \cite{coles1991lognormal}.
All the gravity-induced non-trivial PDF shape deviations at a certain scale can be characterized by a hierarchy of the
central moments. We will use the standard definition of a central moment of $J-th$ order: 
\begin{equation}
    \mu_J=E\left[(\delta-E(\delta))^J\right],
\end{equation}
where $E$ is the expected value, $\delta$ is our random variable, 
and all the variables intrinsically depend on the angular scale $\theta$. In our work, $\theta$ is a radius 
of a circle centered on a particular sky direction $\vec{\Omega}$. We average over many such circles
to obtain our PDFs.

The central moments, $\mu_J$, estimated for a given sky area at some 
angular scale $\theta$  can be considered as area-averages of the full $J$-point angular clustering functions.
They are related with J-point correlation functions by
\cite{bernardeau2002large}:
\begin{equation}
     W_J(\theta)\equiv{\mu_J\over \langle\delta\rangle^J}=\frac{1}{A} \int_{\Omega} d\Omega_1...d\Omega_J w_J(\theta_1,...,\theta_J),
\end{equation}
where $A=2\pi(1-\cos \theta)$ is the sky area enclosed by angle $\theta$.

The $J$-th order moments can be readily estimated using the counts-in-cells method \citep{Bouchet1992,gaztanaga1994high}. 
The moments will be then ensemble averages over all the circular cells (of intrinsic angular scale $\theta$)
cast over the whole area of interest on the sky, $\Omega$:
\begin{equation}
    W_J (\theta)\equiv \left< \delta_{\theta}^J \right>,
\end{equation}
where $\delta_{\theta}$ is a projected angular density fluctuation estimated at scale $\theta$ from angular counts.

\section{Clustering and Moments of Counts in Cells}
\label{sec.calc}

\label{subsec:ang_cluster}
We estimate the moments of the angular clustering using the commonly adopted method of counts-in-cells (CIC) \citep[CIC,][]{gaztanaga1994high}.
We randomly  
place $N_C$ circles of angular radius $\theta$
within the investigated sky area, making sure they are fully within the considered region. Those extending outside the footprint are 
ignored and replaced by new randomly drawn ones.
Then, we 
count
the objects found inside each circle.
The $J$-th central moment
of the CIC distribution is:
\begin{equation}
    m_J(\theta)=\frac{1}{N_{tot}}\sum_{i=0}^{N_{tot}} (N_i-\langle N \rangle)^J,
    \label{eq.moment}
\end{equation}
where $N_i$ stands for
the object count in the $i$-th
cell, 
$\langle N \rangle$ 
is the mean count
over all the circles with a given radius $\theta$, and $N_{tot}$
the total number of circles used.
We
choose $N_{tot}\propto A_{sky}/(2\pi (1-cos(\theta)))$
to scale as
the number of independent circles that we can place within the analyzed sky area.
Since we are interested in the specific shape departures from a normal distribution,
we will work with the connected moments, $\mu_J$. That is, we subtract from the central moments the parts
expected for a Gaussian PDF. The first few connected moments are:
\begin{eqnarray}
    \label{eqn:connected_moments}
    \mu_2 &=& m_2 \, ,\nonumber\\
    \mu_3 &=& m_3 \, ,\nonumber\\
    \mu_4 &=& m_4 - 3m_2^2 \, ,\nonumber\\
    \mu_5 &=& m_5 - 10m_3m_2.
\end{eqnarray}

Since we will work with relatively sparse samples, the mean counts, especially at small $\theta$,
can become small and the impact of the shot-noise will become significant. To reduce it
we follow the procedure of Ref.\ \citep{gaztanaga1994high} and subtract from the connected moments the contribution
expected from a Poisson distribution for a given mean count $\langle N\rangle$ (see eqn.\ A6 therein).

The shot-noise correction is obtained by considering the contribution to the moments from a Poisson distribution
with the same mean number of counts, $\langle N \rangle$, as the studied sample. For reference we only recall the first few shot-noise
corrected moments, $k_J$:
\begin{eqnarray}
    \label{eqn:shot_noise_connected_moments}
    k_2 &=& \mu_2-\langle N \rangle \, , \nonumber\\
    k_3 &=& \mu_3 - 3k_2 - \langle N \rangle \, , \nonumber\\
    k_4 &=& \mu_4 - 7k_2 -6k_3-\langle N \rangle \, , \nonumber\\
    k_5 &=& \mu_5 -15k_2-25k_3-10k_4-\langle N \rangle\,.
\end{eqnarray}

Finally, the $J$-th order corrected and averaged correlation function can be written as:
\begin{equation}
\label{eqn:W_Js}
    W_J=\frac{k_J}{\langle N \rangle^J}\,,
\end{equation}
and the re-scaled cumulants, or
more commonly dubbed in cosmology as \textit{hierarchical amplitudes}, will be
\begin{equation}
\label{eqn:S_Js}
    S_J=\frac{W_J}{W_2^{J-1}}\equiv {W_J\over\sigma^{2J-2}}\,.
\end{equation}

\subsection{Signal significance}\label{sec.ssig}
Following a standard approach we estimate the error on the quantities given by
Eqns.\ \eqref{eqn:W_Js} and \eqref{eqn:S_Js}
as the variance around
the mean, obtained as 
the ensemble average over all equivalent realizations of a given data set (\ie{} a light cone).
In practice, we will be more interested in assessing the differences between each MG model and the fiducial \lcdm\ case.
This is measured by the relative difference of paired observables always taken with respect to the GR case.
Both the fiducial GR and any given MG model sample will be characterized by they own individual variance.
For that reason, comparing clustering moments of different models with different individual variances might be
difficult and not intuitive. To foster a more natural and easy to interpret comparison we will use the signal
significance parameter, $\psi$, defined as:
\begin{equation}
    \psi_J=\frac{X-Y}{\sqrt{\sigma_X^2 +\sigma_Y^2}}
    \label{eq.sig}
\end{equation}
where $X$ and $Y$ are measurements and $\sigma_{X,Y}$ their respective uncertainties.
Here, $J$ indicate the order of the $X$, and $Y$, statistics used to calculate the significance.
So, for example $\psi_3$ can indicate either that $W_3$ or $S_3$ was used.
In this work we will always take $Y$ as GR and any given MG model is taken as $X$. 
The significance $\psi$
gives a simple notion of the direction of the difference w.r.t. the fiducial case preserving the sign of the difference,
and automatically traces the significance of this difference due to the normalization factor in the denominator.

\subsection{Perturbation Theory predictions} 
\label{sec.theoretical_predictions}

Our main results in this work are based on the analysis of $N$-body simulations, which by design
can  probe deeply into the non-linear regime of structure formation. However, it is very informative and beneficial
to provide also analytical predictions with witch the numerical results can be gauged.
For that purpose we use calculations based on the weakly non-linear perturbation theory (PT), which yield predictions for low-order
moments. These are obtained by 

integrating over the matter power spectrum 
with appropriate window and selection functions
\citep[see][]{gaztanaga1997skewness,pollo1997gravitational}.

The second moment is given by:
\begin{equation}
\label{eqn:w2_PT}
    W_2(\theta)=\frac{1}{2\pi} \int_{R_{min}}^{R_{max}} r^4   F^2(r) dr
    \int_{0}^{\infty} k P(k) W_{2D}^2(k\theta r) dk,
\end{equation}
where $P(k)$ is a given model power spectrum , and
\begin{equation}
\label{eqn:2d_tophat}
    W_{2D}(k)=2\frac{J_1(k)}{k}
\end{equation}
is the window function for which we take a circular top-hat in the Fourier space with the first-order spherical Bessel function $J_1$.
The $R_{min/max}$ stand for catalog comoving distance ranges and $F(r)$ is the radial 
selection function
normalized in a such way that:
\begin{equation}
\label{eqn:rmin_rmax}
    \int_{R_{min}}^{R_{max}} r^2F(r) dr =1.
\end{equation}
In our catalogs we do not use specific selections mimicking the observations, hence the selection
function becomes:
\begin{equation}
\label{eqn:Fr_PT}
    F(r)=\frac{3}{R_{max}^3 -R_{min}^3}=const.
\end{equation}

For the third order we have:
\begin{equation}
\begin{aligned}
    W_3(\theta)=6\frac{\theta^{-4}}{(2\pi)^2}
    \int_{R_{min}}^{R^{max}} r^2 F^3(r) dr
    \int_0^{\infty} q W^2_{2D} (q) P(k) 
    dq \times\\
    \left[
    \frac{5}{14}\int_0^{\infty} qW_{2D}^2(q) P(k) 
    dq
    -\frac{1}{4}\int_0^{\infty} q^2 W_{2D}^2(q)\frac{dP(k)}
    {dq} dq
    \right],
\end{aligned}
    \label{eqn:w3_PT}
\end{equation}
where $q=k\theta r$.

The formulas for higher-orders become longer and  recurrently more involved
\citep[see \eg][]{bernardeau2002large}. 
Thus we opt to stop at the third order only, since detailed tests of PT are not our aim here, 
and we will use these predictions for approximate trend comparisons only.
In yielding our PT predictions, we have used both linear and nonlinear (\ie{} the \textsc{Halofit}
\cite{Smith2003stable}
) power spectra models
computed with \textsc{CAMB} software \cite{lewis2000efficient} taken at the effective catalog redshift $z_\mathrm{eff}=0.242$.

\section{Results}\label{sec.results}

Now we are ready to present and investigate the results covering the angular clustering,
distributions of the counts-in-cells and the associated moments. All the presented results  concern
dark matter, halo and galaxy samples extracted from the same depth \elephant-based light cones, as described above.

\subsection{Finding optimal light cone depth}
\label{subsec:lightcone_depth}

Previous studies of clustering have indicated that in the case of MG models considered here,
the magnitude of the deviations from the GR-fiducial case is changing with the cosmic time, in a non-monotonic way 
\citep[see \eg][]{Li2013,hellwing2013hierarchical, hellwing2017revealing}. In case of single redshift snapshots, commonly used in 
the distant observer approximation,
it is straightforward to depict a redshift with maximal deviation from GR. However, in our case, in the light cone projection,
clustering information from the redshift range of the whole light cone is entangled. We want therefore to find
an optimal redshift range for a ligth cone galaxy catalog which maximizes the relative deviation from GR
of the MG clustering signal as far as the moments are concerned.

Considering the redshift ranges we have for our halo ($0<z<1$) and galaxy samples ($0<z<0.5$),
we create 
a $4\times4$ grid of
catalogs with varying minimum redshift 
$z_{min}$ and a defined thickness $\Delta z=z_{max}-z_{min}$.
For the optimization procedure we define our merit parameter to be the $\psi_3$ estimator as defined later
in Eq.\ \eqref{eq.sig}, which measures the relative amplitude of the deviation between a given MG model and the GR case.

Optimization performed over all models and scales would be computationally very expensive. However, since our goal here
is just to find an approximated optimal redshift light cone range, we opt to focus on only two MG models, N1 and F5, and
only one angular scale of $\theta_{opt}=0.08 \degree$. These two MG variants are characterized by the largest
difference in the linear growth-rate with respect to GR. The $\theta_{opt}$ value was selected as a reasonable compromise
between the non-linear regime, where the clustering deviations usually are the largest, and at the same time 
a scale where the shot-noise and simulation resolution effects are not too severe yet.

We have found that the redshift range\footnote{The exact range is $0.1525<z<0.3025$.}
$0.15<z<0.3$ maximizes the MG signal for halos, while the range $0.15<z<0.25$ 
is optimal for the galaxies. Considering the fact that halo light cones
provide wider redshift ranges in comparison with galaxies and the
halos provide stronger signals than galaxies,
we kept $0.15<z<0.3$ 
as the best redshift ranges for all the catalogs.
The final data samples with the imposed redshift cuts have the following characteristic
projected number densities:
\begin{itemize}
    \item DM particles: $\sim 51$ $\deg^{-2}$,
    \item halos: $\sim 40$ $\deg^{-2}$,
    \item galaxies: $\sim 15$ $\deg^{-2}$.
\end{itemize}

Given the effective depth of our light cones, the spatial resolution of the \elephant\ suite, and taking into account that we consider only resolved halos,
we can estimate that our catalogs will be spatially resolved down to $\sim0.5-1$\Mpch{} \citep{alam2020testing}. Within the redshift range we use, 
this sets the minimum angular scales
that we can consider as resolved to be $\theta_{res}\approx 0.05\degree$.

\subsection{Probability density functions}
\label{subsec:PDFs}

\begin{figure}
\centering
\includegraphics[width=\columnwidth]{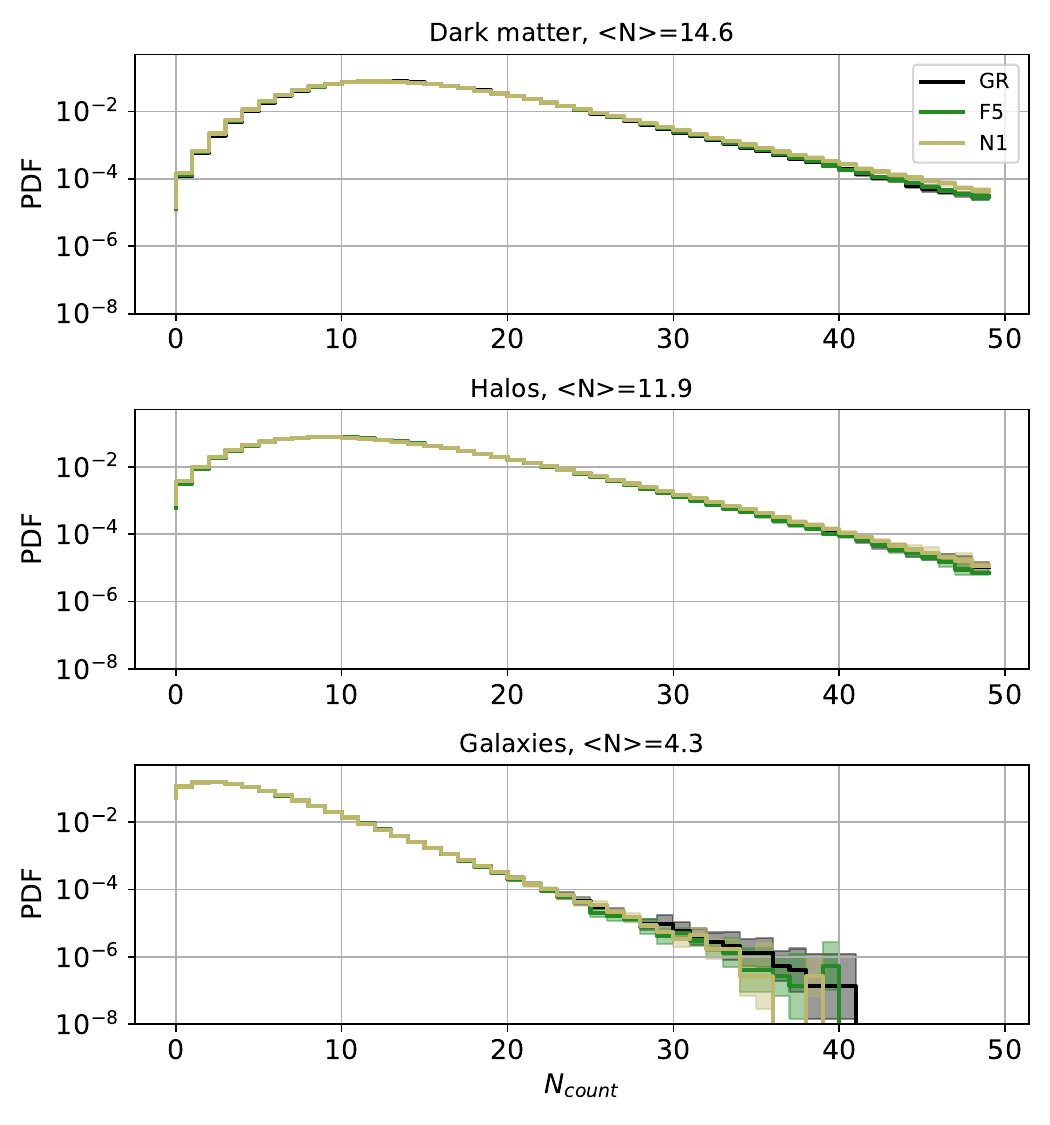}
\caption{Probability density functions of the counts-in-cells from circles of $0.3\degree$ radius, corresponding to $\sim3.6$ \Mpch\ 
physical scales at the effective redshift of our light cones ($z_\mathrm{eff}\sim0.24$). The panels show from top to bottom the results 
for dark matter, halos, and galaxies, derived from the \elephant\ suite for 3 gravity models indicated in the legend. The numbers in the headers indicate
mean counts for each case.
}
\label{fig.rho}
\end{figure}

We begin by showing in Fig.~\ref{fig.rho} an excerpt of angular counts-in-cells distributions for dark matter
(top panel), halos (middle panel) and galaxies (bottom panel). These example PDFs
are measured at the angular scale of $\theta=0.3\degree$. Considering the median redshift of our light cones
(\ie{} $z\sim 0.242$), this scale corresponds to a projected comoving separation of $R=3.6$ \Mpch.
We pick this scale since it constitutes a reasonable compromise between the scales where 
the influence of both the cosmic variance and the sparse sampling remains limited.
For clarity we show only two modified gravity variants, N1 and F5, on top of the GR case.

This example already brings a few interesting observations. First, we can infer that
dropping angular object number density drives the resulting PDF away from a Gaussian and more towards a Poissonian distribution.
Thus, highlighting the importance of the shot-noise corrections for samples with small mean number
counts, $\langle N\rangle.$ The second noticeable feature is that for all three samples 
(\ie{} galaxies, halos, and dark matter) different models arrive at very similar mean number counts.
Furthermore, the associated variances, or the distribution widths, are also comparable. Only, when
we move away from the PDF's centers, towards the tails, the differences between GR and MG models
become more, and more appreciable. This is a clear illustration on how important is to go beyond
central and second order moments, which are much more sensitive to information contained in the distributions
tails (\ie{} the PDF asymmetry and overall shape deviations).

\subsection{Dark Matter}
\label{subsec:dm}

We start our analysis of angular clustering by looking at the projected dark matter density field.
Although this is not directly observable, there is a strong connection
between the underlying smooth projected dark matter distribution and quantities accessible via gravitational
lensing effects. 

These, among others, include convergence and shear power spectra \cite{gaztanaga1997skewness}.

There are also
tomographic techniques to obtain reconstructed 3D dark matter distribution on large scales
\cite{Bacon2003,Jain2003,Massey2007}.
Here, we will focus on simple sky-projected dark matter density fields measured from a sub-sampled 
dark matter $N$-body particle distributions. This distributions and its moments is not directly connected
to observations, but it provides a very good test-bed. Moreover, analyzing the dark matter clustering will 
enable us to compare our CIC results with PT predictions as given in Sec. \ref{sec.theoretical_predictions}, and will provide additional physical insight about
the higher-order angular clustering in MG in linear and non-linear regimes.

\begin{figure}
\centering
\includegraphics[width=\columnwidth]
{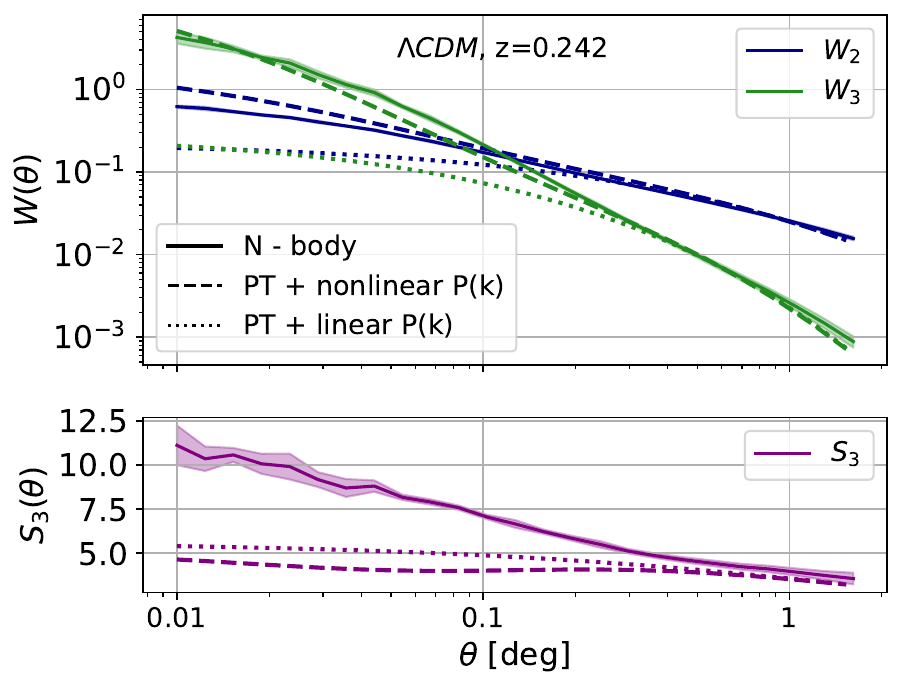}
\caption{First two reduced moments (top) and reduced skewness (bottom) of angular clustering calculated in the \lcdm\ model at an effective 
redshift of $z=0.242$. The solid lines with shaded error ranges show $N$-body simulation results, while the dotted (dashed) lines illustrate 
the perturbation theory predictions derived using linear (non-linear) power spectra as detailed in Sec.\ \ref{sec.theoretical_predictions}.}

\label{fig.PredictionW2W3S3}
\end{figure}

We begin by comparing the $N$-body CIC moments with the PT predictions. This will serve both as a
useful test of our estimators, as well as the indicator of scales where the transition between the non-linear and weakly non-linear
angular clustering regimes occurs. In Fig.~\ref{fig.PredictionW2W3S3} we show the first two moments, 
$W_2(\theta)$ and $W_3(\theta)$ (upper panel), and the reduced skewness, $S_3(\theta)$ (bottom panel). The continuous lines indicate
our $N$-body results, the dotted and dashed curves are the PT prediction obtained using the linear (dotted) and Halofit \cite{Smith2003stable} (dashed) 
dark matter power spectra. The shaded regions indicate $1\sigma$ scatter from the simulation ensemble mean.

The PT predictions agree very well with the $N$-body results at large scales, $\theta \gtrsim 0.2 \degree$. 
For smaller angles the simulation results quickly surpass the values based on the linear theory $P(k)$.
Interestingly, using Halofit as the non-linear power spectrum model readjusts the PT predictions, making them follow
the $N$-body lines much more closely, extending good PT accuracy down to scales of $\theta\sim 0.04 \degree$. 
However, even if the PT predictions for $W_2$ and $W_3$ separately look reasonable, their combination into the reduced skewness, $S_3$,
accumulates the deviation of each individual moment. This is clearly manifested in the bottom panel of Fig.~\ref{fig.PredictionW2W3S3},
where both the PT-based forecasts fail and under-predict the skewness dramatically for $\theta\lesssim0.3\degree$.
To get a better prediction here, one would need to call for higher-order PT templates  \citep[see \eg][]{Hivon95,Okamura2011,Kitaura2012}.
This test indicates that our $N$-body results capture well both the linear and non-linear regimes. In addition, all the significant differences that we might find in GR vs.\ MG clustering above $\theta\simeq0.3\degree$
could be highlighted in future analysis using weakly non-linear PT predictions as detailed in Sec.\ \ref{sec.theoretical_predictions}.

\begin{figure}
\centering
\includegraphics[width=\columnwidth]
{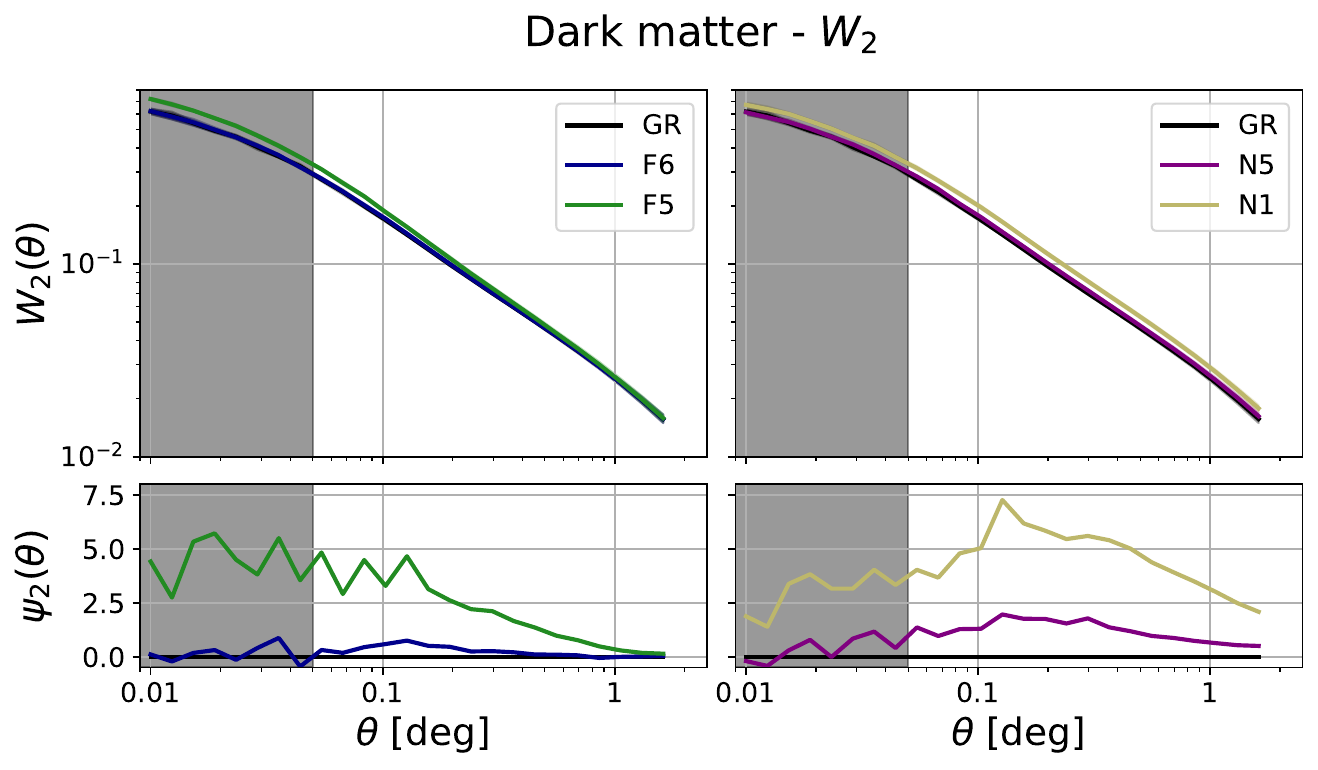}
\caption{Two-point area-averaged angular correlation function of dark matter particles for \lcdm\ (black lines) as compared to two MG scenarios: $f(R)$ in the left-hand column and nDGP in the right-hand one. The MG models have two variants each, as indicated in the legends. Top panels show the correlation function, while the bottom ones illustrate the significance of departure in the MG models from the fiducial GR scenario, as defined in Eq.~\eqref{eq.sig}. The shaded regions cover the $\theta<0.05\degree$ range which we do not use for inferring the signal significance due to the limitations of the \elephant\ simulations.
}
\label{fig.DM_W2}
\end{figure}

\begin{figure}
\centering
\includegraphics[width=\columnwidth]
{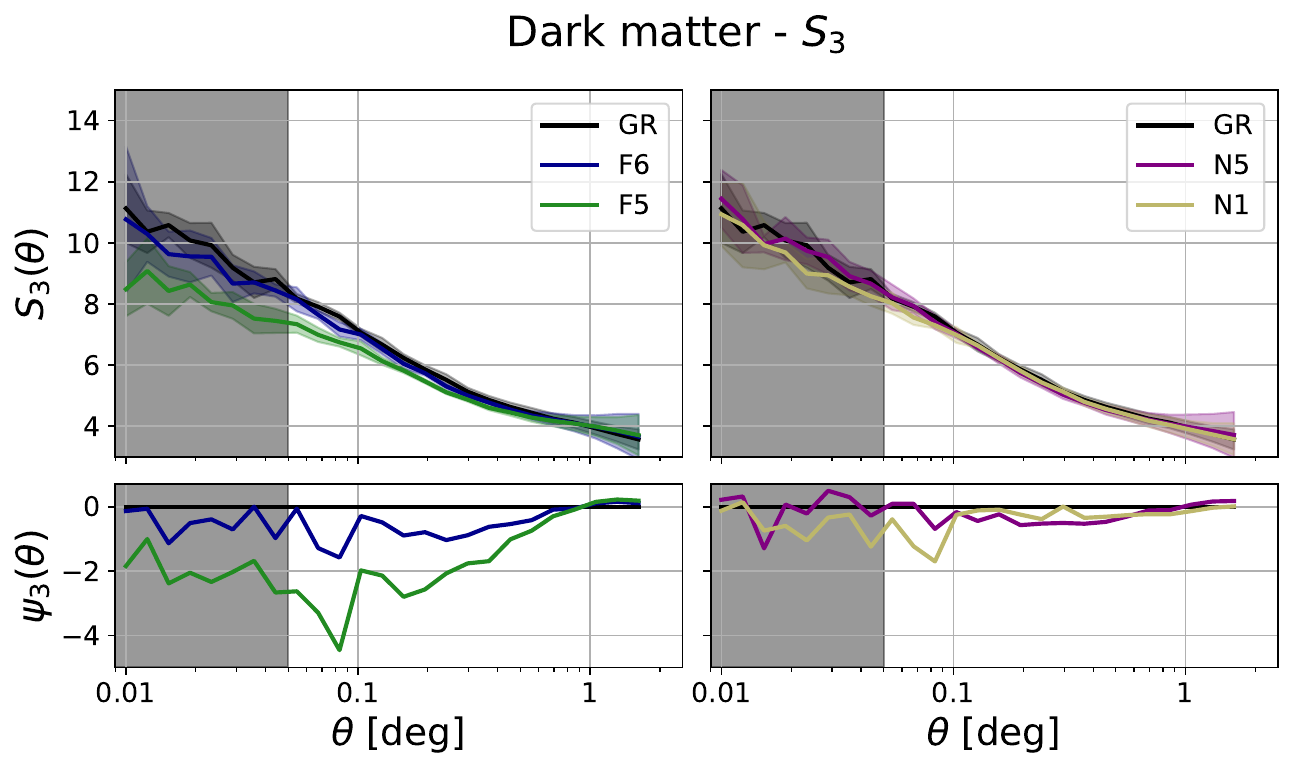}
\caption{Similar to Fig. \ref{fig.DM_W2}, but for the reduced skewness.}
\label{fig.DM_S3}
\end{figure}

Before we move to the main part of the analysis, with the focus on clustering of halos and galaxies,
we take a quick look at the angular variance and reduced skewness of the dark matter projected density field
in our models, shown respectively in Fig.~\ref{fig.DM_W2} and in Fig.~\ref{fig.DM_S3}.
To facilitate easier comparison we group the MG models families, keeping the $f(R)$ models in the left-hand side panels,
and nDGP in the right-hand side. The shaded areas for $\theta<0.05\degree$ indicate the angles lower than the angular convergence scale of \elephant.

Looking first at the angular variance ($W_2$), we can already make a number of
very interesting observations. First, F6 seems to accommodate only minute differences from GR and is virtually
indistinguishable from it for all scales. Second, all the remaining variants, \ie\ F5, N1, and N5, exhibit in $W_2(\theta)$
some unique scale-dependent patterns of their significance signals. The departure of the F5 signal from GR saturates at around $\theta\simlt 0.1\degree$ 
and then decreases to converge to GR at $\theta\simeq 1\degree$. For both the nDGP variants we notice a similar decrease in the signal, 
but now on two sides from the maximal deviation scale of $\theta\simeq0.2\degree$. Thus, the departure of $W_2$ in nDGP from GR
assumes a peaklike shape in angular scaling. 
Moreover, here also the N5 model, generally weakly departing from GR, fosters significant
deviations from the GR case, in contrast to its $f(R)$ cousin, F6. This is a new result, as such features
have not been found in the earlier 3D dark matter clustering studies \citep[see \eg][]{hellwing2013hierarchical,hellwing2017revealing}.

The dark matter results for $S_3$ feature a bit different picture. Here, the noise and errors on both $W_2$ and $W_3$ moments are amplified,
and the resulting signal become generally weaker and more erratic. While for N1 and N5 the  significance is
severely reduced, F5 interestingly still reaches $|\psi_3|\simeq 4$
at $\theta\sim0.1\degree$.
Interestingly enough, for all the MG models their skewness takes values lower than in the fiducial GR, which indicates
the known fact that the relative asymmetry of MG evolved density distributions is lower than in \lcdm\ \citep[see again][]{hellwing2013hierarchical,hellwing2017revealing}.

\subsection{Halos \& Galaxies}
\label{subsec:galaxies_and_halos}

The angular clustering patterns we observed for the dark matter, albeit very interesting,
cannot be easily nor directly translated into expectations for any observables that we
can extract from galaxy surveys. Some potential implications for weak gravitational lensing
could be drawn, but we keep this discussion for later (see sec.~\ref{sec.concl}).
On the other hand, the angular counts and related statistics of discrete objects, such as halos and galaxies,
have much more direct and straightforward connection and interpretation in the context of existing observational catalogs.
Thus we now move to the main part of our analysis and take a look at the hierarchical clustering of halos and galaxies.

\begin{figure*}
\centering
\includegraphics[width=\textwidth]
{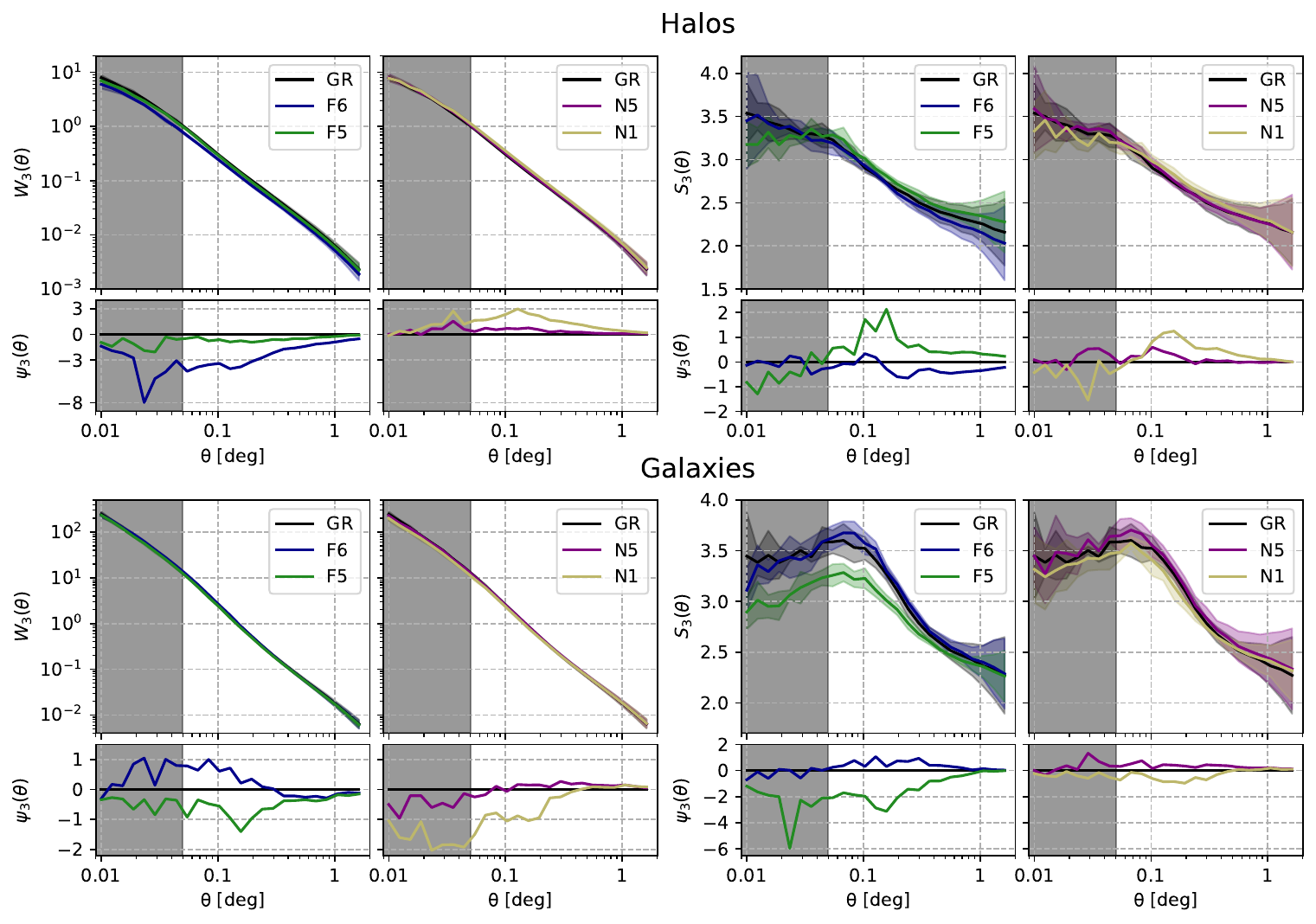}
\caption{Three-point averaged angular correlation function (columns to the left) and reduced skewness (columns to the right) for halos (top blocks of panels) and galaxies (bottom blocks) extracted from \elephant\ simulations for \lcdm\ and two MG scenarios. See caption of Fig.~\ref{fig.DM_W2} for further details.}
\label{fig.Common}
\end{figure*}

In Fig.~\ref{fig.Common} we summarize the third-order statistics ($W_3$, columns to the left; and $S_3$, to the right)
for halos (top blocks of panels) and galaxies (bottom blocks). We focus here on the third-order
statistics, since a detailed analysis for all higher-order moment would be unfeasible, and additionally the moments higher than fourth contain similar information to orders 3 and 4. We choose not to
present here and discuss separately the case for the angular variance $W_2$, as its amplitude
in general is fully degenerate with the first-order angular bias parameter, $b_\theta$:
\begin{equation}
    \label{eqn:angular_bias_w2}
    W_2^{h,g}(\theta)=\left(b_\theta^{h,g}\right)^2 W_2^{DM}(\theta)\,.
\end{equation}
Here, $(h,g)$ stands for halos and galaxies, respectively. In Ref.~\citep{garcia2021probing} it has been shown
that the second-order clustering statistics in MG are affected by this bias degeneracy. For that reason, higher-order
moments and their combinations (like skewness and kurtosis) may contain a more genuine MG signal.
The reason is that, to the first order, the bias degeneracy is reduced for them \citep[see also][]{Fry1993,hellwing2019skewness,alam2020testing}.

Let us first discuss the MG signal for the halo population.

As we have verified  for the case of DM density, the significance of the departure from the GR prediction is higher for $W_3$ alone, compared to the skewness. Again, this is expected given the standard error propagation properties. 
Focusing on the converged scales, \ie{} $\theta\simgt 0.05\degree$, 
we can observe a number of interesting features. Firstly, the weaker F6 variant is characterized by stronger deviations
from the GR case than F5. This might appear as a surprise at first, but can be explained. Although, in terms of the background field
value, the F6 variant should experience weaker scalar-field effects than F5, the former model is actually inherently more non-linear than the latter, 
in terms of the chameleon screening
behavior.
This property  manifests itself especially for the less massive, smaller halos (which actually dominate the sample). This phenomenon
was to some extent already encountered and studied in Ref.~\citep{Shi2015}. This trend is reversed when we look at the skewness, where again
F6 is marginally consistent with GR for the all scales, while F5 also shows small deviations, albeit larger than F6. 

\begin{figure*}
\centering
\includegraphics[width=0.8\textwidth]{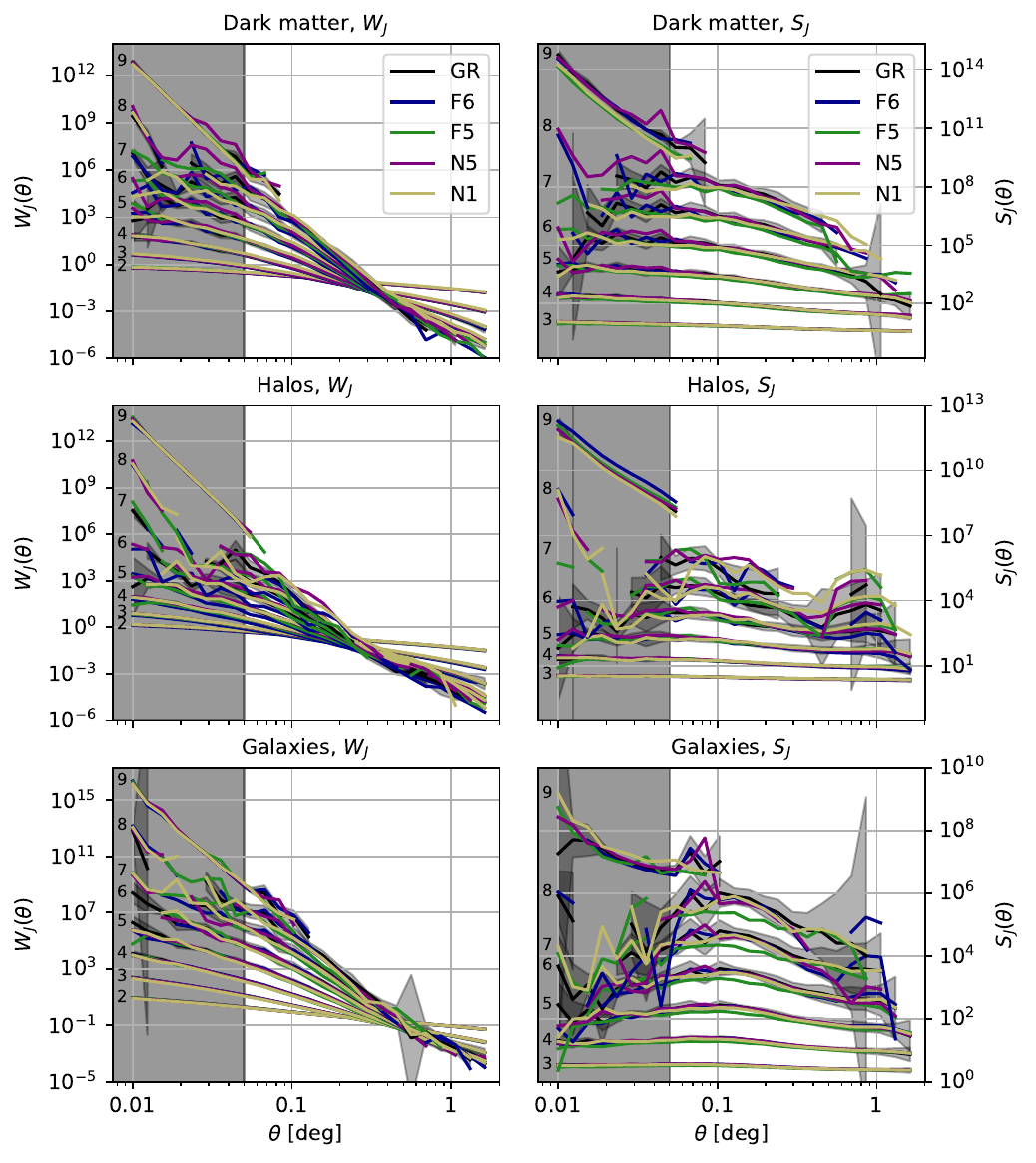}
\caption{Clustering of all orders considered in this work.
From top to bottom, we present the results for dark matter, halos, and galaxies.
Left column shows area-averaged correlation functions $W_J$
for $2\leq J \leq 9$; right column includes hierarchical amplitudes $S_J$ for $3\leq J \leq 9$. The particular orders are indicated with the numbered labels at the left-hand side of each panel. Lines of different colors correspond to the gravity models as in the legend. Light-gray bands illustrate the errors in the GR case; those for MG are comparable. The dark-gray vertical bands cover the angular scales which we do not use for inferring the MG signal.}
\label{fig.Common_combined}
\end{figure*}

In the case of the nDGP models,
we observe behavior that qualitatively agrees with what we have seen already for the variance in the dark matter field. The N1 deviation again assumes
a peaklike scale dependence, with the maximum significance attained at $\theta\simeq0.1\degree$, both in $W_3$ and $S_3$. The departure signal of the N5 variant stays
weak and insignificant for the all scales. The differences in scale and magnitude dependence between the $f(R)$ and nDGP models
clearly indicate that the different physics of their screening mechanisms manifests itself to some degree in different halo clustering.

For the galaxy population we encounter an interestingly different picture. Here we need to recall that the \elephant{}
mock galaxy catalogs were constructed using HOD parameters tuned for each MG model separately, so the resulting distributions have the same
(to within $1-2\%$) projected 2PCFs. Since the real-space 3D clustering was anchored, the higher-order moments
will carry here the genuine residual MG signals. This property of the mocks resulted in quite interesting trends we can single out
in the galaxy statistics. Firstly, we now see that signals for the F5 and F6 models have opposite signs of the effect on $W_3(\theta)$,
but now also F5 reaches comparable deviations in the magnitude as F6. For nDGP, we no longer observe a peaklike shape of the scale dependence,
but instead a saturation of significance at the level of $\sim1$ for N1 at $\theta\simlt0.2\degree$. For N1 the effect is also opposite to the one
we have just noted for halos.
The disappearance of peaklike scale dependence in galaxies for the
N1 scenario is related with the fact that our halo catalogs consist of only central halos
while for galaxies we also consider satellites. A related issue has been already discussed in ref.~\cite{alam2020testing}.

In the case of the skewness, the effect is also flipped (when compared to $W_3$) for F5 and N1, with a clear difference that now
for F5 the maximum signal is marginally stronger than for the halo population.

The remaining higher-order moments  and reduced cumulants up to $J=9$ 
reveal a qualitatively similar picture to
what we have just shown and discussed for the third order.
The general trends are continued, but the scatter in these quantities and associated
errors grow quickly with the order. To obtain a general impression of the trends, we show all the collected $W_J$'s and $S_J$'s
(for $J=2,\dots, 9$)
for dark matter, halos, and galaxies in Fig.~\ref{fig.Common_combined}. Here the top row is for the smooth DM density,
the central one for halos, and the bottom one for galaxies. Column-wise, the plots are organized as the area-averaged 
correlation functions $W_J$'s (left-hand column) and reduced cumulants $S_J$'s (right-hand column). The particular orders are organized as indicated by the labels, while the 
lines are colored according to gravity models in the same way as in Figs.\ \ref{fig.DM_W2}-\ref{fig.Common}.

For the cumulants, but also to some extent for the moments, we can observe that the relative errors
explode at two regimes: for small and large angles. This can be easily attributed to
the shot noise at $\theta\simlt0.01\degree$ scale, and to finite catalog size (\ie~cosmic variance) influencing
the $\theta\simgt2\degree$ range.
Higher moments are more sensitive to these effects,
due to growing powers in Eq.~\eqref{eq.moment}, which multiplies any initial error on the mean counts at a given scale, $\langle{N}\rangle_\theta$.
Such  conditions introduce significant heteroscedasticity into the clustering measurements.
The partly missing measurements in Fig.~\ref{fig.Common_combined},
especially for the eighth- and ninth-order statistics, are due to negative values that cannot be represented in the logarithmic scaling.

\begin{table*}
\caption{The most significant MG deviations from GR in angular clustering in the light cones studied in this work. For a given tracer, we provide the model-statistics pair (columns 2 \& 3) that give the largest signal significance as listed in the fourth column. The fifth column indicates the angular scales at which this maximal signal appears.
}

\begingroup
\setlength{\tabcolsep}{15pt}
\renewcommand{\arraystretch}{1.3}

\begin{tabular}{l c c c c}
\hline
\textbf{Tracer}&\textbf{Model}&\textbf{Statistics}&\textbf{$|\psi_{max}|$}&\textbf{$\theta\,[\deg]$}\\
\hline\hline
\multirow{3}{*}{}&F5&$W_2$&4.8&0.05\\
{Dark matter}&F5&$S_3$&4.5&0.08\\
{}&N1&$W_2$&7.3&0.13\\
\hline
\multirow{3}{*}{}&F6&$W_3$&4.0&0.13\\
{Halos}&F5&$S_3$&2.1&0.16\\
{}&N1&$W_3$&3.0&0.13\\
\hline
\multirow{2}{*}{Galaxies}&F5&$W_5$&2.5&0.16\\
{}&F5&$S_3$&3.1&0.16\\
\hline
\end{tabular}

\endgroup
\label{tab.allmgsignals}
\end{table*}

The precision of our measurements scales roughly as a square root of object number
density (e.g. the relative errors are $\sim1.5-2$ times bigger in the case of galaxies than for DM pseudo
particles catalog; see Section \ref{subsec:lightcone_depth}). This fact makes searching for MG signals
among highest orders inaccessible using these catalogs, since one would need to greatly increase
the sampling or to consider much larger sky coverage.
Overall, we have found that only orders of up-to $J\leq5$
are 
useful for MG signal searches in our catalogs.
Although the deviations from the fiducial GR case are growing with the order, 
the associated scatter
grows even faster. Thus, for the light cone samples used here,  
significant signals
can be extracted only from the statistics up to fifth order.

A noteworthy feature is that galaxies are characterized by larger values of CIC moments than
halos and DM particles. This is a natural consequence of both biased structure formation,
and the HOD catalog construction method used (\ie{} introduction of satellites for the galaxy sample).

One final note about the hierarchical clustering ratios,
shown in the right-hand panels of
Fig.~\ref{fig.Common_combined}, is that their
amplitudes generally do not depend strongly on the
scale. This
is in agreement with both theoretical predictions, as well as with the analyses of observational data
\citep[\eg][]{juszkiewicz1993skewness,gaztanaga1994high,ross2007higher,hellwing2013hierarchical}.

\subsection{Summary}
\label{subsec:summary}
With nine orders of moments, eight of cumulants, and with three different samples of five various gravity models, 
we are dealt with massive and complex data describing angular clustering in our light cones. Detailed analysis
of all potential deviations from GR and associated signal significance would 
be very industrious and at best cumbersome.
Therefore, we chose to present only the most interesting findings from 
this large body of statistics and we summarize them here in
a survey manner.

The most promising features of the MG angular clustering we analyzed are given in Table~\ref{tab.allmgsignals}.
There we quote the absolute values of the deviation from GR significance, $|\psi|$ (Eq.~\ref{eq.sig}), alongside the characteristic angular scales
at which they are noted, the statistics for which the signal is found, and the MG model they are reported for. 

As we have already indicated, the sampling density of \elephant{} allows us to reliably use
up to fifth order statistics, as far as a ratio of MG to GR is concerned. The higher orders become too
noisy, and the associated ratios w.r.t.\ the GR case become scatter-dominated.
When we are concerned with the MG signal significance, the general trends are that the lower orders statistics (\ie{} $J=2,3$) are favored
over the higher ones. However, in just one case (F5 for galaxies) the fourth and fifth orders reach higher significance than the lower moments.
Here, the signal significance  is reaching $|\psi_{3,4,5}|\simeq \{1.4, 2.4, 2.5\}$ for the third, fourth and fifth order, respectively.
Interestingly, we find cases where even if some $W_J$'s reach small values of $|\psi|$,
the associated reduced cumulants $S_J$'s can arrive at much larger significance.
This indicates that the reduced cumulants, due to their unique intrinsic length (or variance) scaling,
contain extra constraining information about the underlying structure formation models.

The statistics we measured for the nDGP models reveal that this MG family
is characterized by much less significant deviations from the fiducial GR case than $f(R)$. Especially if we focus on the hierarchical amplitudes, $S_J$'s.
This would suggest that departures from  GR at higher orders do not differ significantly from the departures at $J=2$.
Indeed, for instance considering DM particles we obtain $|\psi_{2,3}|\simeq\{ 7.3, 6.4\}$
for $N1$ while $F5$ provides $|\psi_{2,3}|\simeq\{4.8, 1.8\}$.
Generally we find that the hierarchical amplitudes of the galaxy samples offer a slightly better sensitivity to deviations from GR than halos and DM.
In contrast, for the halo sample, we find that the area-averaged correlation functions, $W_J$'s, seem to perform marginally better
in differentiating the models.

Finally, our results facilitate a general trend, where the MG models with larger theoretical growth-rate departure from the \lcdm{} case
are characterized by stronger angular clustering deviations as well. There is one notable exception to that trend: the F6 model, whose halo sample offers larger $|\psi|$ than that of the F5 variant, while the latter has theoretically larger growth rate.
This is not a complete surprise, however, as previous studies already found evidence that the F6 model can exhibit more non-linear
behavior than its F5 cousin, owing to the intrinsic non-linear nature of the chameleon screening \citep{Shi2015,alam2020testing}.

\section{Concluding remarks}\label{sec.concl}

In this work we have studied angular clustering by analyzing  the moments of the counts-in-cells for two modified gravity scenarios. 
The literature offers many studies of  three-dimensional clustering,
including  redshift space distortions analyses for such beyond-GR scenarios \citep{Hamilton1998,Arnalte-Mur2017,arnold2019modified,Hernandez-Aguayo2019,garcia2021probing}.
So far, however, very little (to our best knowledge)
was known about the properties of \textit{angular} clustering in the light cone sky-projected density in such scenarios. 
Our work here is the first approach to remedy this lacking. To this end, based on the \elephant{} suite, we have
designed a number of light cones containing dark matter, halo, and galaxy samples. We then proceeded to build an ensemble
of effective sky catalogs that we have used as the main object for our analysis. 

Below, we summarize and recapture the most important results:
\begin{itemize}
    \item The PT predictions for the GR $W_2$ and $W_3$ moments offer a reasonable agreement with the $N$-body dark matter results
    for $\theta\geq0.2\degree$, if only the linear-theory power spectrum was used. A much better agreement is obtained, down to smaller scales, when using a non-linear \textsc{Halofit} model for the power spectrum. For that case the PT and simulations
    agree down to $\theta\sim0.05\degree$. We note, however, that the PT prediction for the reduced skewness, $S_3$, is grossly underestimated,
    when compared with N-body data for $\theta\leq0.5\degree$.
    \item Our light cone  analysis yielded an optimal catalog depth for maximizing the departures from GR in the angular clustering.
    For the case of our simulations and models, this turned-out to be $0.15\leq z\leq0.3$.
    \item On various scales and for various statistics we found up to $20\%$ relative departures from  GR.
     Our data do not allow for probing robustly the scales both smaller than $\theta <0.05\degree$ and larger than $\theta>1\degree$, where
     the effects of the shot noise, and a limited catalog size, respectively dominate.
    \item The reduced skewness, $S_3$, of the galaxy sample has proven to be especially sensitive statistics for the $f(R)$ family models.
    \item We found significant signals even in the catalogs with as low object number density as 15 $deg^{-2}$ (for galaxies); this
    indicates an optimistic outlook for measuring MG signals in real angular galaxy data.
    \item Hierarchy between the second and higher moments is preserved in all the structure formation scenarios,
    with no clear or dramatic changes in weak scale-dependence of the reduced cumulants for all scenarios.
    \item In our data the modified gravity signals can be extracted from higher order statistics up to order $J=5$. In practice,
    we can expect that catalogs with better sampling, in terms of both sky coverage and surface number density, should
    allow measurements that cover even higher orders and larger scales.
\end{itemize}

Our main findings agree with the picture where the angular correlations, due to their 
intrinsic spatial-scale mixing, offer a unique specific
window for clustering analysis, especially in the context of scale-dependent GR modifications. In general, the deviations from GR might not get
as large for angular correlations as in the case of redshift-space distortions and 3D clustering. However, 
the projected counts could gain much in signal significance,
if one could tap the rich potential of much denser sampling stemming from usually many times 
bigger
volumes of photometric galaxy catalogs than for often sparsely sampled redshift surveys. 

Taking into account that, due to the limitations of the simulation suite used, our galaxy and halo catalogs are characterized by much smaller object number densities when
compared to existing and forthcoming imaging sky surveys
\cite{ivezic2009lsst,de2017third,tutusaus2020euclid,abbott2021dark,walmsley2022galaxy},
one can expect that  all the relevant shot-noise and even cosmic-variance effects should be
strongly suppressed in future analysis of observational data. Our mock galaxy 
catalogs contain only $\sim 15$ objects per square degree. 
For comparison, depending on the chosen galaxy sample, the Dark Energy Survey (DES) provides $\sim3.42\times 10^5$ galaxies
at $\sim4200 \deg^2$ for the redshift range similar to ours, $0.15<z<0.35$ \cite{porredon2021dark},
or even $\sim1.7\times 10^6$ galaxies at $0.2<z<0.35$. This gives from 5 up to 27 times higher galaxy surface density 
when confronted with our galaxy catalogs.

The fact that we have found significant, and therefore hopefully detectable, deviations from GR 
in our rather sparse mock galaxy catalogs, which have many 
times smaller object number density 
when compared to
real galaxy samples, offers very promising prospects for 
testing GR and beyond-GR structure formation scenarios in the imaging data  -- an avenue that has not been exploited so far.
However, to fully undertake such endeavor, one will need to account more robustly
for the involved angular and redshift selection effects along with
better galaxy population modeling in beyond-GR. We leave this exciting undertaking for future work.

\acknowledgments
The Authors would like to acknowledge the fruitful discussions with Enrique Gaztañaga at early stages of this project.
This work is supported by National Science Center, Poland under Agreements No.\ 2018/30/E/ST9/00698, No. 2018/31/G/ST9/03388, No.  2020/38/E/ST9/00395,  and No. 2020/39/B/ST9/03494, and by the Polish Ministry of Science and Higher Education through Grant No. DIR/WK/2018/12.
The research presented here also benefited from supercomputer calculations performed at 
the Interdisciplinary Center for Mathematical and Computational Modeling (University of Warsaw) 
under Projects No. GA67-17, No. GA67-16, and No. GB79-7.

\appendix*
\section{Effective volume}
\label{sec.appendix}

The so-called effective volume measures the amount of independent information in the catalog.
As mentioned in Section \ref{sec.models_sims}, calculating and maximizing the effective volume
is critical for dealing with the issue of finite simulation box size.

We pixelize 
each replicated box
(by dividing it into $30^3$ cubic pixels)
and then for each pixel we check
whether its center belongs to the catalog. If this condition is fulfilled, the pixel contributes to $V_\mathrm{eff}$.

The partial effective volume $V_{\mathrm{eff},sh[i]}$ (i.e.  corresponding to the $i$-th shell) is then
calculated from:
\begin{equation}
    V_{\mathrm{eff},sh[i]}=\frac{n_\mathrm{pix}^{(\mathrm{good})}}{n_\mathrm{pix}},
\end{equation}
where $n_\mathrm{pix}$ is the number of pixels that the simulation box was divided into and $n_\mathrm{pix}^{(\mathrm{good})}$ counts the
pixels which fulfill the aforementioned condition. Next, we obtain complete effective volume by summing up
the values from all the shells:
\begin{equation}
    V_\mathrm{eff}=\sum_{i=0}^{n_\mathrm{shells}} V_{\mathrm{eff},sh[i]}.
\end{equation}

\begin{figure}[t!]
\centering
\includegraphics[width=\columnwidth]
{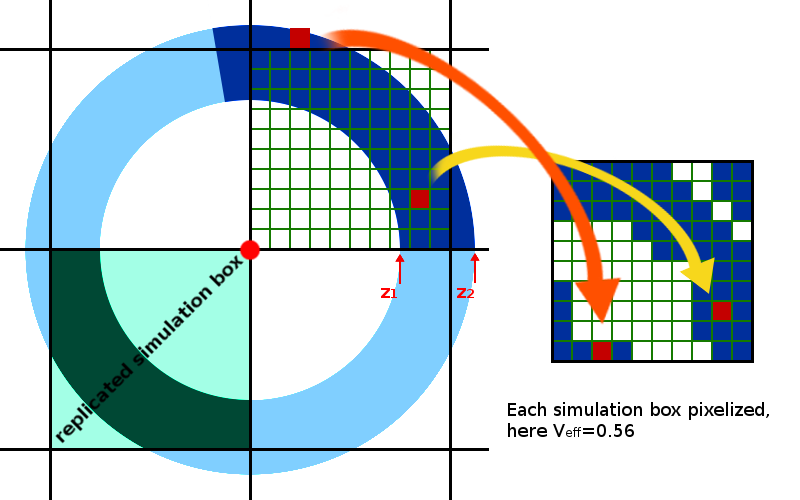}
\caption{Illustration of our method to calculate the effective volume. For clarity we visualize only one
light cone shell $[z_1,z_2]$, included within the catalog
ranges $[z_\mathrm{min},z_\mathrm{max}]$.
}
\label{fig.Veff}
\end{figure}

Fig. \ref{fig.Veff} illustrates briefly our method of calculating $V_\mathrm{eff}$ for just one
light cone shell considered ($[z_1, z_2]$ in the plot).
For simplicity the figure shows the two-dimensional case, but of course we perform the procedure in 3D space.

The left part of the figure symbolizes the light cone with
the redshift cut
(light blue) and also angular cut (dark blue). We marked the observer's position with the red dot
in the center of light cone.  The right part visualizes the counts which would result from the
situation presented in this figure. As an example, we distinguished two pixels.
Note that we do not include two different pixels twice if they occupy the same
position within the output box.

Focusing on the redshift range that we identified as optimal for our study ($0.15 \lesssim z \lesssim 0.30$), 
we maximize $V_\mathrm{eff}$ by drawing
$n_{run}=3000$ 
times a set of 15 
randomly oriented
sky chunks covering $\sim1567$ deg$^2$ each, as described in Sec.\ \ref{sec.models_sims}.
Such numerous collection of sky fragments enabled us to obtain satisfying level of $V_{eff}$ even from such
thin redshift range in the catalog.
From these we chose one set which provides the highest $V_\mathrm{eff}$ value.
We noticed that used $n_{run}$ value is sufficient due to the saturation of largest $V_{eff}$ found.

\begin{figure}[t!]
\vspace{4ex}
\centering
\includegraphics[width=\columnwidth]
{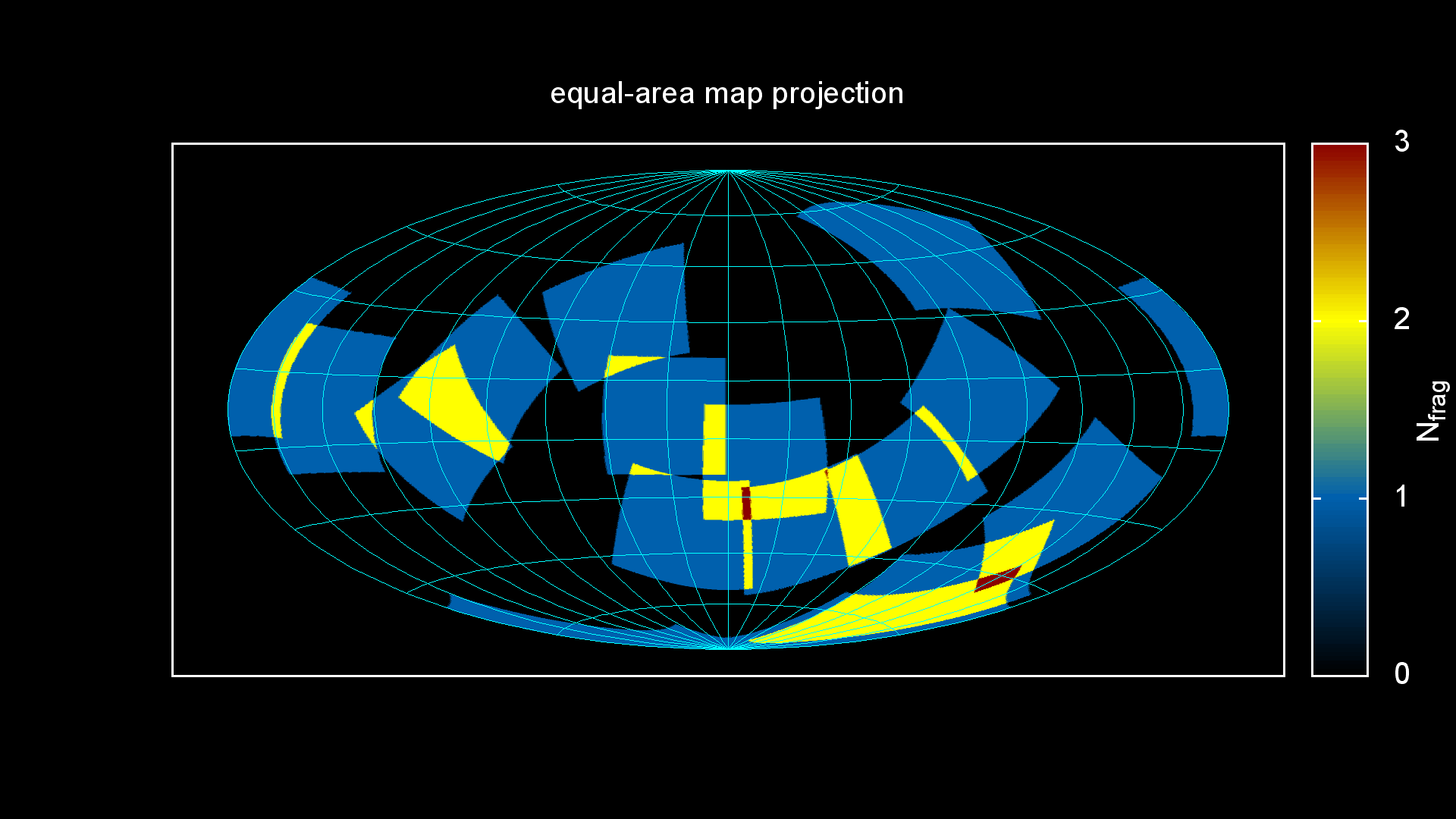}
\caption{Locations of sky fragments jointly maximizing the $V_{eff}$. The $N_{frag}$ value
refers to the number of sky chunks overlapping at certain sky position.}
\label{fig.Veffmax}
\end{figure}

Fig. \ref{fig.Veffmax} shows the positions of sky chunks used in our catalogs
within this work, in Hammer equal-area projection.
Our procedure allowed us to obtain $V_\mathrm{eff}=0.75$. Theoretically, it is possible to obtain
$V_{eff}$ as high as $0.94$ with our redshift ranges. However, it requires a drastic increase
in the number of sky chunks
and becomes computationally ineffective, providing simultaneously a weak increase of independent information in the
catalog.
Note that the value of effective volume
is not limited to 1 by definition. The value $V_\mathrm{eff}=1$ would indicate that potentially constructed catalog
contains all the information from one simulation box, \ie one light cone shell.
Getting $V_{eff}>1$ could be easy achieved e.g. by selecting entire sky and full redshift range of $ELEPHANT$
light cone $z \in [0,0.5]$ -- for that case we would get $V_{eff}\sim2.23$.

\bibliography{baza}

\providecommand{\noopsort}[1]{}\providecommand{\singleletter}[1]{#1}%
\begin{thebibliography}{86}%
\makeatletter
\providecommand \@ifxundefined [1]{%
 \@ifx{#1\undefined}
}%
\providecommand \@ifnum [1]{%
 \ifnum #1\expandafter \@firstoftwo
 \else \expandafter \@secondoftwo
 \fi
}%
\providecommand \@ifx [1]{%
 \ifx #1\expandafter \@firstoftwo
 \else \expandafter \@secondoftwo
 \fi
}%
\providecommand \natexlab [1]{#1}%
\providecommand \enquote  [1]{``#1''}%
\providecommand \bibnamefont  [1]{#1}%
\providecommand \bibfnamefont [1]{#1}%
\providecommand \citenamefont [1]{#1}%
\providecommand \href@noop [0]{\@secondoftwo}%
\providecommand \href [0]{\begingroup \@sanitize@url \@href}%
\providecommand \@href[1]{\@@startlink{#1}\@@href}%
\providecommand \@@href[1]{\endgroup#1\@@endlink}%
\providecommand \@sanitize@url [0]{\catcode `\\12\catcode `\$12\catcode
  `\&12\catcode `\#12\catcode `\^12\catcode `\_12\catcode `\%12\relax}%
\providecommand \@@startlink[1]{}%
\providecommand \@@endlink[0]{}%
\providecommand \url  [0]{\begingroup\@sanitize@url \@url }%
\providecommand \@url [1]{\endgroup\@href {#1}{\urlprefix }}%
\providecommand \urlprefix  [0]{URL }%
\providecommand \Eprint [0]{\href }%
\providecommand \doibase [0]{http://dx.doi.org/}%
\providecommand \selectlanguage [0]{\@gobble}%
\providecommand \bibinfo  [0]{\@secondoftwo}%
\providecommand \bibfield  [0]{\@secondoftwo}%
\providecommand \translation [1]{[#1]}%
\providecommand \BibitemOpen [0]{}%
\providecommand \bibitemStop [0]{}%
\providecommand \bibitemNoStop [0]{.\EOS\space}%
\providecommand \EOS [0]{\spacefactor3000\relax}%
\providecommand \BibitemShut  [1]{\csname bibitem#1\endcsname}%
\let\auto@bib@innerbib\@empty
\bibitem [{\citenamefont {{Planck Collaboration}}\ \emph
  {et~al.}(2020)\citenamefont {{Planck Collaboration}}, \citenamefont
  {{Aghanim}}, \citenamefont {{Akrami}}, \citenamefont {{Ashdown}},
  \citenamefont {{Aumont}}, \citenamefont {{Baccigalupi}}, \citenamefont
  {{Ballardini}}, \citenamefont {{Banday}}, \citenamefont {{Barreiro}},
  \citenamefont {{Bartolo}}, \citenamefont {{Basak}}, \citenamefont {{Battye}},
  \citenamefont {{Benabed}}, \citenamefont {{Bernard}}, \citenamefont
  {{Bersanelli}}, \citenamefont {{Bielewicz}} \emph {et~al.}}]{Planck2020}%
  \BibitemOpen
  \bibfield  {author} {\bibinfo {author} {\bibnamefont {{Planck
  Collaboration}}}, \bibinfo {author} {\bibfnamefont {N.}~\bibnamefont
  {{Aghanim}}}, \bibinfo {author} {\bibfnamefont {Y.}~\bibnamefont {{Akrami}}},
  \bibinfo {author} {\bibfnamefont {M.}~\bibnamefont {{Ashdown}}}, \bibinfo
  {author} {\bibfnamefont {J.}~\bibnamefont {{Aumont}}}, \bibinfo {author}
  {\bibfnamefont {C.}~\bibnamefont {{Baccigalupi}}}, \bibinfo {author}
  {\bibfnamefont {M.}~\bibnamefont {{Ballardini}}}, \bibinfo {author}
  {\bibfnamefont {A.~J.}\ \bibnamefont {{Banday}}}, \bibinfo {author}
  {\bibfnamefont {R.~B.}\ \bibnamefont {{Barreiro}}}, \bibinfo {author}
  {\bibfnamefont {N.}~\bibnamefont {{Bartolo}}}, \bibinfo {author}
  {\bibfnamefont {S.}~\bibnamefont {{Basak}}}, \bibinfo {author} {\bibfnamefont
  {R.}~\bibnamefont {{Battye}}}, \bibinfo {author} {\bibfnamefont
  {K.}~\bibnamefont {{Benabed}}}, \bibinfo {author} {\bibfnamefont {J.~P.}\
  \bibnamefont {{Bernard}}}, \bibinfo {author} {\bibfnamefont {M.}~\bibnamefont
  {{Bersanelli}}}, \bibinfo {author} {\bibnamefont {{Bielewicz}}},  \emph
  {et~al.},\ }\href {\doibase 10.1051/0004-6361/201833910} {\bibfield
  {journal} {\bibinfo  {journal} {\aap}\ }\textbf {\bibinfo {volume} {641}},\
  \bibinfo {eid} {A6} (\bibinfo {year} {2020})},\ \Eprint
  {http://arxiv.org/abs/1807.06209} {arXiv:1807.06209 [astro-ph.CO]}
  \BibitemShut {NoStop}%
\bibitem [{\citenamefont {{Alam}}\ \emph {et~al.}(2021)\citenamefont {{Alam}},
  \citenamefont {{Aubert}}, \citenamefont {{Avila}}, \citenamefont {{Balland}},
  \citenamefont {{Bautista}}, \citenamefont {{Bershady}}, \citenamefont
  {{Bizyaev}}, \citenamefont {{Blanton}}, \citenamefont {{Bolton}},
  \citenamefont {{Bovy}}, \citenamefont {{Brinkmann}}, \citenamefont
  {{Brownstein}}, \citenamefont {{Burtin}}, \citenamefont {{Chabanier}},
  \citenamefont {{Chapman}} \emph {et~al.}}]{Alam2021}%
  \BibitemOpen
  \bibfield  {author} {\bibinfo {author} {\bibfnamefont {S.}~\bibnamefont
  {{Alam}}}, \bibinfo {author} {\bibfnamefont {M.}~\bibnamefont {{Aubert}}},
  \bibinfo {author} {\bibfnamefont {S.}~\bibnamefont {{Avila}}}, \bibinfo
  {author} {\bibfnamefont {C.}~\bibnamefont {{Balland}}}, \bibinfo {author}
  {\bibfnamefont {J.~E.}\ \bibnamefont {{Bautista}}}, \bibinfo {author}
  {\bibfnamefont {M.~A.}\ \bibnamefont {{Bershady}}}, \bibinfo {author}
  {\bibfnamefont {D.}~\bibnamefont {{Bizyaev}}}, \bibinfo {author}
  {\bibfnamefont {M.~R.}\ \bibnamefont {{Blanton}}}, \bibinfo {author}
  {\bibfnamefont {A.~S.}\ \bibnamefont {{Bolton}}}, \bibinfo {author}
  {\bibfnamefont {J.}~\bibnamefont {{Bovy}}}, \bibinfo {author} {\bibfnamefont
  {J.}~\bibnamefont {{Brinkmann}}}, \bibinfo {author} {\bibfnamefont {J.~R.}\
  \bibnamefont {{Brownstein}}}, \bibinfo {author} {\bibfnamefont
  {E.}~\bibnamefont {{Burtin}}}, \bibinfo {author} {\bibfnamefont
  {S.}~\bibnamefont {{Chabanier}}}, \bibinfo {author} {\bibfnamefont {M.~J.}\
  \bibnamefont {{Chapman}}},  \emph {et~al.},\ }\href {\doibase
  10.1103/PhysRevD.103.083533} {\bibfield  {journal} {\bibinfo  {journal}
  {\prd}\ }\textbf {\bibinfo {volume} {103}},\ \bibinfo {eid} {083533}
  (\bibinfo {year} {2021})},\ \Eprint {http://arxiv.org/abs/2007.08991}
  {arXiv:2007.08991 [astro-ph.CO]} \BibitemShut {NoStop}%
\bibitem [{\citenamefont {{Abbott}}\ \emph {et~al.}(2022)\citenamefont
  {{Abbott}}, \citenamefont {{Aguena}}, \citenamefont {{Alarcon}},
  \citenamefont {{Allam}}, \citenamefont {{Alves}}, \citenamefont {{Amon}},
  \citenamefont {{Andrade-Oliveira}}, \citenamefont {{Annis}}, \citenamefont
  {{Avila}}, \citenamefont {{Bacon}}, \citenamefont {{Baxter}}, \citenamefont
  {{Bechtol}}, \citenamefont {{Becker}}, \citenamefont {{Bernstein}},
  \citenamefont {{Bhargava}},\ and\ \citenamefont {{DES
  Collaboration}}}]{Abbott2022}%
  \BibitemOpen
  \bibfield  {author} {\bibinfo {author} {\bibfnamefont {T.~M.~C.}\
  \bibnamefont {{Abbott}}}, \bibinfo {author} {\bibfnamefont {M.}~\bibnamefont
  {{Aguena}}}, \bibinfo {author} {\bibfnamefont {A.}~\bibnamefont {{Alarcon}}},
  \bibinfo {author} {\bibfnamefont {S.}~\bibnamefont {{Allam}}}, \bibinfo
  {author} {\bibfnamefont {O.}~\bibnamefont {{Alves}}}, \bibinfo {author}
  {\bibfnamefont {A.}~\bibnamefont {{Amon}}}, \bibinfo {author} {\bibfnamefont
  {F.}~\bibnamefont {{Andrade-Oliveira}}}, \bibinfo {author} {\bibfnamefont
  {J.}~\bibnamefont {{Annis}}}, \bibinfo {author} {\bibfnamefont
  {S.}~\bibnamefont {{Avila}}}, \bibinfo {author} {\bibfnamefont
  {D.}~\bibnamefont {{Bacon}}}, \bibinfo {author} {\bibfnamefont
  {E.}~\bibnamefont {{Baxter}}}, \bibinfo {author} {\bibfnamefont
  {K.}~\bibnamefont {{Bechtol}}}, \bibinfo {author} {\bibfnamefont {M.~R.}\
  \bibnamefont {{Becker}}}, \bibinfo {author} {\bibfnamefont {G.~M.}\
  \bibnamefont {{Bernstein}}}, \bibinfo {author} {\bibfnamefont
  {S.}~\bibnamefont {{Bhargava}}}, \ and\ \bibinfo {author} {\bibnamefont {{DES
  Collaboration}}},\ }\href {\doibase 10.1103/PhysRevD.105.023520} {\bibfield
  {journal} {\bibinfo  {journal} {\prd}\ }\textbf {\bibinfo {volume} {105}},\
  \bibinfo {eid} {023520} (\bibinfo {year} {2022})},\ \Eprint
  {http://arxiv.org/abs/2105.13549} {arXiv:2105.13549 [astro-ph.CO]}
  \BibitemShut {NoStop}%
\bibitem [{\citenamefont {{Brout}}\ \emph {et~al.}(2022)\citenamefont
  {{Brout}}, \citenamefont {{Scolnic}}, \citenamefont {{Popovic}},
  \citenamefont {{Riess}}, \citenamefont {{Zuntz}}, \citenamefont {{Kessler}},
  \citenamefont {{Carr}}, \citenamefont {{Davis}}, \citenamefont {{Hinton}},
  \citenamefont {{Jones}}, \citenamefont {{Kenworthy}}, \citenamefont
  {{Peterson}}, \citenamefont {{Said}}, \citenamefont {{Taylor}}, \citenamefont
  {{Ali}} \emph {et~al.}}]{Brout2022}%
  \BibitemOpen
  \bibfield  {author} {\bibinfo {author} {\bibfnamefont {D.}~\bibnamefont
  {{Brout}}}, \bibinfo {author} {\bibfnamefont {D.}~\bibnamefont {{Scolnic}}},
  \bibinfo {author} {\bibfnamefont {B.}~\bibnamefont {{Popovic}}}, \bibinfo
  {author} {\bibfnamefont {A.~G.}\ \bibnamefont {{Riess}}}, \bibinfo {author}
  {\bibfnamefont {J.}~\bibnamefont {{Zuntz}}}, \bibinfo {author} {\bibfnamefont
  {R.}~\bibnamefont {{Kessler}}}, \bibinfo {author} {\bibfnamefont
  {A.}~\bibnamefont {{Carr}}}, \bibinfo {author} {\bibfnamefont {T.~M.}\
  \bibnamefont {{Davis}}}, \bibinfo {author} {\bibfnamefont {S.}~\bibnamefont
  {{Hinton}}}, \bibinfo {author} {\bibfnamefont {D.}~\bibnamefont {{Jones}}},
  \bibinfo {author} {\bibfnamefont {W.~D.}\ \bibnamefont {{Kenworthy}}},
  \bibinfo {author} {\bibfnamefont {E.~R.}\ \bibnamefont {{Peterson}}},
  \bibinfo {author} {\bibfnamefont {K.}~\bibnamefont {{Said}}}, \bibinfo
  {author} {\bibfnamefont {G.}~\bibnamefont {{Taylor}}}, \bibinfo {author}
  {\bibfnamefont {N.}~\bibnamefont {{Ali}}},  \emph {et~al.},\ }\href@noop {}
  {\bibfield  {journal} {\bibinfo  {journal} {arXiv e-prints}\ ,\ \bibinfo
  {eid} {arXiv:2202.04077}} (\bibinfo {year} {2022})},\ \Eprint
  {http://arxiv.org/abs/2202.04077} {arXiv:2202.04077 [astro-ph.CO]}
  \BibitemShut {NoStop}%
\bibitem [{\citenamefont {{Peebles}}(1980)}]{peebles1980large}%
  \BibitemOpen
  \bibfield  {author} {\bibinfo {author} {\bibfnamefont {P.~J.~E.}\
  \bibnamefont {{Peebles}}},\ }\href@noop {} {\emph {\bibinfo {title} {{The
  large-scale structure of the universe}}}}\ (\bibinfo {year}
  {1980})\BibitemShut {NoStop}%
\bibitem [{\citenamefont {{Bond}}\ \emph {et~al.}(1996)\citenamefont {{Bond}},
  \citenamefont {{Kofman}},\ and\ \citenamefont
  {{Pogosyan}}}]{bond1996filaments}%
  \BibitemOpen
  \bibfield  {author} {\bibinfo {author} {\bibfnamefont {J.~R.}\ \bibnamefont
  {{Bond}}}, \bibinfo {author} {\bibfnamefont {L.}~\bibnamefont {{Kofman}}}, \
  and\ \bibinfo {author} {\bibfnamefont {D.}~\bibnamefont {{Pogosyan}}},\
  }\href {\doibase 10.1038/380603a0} {\bibfield  {journal} {\bibinfo  {journal}
  {\nat}\ }\textbf {\bibinfo {volume} {380}},\ \bibinfo {pages} {603} (\bibinfo
  {year} {1996})},\ \Eprint {http://arxiv.org/abs/astro-ph/9512141}
  {arXiv:astro-ph/9512141 [astro-ph]} \BibitemShut {NoStop}%
\bibitem [{\citenamefont {{Bernardeau}}\ \emph {et~al.}(2002)\citenamefont
  {{Bernardeau}}, \citenamefont {{Colombi}}, \citenamefont {{Gazta{\~n}aga}},\
  and\ \citenamefont {{Scoccimarro}}}]{bernardeau2002large}%
  \BibitemOpen
  \bibfield  {author} {\bibinfo {author} {\bibfnamefont {F.}~\bibnamefont
  {{Bernardeau}}}, \bibinfo {author} {\bibfnamefont {S.}~\bibnamefont
  {{Colombi}}}, \bibinfo {author} {\bibfnamefont {E.}~\bibnamefont
  {{Gazta{\~n}aga}}}, \ and\ \bibinfo {author} {\bibfnamefont {R.}~\bibnamefont
  {{Scoccimarro}}},\ }\href {\doibase 10.1016/S0370-1573(02)00135-7} {\bibfield
   {journal} {\bibinfo  {journal} {\physrep}\ }\textbf {\bibinfo {volume}
  {367}},\ \bibinfo {pages} {1} (\bibinfo {year} {2002})},\ \Eprint
  {http://arxiv.org/abs/astro-ph/0112551} {arXiv:astro-ph/0112551 [astro-ph]}
  \BibitemShut {NoStop}%
\bibitem [{\citenamefont {{Hawkins}}\ \emph {et~al.}(2003)\citenamefont
  {{Hawkins}}, \citenamefont {{Maddox}}, \citenamefont {{Cole}}, \citenamefont
  {{Lahav}}, \citenamefont {{Madgwick}}, \citenamefont {{Norberg}},
  \citenamefont {{Peacock}}, \citenamefont {{Baldry}}, \citenamefont {{Baugh}},
  \citenamefont {{Bland-Hawthorn}}, \citenamefont {{Bridges}}, \citenamefont
  {{Cannon}}, \citenamefont {{Colless}}, \citenamefont {{Collins}},
  \citenamefont {{Couch}}, \citenamefont {{Dalton}}, \citenamefont {{De
  Propris}}, \citenamefont {{Driver}}, \citenamefont {{Efstathiou}},
  \citenamefont {{Ellis}}, \citenamefont {{Frenk}}, \citenamefont
  {{Glazebrook}}, \citenamefont {{Jackson}}, \citenamefont {{Jones}},
  \citenamefont {{Lewis}}, \citenamefont {{Lumsden}}, \citenamefont
  {{Percival}}, \citenamefont {{Peterson}}, \citenamefont {{Sutherland}},\ and\
  \citenamefont {{Taylor}}}]{hawkins20032df}%
  \BibitemOpen
  \bibfield  {author} {\bibinfo {author} {\bibfnamefont {E.}~\bibnamefont
  {{Hawkins}}}, \bibinfo {author} {\bibfnamefont {S.}~\bibnamefont {{Maddox}}},
  \bibinfo {author} {\bibfnamefont {S.}~\bibnamefont {{Cole}}}, \bibinfo
  {author} {\bibfnamefont {O.}~\bibnamefont {{Lahav}}}, \bibinfo {author}
  {\bibfnamefont {D.~S.}\ \bibnamefont {{Madgwick}}}, \bibinfo {author}
  {\bibfnamefont {P.}~\bibnamefont {{Norberg}}}, \bibinfo {author}
  {\bibfnamefont {J.~A.}\ \bibnamefont {{Peacock}}}, \bibinfo {author}
  {\bibfnamefont {I.~K.}\ \bibnamefont {{Baldry}}}, \bibinfo {author}
  {\bibfnamefont {C.~M.}\ \bibnamefont {{Baugh}}}, \bibinfo {author}
  {\bibfnamefont {J.}~\bibnamefont {{Bland-Hawthorn}}}, \bibinfo {author}
  {\bibfnamefont {T.}~\bibnamefont {{Bridges}}}, \bibinfo {author}
  {\bibfnamefont {R.}~\bibnamefont {{Cannon}}}, \bibinfo {author}
  {\bibfnamefont {M.}~\bibnamefont {{Colless}}}, \bibinfo {author}
  {\bibfnamefont {C.}~\bibnamefont {{Collins}}}, \bibinfo {author}
  {\bibfnamefont {W.}~\bibnamefont {{Couch}}}, \bibinfo {author} {\bibfnamefont
  {G.}~\bibnamefont {{Dalton}}}, \bibinfo {author} {\bibfnamefont
  {R.}~\bibnamefont {{De Propris}}}, \bibinfo {author} {\bibfnamefont {S.~P.}\
  \bibnamefont {{Driver}}}, \bibinfo {author} {\bibfnamefont {G.}~\bibnamefont
  {{Efstathiou}}}, \bibinfo {author} {\bibfnamefont {R.~S.}\ \bibnamefont
  {{Ellis}}}, \bibinfo {author} {\bibfnamefont {C.~S.}\ \bibnamefont
  {{Frenk}}}, \bibinfo {author} {\bibfnamefont {K.}~\bibnamefont
  {{Glazebrook}}}, \bibinfo {author} {\bibfnamefont {C.}~\bibnamefont
  {{Jackson}}}, \bibinfo {author} {\bibfnamefont {B.}~\bibnamefont {{Jones}}},
  \bibinfo {author} {\bibfnamefont {I.}~\bibnamefont {{Lewis}}}, \bibinfo
  {author} {\bibfnamefont {S.}~\bibnamefont {{Lumsden}}}, \bibinfo {author}
  {\bibfnamefont {W.}~\bibnamefont {{Percival}}}, \bibinfo {author}
  {\bibfnamefont {B.~A.}\ \bibnamefont {{Peterson}}}, \bibinfo {author}
  {\bibfnamefont {W.}~\bibnamefont {{Sutherland}}}, \ and\ \bibinfo {author}
  {\bibfnamefont {K.}~\bibnamefont {{Taylor}}},\ }\href {\doibase
  10.1046/j.1365-2966.2003.07063.x} {\bibfield  {journal} {\bibinfo  {journal}
  {\mnras}\ }\textbf {\bibinfo {volume} {346}},\ \bibinfo {pages} {78}
  (\bibinfo {year} {2003})},\ \Eprint {http://arxiv.org/abs/astro-ph/0212375}
  {arXiv:astro-ph/0212375 [astro-ph]} \BibitemShut {NoStop}%
\bibitem [{\citenamefont {{Zehavi}}\ \emph {et~al.}(2011)\citenamefont
  {{Zehavi}}, \citenamefont {{Zheng}}, \citenamefont {{Weinberg}},
  \citenamefont {{Blanton}}, \citenamefont {{Bahcall}}, \citenamefont
  {{Berlind}}, \citenamefont {{Brinkmann}}, \citenamefont {{Frieman}},
  \citenamefont {{Gunn}}, \citenamefont {{Lupton}}, \citenamefont {{Nichol}},
  \citenamefont {{Percival}}, \citenamefont {{Schneider}}, \citenamefont
  {{Skibba}}, \citenamefont {{Strauss}} \emph {et~al.}}]{zehavi2011galaxy}%
  \BibitemOpen
  \bibfield  {author} {\bibinfo {author} {\bibfnamefont {I.}~\bibnamefont
  {{Zehavi}}}, \bibinfo {author} {\bibfnamefont {Z.}~\bibnamefont {{Zheng}}},
  \bibinfo {author} {\bibfnamefont {D.~H.}\ \bibnamefont {{Weinberg}}},
  \bibinfo {author} {\bibfnamefont {M.~R.}\ \bibnamefont {{Blanton}}}, \bibinfo
  {author} {\bibfnamefont {N.~A.}\ \bibnamefont {{Bahcall}}}, \bibinfo {author}
  {\bibfnamefont {A.~A.}\ \bibnamefont {{Berlind}}}, \bibinfo {author}
  {\bibfnamefont {J.}~\bibnamefont {{Brinkmann}}}, \bibinfo {author}
  {\bibfnamefont {J.~A.}\ \bibnamefont {{Frieman}}}, \bibinfo {author}
  {\bibfnamefont {J.~E.}\ \bibnamefont {{Gunn}}}, \bibinfo {author}
  {\bibfnamefont {R.~H.}\ \bibnamefont {{Lupton}}}, \bibinfo {author}
  {\bibfnamefont {R.~C.}\ \bibnamefont {{Nichol}}}, \bibinfo {author}
  {\bibfnamefont {W.~J.}\ \bibnamefont {{Percival}}}, \bibinfo {author}
  {\bibfnamefont {D.~P.}\ \bibnamefont {{Schneider}}}, \bibinfo {author}
  {\bibfnamefont {R.~A.}\ \bibnamefont {{Skibba}}}, \bibinfo {author}
  {\bibnamefont {{Strauss}}},  \emph {et~al.},\ }\href {\doibase
  10.1088/0004-637X/736/1/59} {\bibfield  {journal} {\bibinfo  {journal}
  {\apj}\ }\textbf {\bibinfo {volume} {736}},\ \bibinfo {eid} {59} (\bibinfo
  {year} {2011})},\ \Eprint {http://arxiv.org/abs/1005.2413} {arXiv:1005.2413
  [astro-ph.CO]} \BibitemShut {NoStop}%
\bibitem [{\citenamefont {{Piscionere}}\ \emph {et~al.}(2015)\citenamefont
  {{Piscionere}}, \citenamefont {{Berlind}}, \citenamefont {{McBride}},\ and\
  \citenamefont {{Scoccimarro}}}]{Piscionere2015}%
  \BibitemOpen
  \bibfield  {author} {\bibinfo {author} {\bibfnamefont {J.~A.}\ \bibnamefont
  {{Piscionere}}}, \bibinfo {author} {\bibfnamefont {A.~A.}\ \bibnamefont
  {{Berlind}}}, \bibinfo {author} {\bibfnamefont {C.~K.}\ \bibnamefont
  {{McBride}}}, \ and\ \bibinfo {author} {\bibfnamefont {R.}~\bibnamefont
  {{Scoccimarro}}},\ }\href {\doibase 10.1088/0004-637X/806/1/125} {\bibfield
  {journal} {\bibinfo  {journal} {\apj}\ }\textbf {\bibinfo {volume} {806}},\
  \bibinfo {eid} {125} (\bibinfo {year} {2015})},\ \Eprint
  {http://arxiv.org/abs/1407.6740} {arXiv:1407.6740 [astro-ph.GA]} \BibitemShut
  {NoStop}%
\bibitem [{\citenamefont {{Beutler}}\ \emph {et~al.}(2017)\citenamefont
  {{Beutler}}, \citenamefont {{Seo}}, \citenamefont {{Saito}}, \citenamefont
  {{Chuang}}, \citenamefont {{Cuesta}}, \citenamefont {{Eisenstein}},
  \citenamefont {{Gil-Mar{\'\i}n}}, \citenamefont {{Grieb}}, \citenamefont
  {{Hand}}, \citenamefont {{Kitaura}}, \citenamefont {{Modi}}, \citenamefont
  {{Nichol}}, \citenamefont {{Olmstead}}, \citenamefont {{Percival}} \emph
  {et~al.}}]{Beutler2017}%
  \BibitemOpen
  \bibfield  {author} {\bibinfo {author} {\bibfnamefont {F.}~\bibnamefont
  {{Beutler}}}, \bibinfo {author} {\bibfnamefont {H.-J.}\ \bibnamefont
  {{Seo}}}, \bibinfo {author} {\bibfnamefont {S.}~\bibnamefont {{Saito}}},
  \bibinfo {author} {\bibfnamefont {C.-H.}\ \bibnamefont {{Chuang}}}, \bibinfo
  {author} {\bibfnamefont {A.~J.}\ \bibnamefont {{Cuesta}}}, \bibinfo {author}
  {\bibfnamefont {D.~J.}\ \bibnamefont {{Eisenstein}}}, \bibinfo {author}
  {\bibfnamefont {H.}~\bibnamefont {{Gil-Mar{\'\i}n}}}, \bibinfo {author}
  {\bibfnamefont {J.~N.}\ \bibnamefont {{Grieb}}}, \bibinfo {author}
  {\bibfnamefont {N.}~\bibnamefont {{Hand}}}, \bibinfo {author} {\bibfnamefont
  {F.-S.}\ \bibnamefont {{Kitaura}}}, \bibinfo {author} {\bibfnamefont
  {C.}~\bibnamefont {{Modi}}}, \bibinfo {author} {\bibfnamefont {R.~C.}\
  \bibnamefont {{Nichol}}}, \bibinfo {author} {\bibfnamefont {M.~D.}\
  \bibnamefont {{Olmstead}}}, \bibinfo {author} {\bibfnamefont {W.~J.}\
  \bibnamefont {{Percival}}},  \emph {et~al.},\ }\href {\doibase
  10.1093/mnras/stw3298} {\bibfield  {journal} {\bibinfo  {journal} {\mnras}\
  }\textbf {\bibinfo {volume} {466}},\ \bibinfo {pages} {2242} (\bibinfo {year}
  {2017})},\ \Eprint {http://arxiv.org/abs/1607.03150} {arXiv:1607.03150
  [astro-ph.CO]} \BibitemShut {NoStop}%
\bibitem [{\citenamefont {{Maddox}}\ \emph {et~al.}(1995)\citenamefont
  {{Maddox}}, \citenamefont {{Efstathiou}},\ and\ \citenamefont
  {{Sutherland}}}]{maddox1996apm}%
  \BibitemOpen
  \bibfield  {author} {\bibinfo {author} {\bibfnamefont {S.~J.}\ \bibnamefont
  {{Maddox}}}, \bibinfo {author} {\bibfnamefont {G.}~\bibnamefont
  {{Efstathiou}}}, \ and\ \bibinfo {author} {\bibfnamefont {W.~J.}\
  \bibnamefont {{Sutherland}}},\ }\href@noop {} {\ ,\ \bibinfo {pages} {27602}
  (\bibinfo {year} {1995})}\BibitemShut {NoStop}%
\bibitem [{\citenamefont {Connolly}\ \emph {et~al.}(2002)\citenamefont
  {Connolly}, \citenamefont {Scranton}, \citenamefont {Johnston}, \citenamefont
  {Dodelson}, \citenamefont {Eisenstein}, \citenamefont {Frieman},
  \citenamefont {Gunn}, \citenamefont {Hui}, \citenamefont {Jain},
  \citenamefont {Kent} \emph {et~al.}}]{connolly2002angular}%
  \BibitemOpen
  \bibfield  {author} {\bibinfo {author} {\bibfnamefont {A.~J.}\ \bibnamefont
  {Connolly}}, \bibinfo {author} {\bibfnamefont {R.}~\bibnamefont {Scranton}},
  \bibinfo {author} {\bibfnamefont {D.}~\bibnamefont {Johnston}}, \bibinfo
  {author} {\bibfnamefont {S.}~\bibnamefont {Dodelson}}, \bibinfo {author}
  {\bibfnamefont {D.~J.}\ \bibnamefont {Eisenstein}}, \bibinfo {author}
  {\bibfnamefont {J.~A.}\ \bibnamefont {Frieman}}, \bibinfo {author}
  {\bibfnamefont {J.~E.}\ \bibnamefont {Gunn}}, \bibinfo {author}
  {\bibfnamefont {L.}~\bibnamefont {Hui}}, \bibinfo {author} {\bibfnamefont
  {B.}~\bibnamefont {Jain}}, \bibinfo {author} {\bibfnamefont {S.}~\bibnamefont
  {Kent}},  \emph {et~al.},\ }\href {\doibase 10.1086/342787} {\bibfield
  {journal} {\bibinfo  {journal} {\apj}\ }\textbf {\bibinfo {volume} {579}},\
  \bibinfo {pages} {42} (\bibinfo {year} {2002})},\ \Eprint
  {http://arxiv.org/abs/astro-ph/0107417} {arXiv:astro-ph/0107417 [astro-ph]}
  \BibitemShut {NoStop}%
\bibitem [{\citenamefont {{Maller}}\ \emph {et~al.}(2005)\citenamefont
  {{Maller}}, \citenamefont {{McIntosh}}, \citenamefont {{Katz}},\ and\
  \citenamefont {{Weinberg}}}]{maller2005galaxy}%
  \BibitemOpen
  \bibfield  {author} {\bibinfo {author} {\bibfnamefont {A.~H.}\ \bibnamefont
  {{Maller}}}, \bibinfo {author} {\bibfnamefont {D.~H.}\ \bibnamefont
  {{McIntosh}}}, \bibinfo {author} {\bibfnamefont {N.}~\bibnamefont {{Katz}}},
  \ and\ \bibinfo {author} {\bibfnamefont {M.~D.}\ \bibnamefont {{Weinberg}}},\
  }\href {\doibase 10.1086/426181} {\bibfield  {journal} {\bibinfo  {journal}
  {\apj}\ }\textbf {\bibinfo {volume} {619}},\ \bibinfo {pages} {147} (\bibinfo
  {year} {2005})},\ \Eprint {http://arxiv.org/abs/astro-ph/0304005}
  {arXiv:astro-ph/0304005 [astro-ph]} \BibitemShut {NoStop}%
\bibitem [{\citenamefont {{Crocce}}\ \emph {et~al.}(2011)\citenamefont
  {{Crocce}}, \citenamefont {{Cabr{\'e}}},\ and\ \citenamefont
  {{Gazta{\~n}aga}}}]{crocce2011modelling}%
  \BibitemOpen
  \bibfield  {author} {\bibinfo {author} {\bibfnamefont {M.}~\bibnamefont
  {{Crocce}}}, \bibinfo {author} {\bibfnamefont {A.}~\bibnamefont
  {{Cabr{\'e}}}}, \ and\ \bibinfo {author} {\bibfnamefont {E.}~\bibnamefont
  {{Gazta{\~n}aga}}},\ }\href {\doibase 10.1111/j.1365-2966.2011.18393.x}
  {\bibfield  {journal} {\bibinfo  {journal} {\mnras}\ }\textbf {\bibinfo
  {volume} {414}},\ \bibinfo {pages} {329} (\bibinfo {year} {2011})},\ \Eprint
  {http://arxiv.org/abs/1004.4640} {arXiv:1004.4640 [astro-ph.CO]} \BibitemShut
  {NoStop}%
\bibitem [{\citenamefont {{Wang}}\ \emph {et~al.}(2013)\citenamefont {{Wang}},
  \citenamefont {{Brunner}},\ and\ \citenamefont {{Dolence}}}]{wang2013sdss}%
  \BibitemOpen
  \bibfield  {author} {\bibinfo {author} {\bibfnamefont {Y.}~\bibnamefont
  {{Wang}}}, \bibinfo {author} {\bibfnamefont {R.~J.}\ \bibnamefont
  {{Brunner}}}, \ and\ \bibinfo {author} {\bibfnamefont {J.~C.}\ \bibnamefont
  {{Dolence}}},\ }\href {\doibase 10.1093/mnras/stt450} {\bibfield  {journal}
  {\bibinfo  {journal} {\mnras}\ }\textbf {\bibinfo {volume} {432}},\ \bibinfo
  {pages} {1961} (\bibinfo {year} {2013})},\ \Eprint
  {http://arxiv.org/abs/1303.2432} {arXiv:1303.2432 [astro-ph.CO]} \BibitemShut
  {NoStop}%
\bibitem [{\citenamefont {{Takada}}\ and\ \citenamefont
  {{Jain}}(2003)}]{Takada2003}%
  \BibitemOpen
  \bibfield  {author} {\bibinfo {author} {\bibfnamefont {M.}~\bibnamefont
  {{Takada}}}\ and\ \bibinfo {author} {\bibfnamefont {B.}~\bibnamefont
  {{Jain}}},\ }\href {\doibase 10.1046/j.1365-8711.2003.06321.x} {\bibfield
  {journal} {\bibinfo  {journal} {\mnras}\ }\textbf {\bibinfo {volume} {340}},\
  \bibinfo {pages} {580} (\bibinfo {year} {2003})},\ \Eprint
  {http://arxiv.org/abs/astro-ph/0209167} {arXiv:astro-ph/0209167 [astro-ph]}
  \BibitemShut {NoStop}%
\bibitem [{\citenamefont {{Nichol}}\ \emph {et~al.}(2006)\citenamefont
  {{Nichol}}, \citenamefont {{Sheth}}, \citenamefont {{Suto}}, \citenamefont
  {{Gray}}, \citenamefont {{Kayo}}, \citenamefont {{Wechsler}}, \citenamefont
  {{Marin}}, \citenamefont {{Kulkarni}}, \citenamefont {{Blanton}},
  \citenamefont {{Connolly}}, \citenamefont {{Gardner}}, \citenamefont
  {{Jain}}, \citenamefont {{Miller}}, \citenamefont {{Moore}}, \citenamefont
  {{Pope}} \emph {et~al.}}]{Nichol2006}%
  \BibitemOpen
  \bibfield  {author} {\bibinfo {author} {\bibfnamefont {R.~C.}\ \bibnamefont
  {{Nichol}}}, \bibinfo {author} {\bibfnamefont {R.~K.}\ \bibnamefont
  {{Sheth}}}, \bibinfo {author} {\bibfnamefont {Y.}~\bibnamefont {{Suto}}},
  \bibinfo {author} {\bibfnamefont {A.~J.}\ \bibnamefont {{Gray}}}, \bibinfo
  {author} {\bibfnamefont {I.}~\bibnamefont {{Kayo}}}, \bibinfo {author}
  {\bibfnamefont {R.~H.}\ \bibnamefont {{Wechsler}}}, \bibinfo {author}
  {\bibfnamefont {F.}~\bibnamefont {{Marin}}}, \bibinfo {author} {\bibfnamefont
  {G.}~\bibnamefont {{Kulkarni}}}, \bibinfo {author} {\bibfnamefont
  {M.}~\bibnamefont {{Blanton}}}, \bibinfo {author} {\bibfnamefont {A.~J.}\
  \bibnamefont {{Connolly}}}, \bibinfo {author} {\bibfnamefont {J.~P.}\
  \bibnamefont {{Gardner}}}, \bibinfo {author} {\bibfnamefont {B.}~\bibnamefont
  {{Jain}}}, \bibinfo {author} {\bibfnamefont {C.~J.}\ \bibnamefont
  {{Miller}}}, \bibinfo {author} {\bibfnamefont {A.~W.}\ \bibnamefont
  {{Moore}}}, \bibinfo {author} {\bibfnamefont {A.}~\bibnamefont {{Pope}}},
  \emph {et~al.},\ }\href {\doibase 10.1111/j.1365-2966.2006.10239.x}
  {\bibfield  {journal} {\bibinfo  {journal} {\mnras}\ }\textbf {\bibinfo
  {volume} {368}},\ \bibinfo {pages} {1507} (\bibinfo {year} {2006})},\ \Eprint
  {http://arxiv.org/abs/astro-ph/0602548} {arXiv:astro-ph/0602548 [astro-ph]}
  \BibitemShut {NoStop}%
\bibitem [{\citenamefont {{Guo}}\ \emph {et~al.}(2015)\citenamefont {{Guo}},
  \citenamefont {{Zheng}}, \citenamefont {{Jing}}, \citenamefont {{Zehavi}},
  \citenamefont {{Li}}, \citenamefont {{Weinberg}}, \citenamefont {{Skibba}},
  \citenamefont {{Nichol}}, \citenamefont {{Rossi}}, \citenamefont {{Sabiu}},
  \citenamefont {{Schneider}},\ and\ \citenamefont {{McBride}}}]{Guo2015}%
  \BibitemOpen
  \bibfield  {author} {\bibinfo {author} {\bibfnamefont {H.}~\bibnamefont
  {{Guo}}}, \bibinfo {author} {\bibfnamefont {Z.}~\bibnamefont {{Zheng}}},
  \bibinfo {author} {\bibfnamefont {Y.~P.}\ \bibnamefont {{Jing}}}, \bibinfo
  {author} {\bibfnamefont {I.}~\bibnamefont {{Zehavi}}}, \bibinfo {author}
  {\bibfnamefont {C.}~\bibnamefont {{Li}}}, \bibinfo {author} {\bibfnamefont
  {D.~H.}\ \bibnamefont {{Weinberg}}}, \bibinfo {author} {\bibfnamefont
  {R.~A.}\ \bibnamefont {{Skibba}}}, \bibinfo {author} {\bibfnamefont {R.~C.}\
  \bibnamefont {{Nichol}}}, \bibinfo {author} {\bibfnamefont {G.}~\bibnamefont
  {{Rossi}}}, \bibinfo {author} {\bibfnamefont {C.~G.}\ \bibnamefont
  {{Sabiu}}}, \bibinfo {author} {\bibfnamefont {D.~P.}\ \bibnamefont
  {{Schneider}}}, \ and\ \bibinfo {author} {\bibfnamefont {C.~K.}\ \bibnamefont
  {{McBride}}},\ }\href {\doibase 10.1093/mnrasl/slv020} {\bibfield  {journal}
  {\bibinfo  {journal} {\mnras}\ }\textbf {\bibinfo {volume} {449}},\ \bibinfo
  {pages} {L95} (\bibinfo {year} {2015})},\ \Eprint
  {http://arxiv.org/abs/1409.7389} {arXiv:1409.7389 [astro-ph.CO]} \BibitemShut
  {NoStop}%
\bibitem [{\citenamefont {{Slepian}}\ \emph {et~al.}(2017)\citenamefont
  {{Slepian}}, \citenamefont {{Eisenstein}}, \citenamefont {{Beutler}},
  \citenamefont {{Chuang}}, \citenamefont {{Cuesta}}, \citenamefont {{Ge}},
  \citenamefont {{Gil-Mar{\'\i}n}}, \citenamefont {{Ho}}, \citenamefont
  {{Kitaura}}, \citenamefont {{McBride}}, \citenamefont {{Nichol}},
  \citenamefont {{Percival}}, \citenamefont {{Rodr{\'\i}guez-Torres}},
  \citenamefont {{Ross}}, \citenamefont {{Scoccimarro}} \emph
  {et~al.}}]{Slepian2017}%
  \BibitemOpen
  \bibfield  {author} {\bibinfo {author} {\bibfnamefont {Z.}~\bibnamefont
  {{Slepian}}}, \bibinfo {author} {\bibfnamefont {D.~J.}\ \bibnamefont
  {{Eisenstein}}}, \bibinfo {author} {\bibfnamefont {F.}~\bibnamefont
  {{Beutler}}}, \bibinfo {author} {\bibfnamefont {C.-H.}\ \bibnamefont
  {{Chuang}}}, \bibinfo {author} {\bibfnamefont {A.~J.}\ \bibnamefont
  {{Cuesta}}}, \bibinfo {author} {\bibfnamefont {J.}~\bibnamefont {{Ge}}},
  \bibinfo {author} {\bibfnamefont {H.}~\bibnamefont {{Gil-Mar{\'\i}n}}},
  \bibinfo {author} {\bibfnamefont {S.}~\bibnamefont {{Ho}}}, \bibinfo {author}
  {\bibfnamefont {F.-S.}\ \bibnamefont {{Kitaura}}}, \bibinfo {author}
  {\bibfnamefont {C.~K.}\ \bibnamefont {{McBride}}}, \bibinfo {author}
  {\bibfnamefont {R.~C.}\ \bibnamefont {{Nichol}}}, \bibinfo {author}
  {\bibfnamefont {W.~J.}\ \bibnamefont {{Percival}}}, \bibinfo {author}
  {\bibfnamefont {S.}~\bibnamefont {{Rodr{\'\i}guez-Torres}}}, \bibinfo
  {author} {\bibfnamefont {A.~J.}\ \bibnamefont {{Ross}}}, \bibinfo {author}
  {\bibfnamefont {R.}~\bibnamefont {{Scoccimarro}}},  \emph {et~al.},\ }\href
  {\doibase 10.1093/mnras/stw3234} {\bibfield  {journal} {\bibinfo  {journal}
  {\mnras}\ }\textbf {\bibinfo {volume} {468}},\ \bibinfo {pages} {1070}
  (\bibinfo {year} {2017})},\ \Eprint {http://arxiv.org/abs/1512.02231}
  {arXiv:1512.02231 [astro-ph.CO]} \BibitemShut {NoStop}%
\bibitem [{\citenamefont {{Sosa Nu{\~n}ez}}\ and\ \citenamefont
  {{Niz}}(2020)}]{nunez2020fast}%
  \BibitemOpen
  \bibfield  {author} {\bibinfo {author} {\bibfnamefont {F.}~\bibnamefont
  {{Sosa Nu{\~n}ez}}}\ and\ \bibinfo {author} {\bibfnamefont {G.}~\bibnamefont
  {{Niz}}},\ }\href {\doibase 10.1088/1475-7516/2020/12/021} {\bibfield
  {journal} {\bibinfo  {journal} {\jcap}\ }\textbf {\bibinfo {volume} {2020}},\
  \bibinfo {eid} {021} (\bibinfo {year} {2020})},\ \Eprint
  {http://arxiv.org/abs/2006.05434} {arXiv:2006.05434 [astro-ph.CO]}
  \BibitemShut {NoStop}%
\bibitem [{\citenamefont {{Slepian}}\ and\ \citenamefont
  {{Eisenstein}}(2018)}]{slepian2018practical}%
  \BibitemOpen
  \bibfield  {author} {\bibinfo {author} {\bibfnamefont {Z.}~\bibnamefont
  {{Slepian}}}\ and\ \bibinfo {author} {\bibfnamefont {D.~J.}\ \bibnamefont
  {{Eisenstein}}},\ }\href {\doibase 10.1093/mnras/sty1063} {\bibfield
  {journal} {\bibinfo  {journal} {\mnras}\ }\textbf {\bibinfo {volume} {478}},\
  \bibinfo {pages} {1468} (\bibinfo {year} {2018})},\ \Eprint
  {http://arxiv.org/abs/1709.10150} {arXiv:1709.10150 [astro-ph.CO]}
  \BibitemShut {NoStop}%
\bibitem [{\citenamefont {{Umeh}}(2021)}]{umeh2021optimal}%
  \BibitemOpen
  \bibfield  {author} {\bibinfo {author} {\bibfnamefont {O.}~\bibnamefont
  {{Umeh}}},\ }\href {\doibase 10.1088/1475-7516/2021/05/035} {\bibfield
  {journal} {\bibinfo  {journal} {\jcap}\ }\textbf {\bibinfo {volume} {2021}},\
  \bibinfo {eid} {035} (\bibinfo {year} {2021})},\ \Eprint
  {http://arxiv.org/abs/2011.05889} {arXiv:2011.05889 [astro-ph.CO]}
  \BibitemShut {NoStop}%
\bibitem [{\citenamefont {{White}}(1979)}]{white1979hierarchy}%
  \BibitemOpen
  \bibfield  {author} {\bibinfo {author} {\bibfnamefont {S.~D.~M.}\
  \bibnamefont {{White}}},\ }\href {\doibase 10.1093/mnras/186.2.145}
  {\bibfield  {journal} {\bibinfo  {journal} {\mnras}\ }\textbf {\bibinfo
  {volume} {186}},\ \bibinfo {pages} {145} (\bibinfo {year}
  {1979})}\BibitemShut {NoStop}%
\bibitem [{\citenamefont {{Gaztanaga}}(1994)}]{gaztanaga1994high}%
  \BibitemOpen
  \bibfield  {author} {\bibinfo {author} {\bibfnamefont {E.}~\bibnamefont
  {{Gaztanaga}}},\ }\href {\doibase 10.1093/mnras/268.4.913} {\bibfield
  {journal} {\bibinfo  {journal} {\mnras}\ }\textbf {\bibinfo {volume} {268}},\
  \bibinfo {pages} {913} (\bibinfo {year} {1994})},\ \Eprint
  {http://arxiv.org/abs/astro-ph/9309019} {arXiv:astro-ph/9309019 [astro-ph]}
  \BibitemShut {NoStop}%
\bibitem [{\citenamefont {Croton}\ \emph {et~al.}(2004)\citenamefont {Croton},
  \citenamefont {Gaztanaga}, \citenamefont {Baugh}, \citenamefont {Norberg},
  \citenamefont {Colless}, \citenamefont {Baldry}, \citenamefont
  {Bland-Hawthorn}, \citenamefont {Bridges}, \citenamefont {Cannon},
  \citenamefont {Cole} \emph {et~al.}}]{croton20042df}%
  \BibitemOpen
  \bibfield  {author} {\bibinfo {author} {\bibfnamefont {D.~J.}\ \bibnamefont
  {Croton}}, \bibinfo {author} {\bibfnamefont {E.}~\bibnamefont {Gaztanaga}},
  \bibinfo {author} {\bibfnamefont {C.~M.}\ \bibnamefont {Baugh}}, \bibinfo
  {author} {\bibfnamefont {P.}~\bibnamefont {Norberg}}, \bibinfo {author}
  {\bibfnamefont {M.}~\bibnamefont {Colless}}, \bibinfo {author} {\bibfnamefont
  {I.~K.}\ \bibnamefont {Baldry}}, \bibinfo {author} {\bibfnamefont
  {J.}~\bibnamefont {Bland-Hawthorn}}, \bibinfo {author} {\bibfnamefont
  {T.}~\bibnamefont {Bridges}}, \bibinfo {author} {\bibfnamefont
  {R.}~\bibnamefont {Cannon}}, \bibinfo {author} {\bibfnamefont
  {S.}~\bibnamefont {Cole}},  \emph {et~al.},\ }\href {\doibase
  10.1111/j.1365-2966.2004.08017.x} {\bibfield  {journal} {\bibinfo  {journal}
  {\mnras}\ }\textbf {\bibinfo {volume} {352}},\ \bibinfo {pages} {1232}
  (\bibinfo {year} {2004})},\ \Eprint {http://arxiv.org/abs/astro-ph/0401434}
  {arXiv:astro-ph/0401434 [astro-ph]} \BibitemShut {NoStop}%
\bibitem [{\citenamefont {{Juszkiewicz}}\ \emph {et~al.}(1993)\citenamefont
  {{Juszkiewicz}}, \citenamefont {{Bouchet}},\ and\ \citenamefont
  {{Colombi}}}]{juszkiewicz1993skewness}%
  \BibitemOpen
  \bibfield  {author} {\bibinfo {author} {\bibfnamefont {R.}~\bibnamefont
  {{Juszkiewicz}}}, \bibinfo {author} {\bibfnamefont {F.~R.}\ \bibnamefont
  {{Bouchet}}}, \ and\ \bibinfo {author} {\bibfnamefont {S.}~\bibnamefont
  {{Colombi}}},\ }\href {\doibase 10.1086/186927} {\bibfield  {journal}
  {\bibinfo  {journal} {\apjl}\ }\textbf {\bibinfo {volume} {412}},\ \bibinfo
  {pages} {L9} (\bibinfo {year} {1993})},\ \Eprint
  {http://arxiv.org/abs/astro-ph/9306003} {arXiv:astro-ph/9306003 [astro-ph]}
  \BibitemShut {NoStop}%
\bibitem [{\citenamefont {{Lokas}}\ \emph {et~al.}(1995)\citenamefont
  {{Lokas}}, \citenamefont {{Juszkiewicz}}, \citenamefont {{Weinberg}},\ and\
  \citenamefont {{Bouchet}}}]{lokas1995kurtosis}%
  \BibitemOpen
  \bibfield  {author} {\bibinfo {author} {\bibfnamefont {E.~L.}\ \bibnamefont
  {{Lokas}}}, \bibinfo {author} {\bibfnamefont {R.}~\bibnamefont
  {{Juszkiewicz}}}, \bibinfo {author} {\bibfnamefont {D.~H.}\ \bibnamefont
  {{Weinberg}}}, \ and\ \bibinfo {author} {\bibfnamefont {F.~R.}\ \bibnamefont
  {{Bouchet}}},\ }\href {\doibase 10.1093/mnras/274.3.730} {\bibfield
  {journal} {\bibinfo  {journal} {\mnras}\ }\textbf {\bibinfo {volume} {274}},\
  \bibinfo {pages} {730} (\bibinfo {year} {1995})},\ \Eprint
  {http://arxiv.org/abs/astro-ph/9407095} {arXiv:astro-ph/9407095 [astro-ph]}
  \BibitemShut {NoStop}%
\bibitem [{\citenamefont {{Clifton}}\ \emph {et~al.}(2012)\citenamefont
  {{Clifton}}, \citenamefont {{Ferreira}}, \citenamefont {{Padilla}},\ and\
  \citenamefont {{Skordis}}}]{Clifton2012}%
  \BibitemOpen
  \bibfield  {author} {\bibinfo {author} {\bibfnamefont {T.}~\bibnamefont
  {{Clifton}}}, \bibinfo {author} {\bibfnamefont {P.~G.}\ \bibnamefont
  {{Ferreira}}}, \bibinfo {author} {\bibfnamefont {A.}~\bibnamefont
  {{Padilla}}}, \ and\ \bibinfo {author} {\bibfnamefont {C.}~\bibnamefont
  {{Skordis}}},\ }\href {\doibase 10.1016/j.physrep.2012.01.001} {\bibfield
  {journal} {\bibinfo  {journal} {\physrep}\ }\textbf {\bibinfo {volume}
  {513}},\ \bibinfo {pages} {1} (\bibinfo {year} {2012})},\ \Eprint
  {http://arxiv.org/abs/1106.2476} {arXiv:1106.2476 [astro-ph.CO]} \BibitemShut
  {NoStop}%
\bibitem [{\citenamefont {{Hu}}\ and\ \citenamefont
  {{Sawicki}}(2007)}]{hu2007models}%
  \BibitemOpen
  \bibfield  {author} {\bibinfo {author} {\bibfnamefont {W.}~\bibnamefont
  {{Hu}}}\ and\ \bibinfo {author} {\bibfnamefont {I.}~\bibnamefont
  {{Sawicki}}},\ }\href {\doibase 10.1103/PhysRevD.76.064004} {\bibfield
  {journal} {\bibinfo  {journal} {\prd}\ }\textbf {\bibinfo {volume} {76}},\
  \bibinfo {eid} {064004} (\bibinfo {year} {2007})},\ \Eprint
  {http://arxiv.org/abs/0705.1158} {arXiv:0705.1158 [astro-ph]} \BibitemShut
  {NoStop}%
\bibitem [{\citenamefont {{Sotiriou}}\ and\ \citenamefont
  {{Faraoni}}(2010)}]{sotiriou2010f}%
  \BibitemOpen
  \bibfield  {author} {\bibinfo {author} {\bibfnamefont {T.~P.}\ \bibnamefont
  {{Sotiriou}}}\ and\ \bibinfo {author} {\bibfnamefont {V.}~\bibnamefont
  {{Faraoni}}},\ }\href {\doibase 10.1103/RevModPhys.82.451} {\bibfield
  {journal} {\bibinfo  {journal} {Reviews of Modern Physics}\ }\textbf
  {\bibinfo {volume} {82}},\ \bibinfo {pages} {451} (\bibinfo {year} {2010})},\
  \Eprint {http://arxiv.org/abs/0805.1726} {arXiv:0805.1726 [gr-qc]}
  \BibitemShut {NoStop}%
\bibitem [{\citenamefont {{Buchdahl}}(1970)}]{Buchdahl1970}%
  \BibitemOpen
  \bibfield  {author} {\bibinfo {author} {\bibfnamefont {H.~A.}\ \bibnamefont
  {{Buchdahl}}},\ }\href {\doibase 10.1093/mnras/150.1.1} {\bibfield  {journal}
  {\bibinfo  {journal} {\mnras}\ }\textbf {\bibinfo {volume} {150}},\ \bibinfo
  {pages} {1} (\bibinfo {year} {1970})}\BibitemShut {NoStop}%
\bibitem [{\citenamefont {{Vainshtein}}(1972)}]{vainshtein1972problem}%
  \BibitemOpen
  \bibfield  {author} {\bibinfo {author} {\bibfnamefont {A.~I.}\ \bibnamefont
  {{Vainshtein}}},\ }\href {\doibase 10.1016/0370-2693(72)90147-5} {\bibfield
  {journal} {\bibinfo  {journal} {Physics Letters B}\ }\textbf {\bibinfo
  {volume} {39}},\ \bibinfo {pages} {393} (\bibinfo {year} {1972})}\BibitemShut
  {NoStop}%
\bibitem [{\citenamefont {{Dvali}}\ \emph {et~al.}(2000)\citenamefont
  {{Dvali}}, \citenamefont {{Gabadadze}},\ and\ \citenamefont
  {{Porrati}}}]{dvali20004d}%
  \BibitemOpen
  \bibfield  {author} {\bibinfo {author} {\bibfnamefont {G.}~\bibnamefont
  {{Dvali}}}, \bibinfo {author} {\bibfnamefont {G.}~\bibnamefont
  {{Gabadadze}}}, \ and\ \bibinfo {author} {\bibfnamefont {M.}~\bibnamefont
  {{Porrati}}},\ }\href {\doibase 10.1016/S0370-2693(00)00669-9} {\bibfield
  {journal} {\bibinfo  {journal} {Physics Letters B}\ }\textbf {\bibinfo
  {volume} {485}},\ \bibinfo {pages} {208} (\bibinfo {year} {2000})},\ \Eprint
  {http://arxiv.org/abs/hep-th/0005016} {arXiv:hep-th/0005016 [hep-th]}
  \BibitemShut {NoStop}%
\bibitem [{\citenamefont {{Sahni}}\ and\ \citenamefont
  {{Shtanov}}(2003)}]{Sahni2003}%
  \BibitemOpen
  \bibfield  {author} {\bibinfo {author} {\bibfnamefont {V.}~\bibnamefont
  {{Sahni}}}\ and\ \bibinfo {author} {\bibfnamefont {Y.}~\bibnamefont
  {{Shtanov}}},\ }\href {\doibase 10.1088/1475-7516/2003/11/014} {\bibfield
  {journal} {\bibinfo  {journal} {\jcap}\ }\textbf {\bibinfo {volume} {2003}},\
  \bibinfo {eid} {014} (\bibinfo {year} {2003})},\ \Eprint
  {http://arxiv.org/abs/astro-ph/0202346} {arXiv:astro-ph/0202346 [astro-ph]}
  \BibitemShut {NoStop}%
\bibitem [{\citenamefont {{Li}}\ and\ \citenamefont {{Zhao}}(2009)}]{Li2009}%
  \BibitemOpen
  \bibfield  {author} {\bibinfo {author} {\bibfnamefont {B.}~\bibnamefont
  {{Li}}}\ and\ \bibinfo {author} {\bibfnamefont {H.}~\bibnamefont {{Zhao}}},\
  }\href {\doibase 10.1103/PhysRevD.80.044027} {\bibfield  {journal} {\bibinfo
  {journal} {\prd}\ }\textbf {\bibinfo {volume} {80}},\ \bibinfo {eid} {044027}
  (\bibinfo {year} {2009})},\ \Eprint {http://arxiv.org/abs/0906.3880}
  {arXiv:0906.3880 [astro-ph.CO]} \BibitemShut {NoStop}%
\bibitem [{\citenamefont {{Liu}}\ \emph {et~al.}(2014)\citenamefont {{Liu}},
  \citenamefont {{Eatough}}, \citenamefont {{Wex}},\ and\ \citenamefont
  {{Kramer}}}]{Liu2014}%
  \BibitemOpen
  \bibfield  {author} {\bibinfo {author} {\bibfnamefont {K.}~\bibnamefont
  {{Liu}}}, \bibinfo {author} {\bibfnamefont {R.~P.}\ \bibnamefont
  {{Eatough}}}, \bibinfo {author} {\bibfnamefont {N.}~\bibnamefont {{Wex}}}, \
  and\ \bibinfo {author} {\bibfnamefont {M.}~\bibnamefont {{Kramer}}},\ }\href
  {\doibase 10.1093/mnras/stu1913} {\bibfield  {journal} {\bibinfo  {journal}
  {\mnras}\ }\textbf {\bibinfo {volume} {445}},\ \bibinfo {pages} {3115}
  (\bibinfo {year} {2014})},\ \Eprint {http://arxiv.org/abs/1409.3882}
  {arXiv:1409.3882 [astro-ph.GA]} \BibitemShut {NoStop}%
\bibitem [{\citenamefont {{Wei}}\ \emph {et~al.}(2017)\citenamefont {{Wei}},
  \citenamefont {{Zhang}}, \citenamefont {{Wu}}, \citenamefont {{Gao}},
  \citenamefont {{M{\'e}sz{\'a}ros}}, \citenamefont {{Zhang}}, \citenamefont
  {{Dai}}, \citenamefont {{Zhang}},\ and\ \citenamefont {{Zhu}}}]{Wei2017}%
  \BibitemOpen
  \bibfield  {author} {\bibinfo {author} {\bibfnamefont {J.-J.}\ \bibnamefont
  {{Wei}}}, \bibinfo {author} {\bibfnamefont {B.-B.}\ \bibnamefont {{Zhang}}},
  \bibinfo {author} {\bibfnamefont {X.-F.}\ \bibnamefont {{Wu}}}, \bibinfo
  {author} {\bibfnamefont {H.}~\bibnamefont {{Gao}}}, \bibinfo {author}
  {\bibfnamefont {P.}~\bibnamefont {{M{\'e}sz{\'a}ros}}}, \bibinfo {author}
  {\bibfnamefont {B.}~\bibnamefont {{Zhang}}}, \bibinfo {author} {\bibfnamefont
  {Z.-G.}\ \bibnamefont {{Dai}}}, \bibinfo {author} {\bibfnamefont {S.-N.}\
  \bibnamefont {{Zhang}}}, \ and\ \bibinfo {author} {\bibfnamefont {Z.-H.}\
  \bibnamefont {{Zhu}}},\ }\href {\doibase 10.1088/1475-7516/2017/11/035}
  {\bibfield  {journal} {\bibinfo  {journal} {\jcap}\ }\textbf {\bibinfo
  {volume} {2017}},\ \bibinfo {eid} {035} (\bibinfo {year} {2017})},\ \Eprint
  {http://arxiv.org/abs/1710.05860} {arXiv:1710.05860 [astro-ph.HE]}
  \BibitemShut {NoStop}%
\bibitem [{\citenamefont {{Abbott}}\ \emph {et~al.}(2020)\citenamefont
  {{Abbott}}, \citenamefont {{Abbott}}, \citenamefont {{Abraham}},
  \citenamefont {{Acernese}}, \citenamefont {{Ackley}}, \citenamefont
  {{Adams}}, \citenamefont {{Adhikari}}, \citenamefont {{Adya}}, \citenamefont
  {{Affeldt}}, \citenamefont {{Agathos}}, \citenamefont {{Agatsuma}},
  \citenamefont {{Aggarwal}} \emph {et~al.}}]{Abbott2020}%
  \BibitemOpen
  \bibfield  {author} {\bibinfo {author} {\bibfnamefont {R.}~\bibnamefont
  {{Abbott}}}, \bibinfo {author} {\bibfnamefont {T.~D.}\ \bibnamefont
  {{Abbott}}}, \bibinfo {author} {\bibfnamefont {S.}~\bibnamefont {{Abraham}}},
  \bibinfo {author} {\bibfnamefont {F.}~\bibnamefont {{Acernese}}}, \bibinfo
  {author} {\bibfnamefont {K.}~\bibnamefont {{Ackley}}}, \bibinfo {author}
  {\bibfnamefont {C.}~\bibnamefont {{Adams}}}, \bibinfo {author} {\bibfnamefont
  {R.~X.}\ \bibnamefont {{Adhikari}}}, \bibinfo {author} {\bibfnamefont
  {V.~B.}\ \bibnamefont {{Adya}}}, \bibinfo {author} {\bibfnamefont
  {C.}~\bibnamefont {{Affeldt}}}, \bibinfo {author} {\bibfnamefont
  {M.}~\bibnamefont {{Agathos}}}, \bibinfo {author} {\bibfnamefont
  {K.}~\bibnamefont {{Agatsuma}}}, \bibinfo {author} {\bibnamefont
  {{Aggarwal}}},  \emph {et~al.},\ }\href {\doibase 10.3847/2041-8213/ab960f}
  {\bibfield  {journal} {\bibinfo  {journal} {\apjl}\ }\textbf {\bibinfo
  {volume} {896}},\ \bibinfo {eid} {L44} (\bibinfo {year} {2020})},\ \Eprint
  {http://arxiv.org/abs/2006.12611} {arXiv:2006.12611 [astro-ph.HE]}
  \BibitemShut {NoStop}%
\bibitem [{\citenamefont {{Stairs}}(2003)}]{Ingrid2003}%
  \BibitemOpen
  \bibfield  {author} {\bibinfo {author} {\bibfnamefont {I.~H.}\ \bibnamefont
  {{Stairs}}},\ }\href {\doibase 10.12942/lrr-2003-5} {\bibfield  {journal}
  {\bibinfo  {journal} {Living Reviews in Relativity}\ }\textbf {\bibinfo
  {volume} {6}},\ \bibinfo {eid} {5} (\bibinfo {year} {2003})},\ \Eprint
  {http://arxiv.org/abs/astro-ph/0307536} {arXiv:astro-ph/0307536 [astro-ph]}
  \BibitemShut {NoStop}%
\bibitem [{\citenamefont {{De Marchi}}\ and\ \citenamefont
  {{Cascioli}}(2020)}]{DeMarchi2020}%
  \BibitemOpen
  \bibfield  {author} {\bibinfo {author} {\bibfnamefont {F.}~\bibnamefont {{De
  Marchi}}}\ and\ \bibinfo {author} {\bibfnamefont {G.}~\bibnamefont
  {{Cascioli}}},\ }\href {\doibase 10.1088/1361-6382/ab6ae0} {\bibfield
  {journal} {\bibinfo  {journal} {Classical and Quantum Gravity}\ }\textbf
  {\bibinfo {volume} {37}},\ \bibinfo {eid} {095007} (\bibinfo {year}
  {2020})},\ \Eprint {http://arxiv.org/abs/1911.05561} {arXiv:1911.05561
  [gr-qc]} \BibitemShut {NoStop}%
\bibitem [{\citenamefont {{Bose}}\ \emph {et~al.}(2017)\citenamefont {{Bose}},
  \citenamefont {{Koyama}}, \citenamefont {{Hellwing}}, \citenamefont
  {{Zhao}},\ and\ \citenamefont {{Winther}}}]{Bose2017}%
  \BibitemOpen
  \bibfield  {author} {\bibinfo {author} {\bibfnamefont {B.}~\bibnamefont
  {{Bose}}}, \bibinfo {author} {\bibfnamefont {K.}~\bibnamefont {{Koyama}}},
  \bibinfo {author} {\bibfnamefont {W.~A.}\ \bibnamefont {{Hellwing}}},
  \bibinfo {author} {\bibfnamefont {G.-B.}\ \bibnamefont {{Zhao}}}, \ and\
  \bibinfo {author} {\bibfnamefont {H.~A.}\ \bibnamefont {{Winther}}},\ }\href
  {\doibase 10.1103/PhysRevD.96.023519} {\bibfield  {journal} {\bibinfo
  {journal} {\prd}\ }\textbf {\bibinfo {volume} {96}},\ \bibinfo {eid} {023519}
  (\bibinfo {year} {2017})},\ \Eprint {http://arxiv.org/abs/1702.02348}
  {arXiv:1702.02348 [astro-ph.CO]} \BibitemShut {NoStop}%
\bibitem [{\citenamefont {{Bose}}\ \emph {et~al.}(2019)\citenamefont {{Bose}},
  \citenamefont {{Koyama}},\ and\ \citenamefont {{Winther}}}]{Bose2019}%
  \BibitemOpen
  \bibfield  {author} {\bibinfo {author} {\bibfnamefont {B.}~\bibnamefont
  {{Bose}}}, \bibinfo {author} {\bibfnamefont {K.}~\bibnamefont {{Koyama}}}, \
  and\ \bibinfo {author} {\bibfnamefont {H.~A.}\ \bibnamefont {{Winther}}},\
  }\href {\doibase 10.1088/1475-7516/2019/10/021} {\bibfield  {journal}
  {\bibinfo  {journal} {\jcap}\ }\textbf {\bibinfo {volume} {2019}},\ \bibinfo
  {eid} {021} (\bibinfo {year} {2019})},\ \Eprint
  {http://arxiv.org/abs/1905.05135} {arXiv:1905.05135 [astro-ph.CO]}
  \BibitemShut {NoStop}%
\bibitem [{\citenamefont {{Garc{\'\i}a-Farieta}}\ \emph
  {et~al.}(2021)\citenamefont {{Garc{\'\i}a-Farieta}}, \citenamefont
  {{Hellwing}}, \citenamefont {{Gupta}},\ and\ \citenamefont
  {{Bilicki}}}]{garcia2021probing}%
  \BibitemOpen
  \bibfield  {author} {\bibinfo {author} {\bibfnamefont {J.~E.}\ \bibnamefont
  {{Garc{\'\i}a-Farieta}}}, \bibinfo {author} {\bibfnamefont {W.~A.}\
  \bibnamefont {{Hellwing}}}, \bibinfo {author} {\bibfnamefont
  {S.}~\bibnamefont {{Gupta}}}, \ and\ \bibinfo {author} {\bibfnamefont
  {M.}~\bibnamefont {{Bilicki}}},\ }\href {\doibase
  10.1103/PhysRevD.103.103524} {\bibfield  {journal} {\bibinfo  {journal}
  {\prd}\ }\textbf {\bibinfo {volume} {103}},\ \bibinfo {eid} {103524}
  (\bibinfo {year} {2021})},\ \Eprint {http://arxiv.org/abs/2103.14019}
  {arXiv:2103.14019 [astro-ph.CO]} \BibitemShut {NoStop}%
\bibitem [{\citenamefont {Ivezic}\ \emph {et~al.}(2009)\citenamefont {Ivezic},
  \citenamefont {Tyson}, \citenamefont {Axelrod}, \citenamefont {Burke},
  \citenamefont {Claver}, \citenamefont {Cook}, \citenamefont {Kahn},
  \citenamefont {Lupton}, \citenamefont {Monet}, \citenamefont {Pinto} \emph
  {et~al.}}]{ivezic2009lsst}%
  \BibitemOpen
  \bibfield  {author} {\bibinfo {author} {\bibfnamefont {Z.}~\bibnamefont
  {Ivezic}}, \bibinfo {author} {\bibfnamefont {J.}~\bibnamefont {Tyson}},
  \bibinfo {author} {\bibfnamefont {T.}~\bibnamefont {Axelrod}}, \bibinfo
  {author} {\bibfnamefont {D.}~\bibnamefont {Burke}}, \bibinfo {author}
  {\bibfnamefont {C.}~\bibnamefont {Claver}}, \bibinfo {author} {\bibfnamefont
  {K.}~\bibnamefont {Cook}}, \bibinfo {author} {\bibfnamefont {S.}~\bibnamefont
  {Kahn}}, \bibinfo {author} {\bibfnamefont {R.}~\bibnamefont {Lupton}},
  \bibinfo {author} {\bibfnamefont {D.}~\bibnamefont {Monet}}, \bibinfo
  {author} {\bibfnamefont {P.}~\bibnamefont {Pinto}},  \emph {et~al.},\ }in\
  \href@noop {} {\emph {\bibinfo {booktitle} {American Astronomical Society
  Meeting Abstracts \#213}}},\ \bibinfo {series} {American Astronomical Society
  Meeting Abstracts}, Vol.\ \bibinfo {volume} {213}\ (\bibinfo {year} {2009})\
  p.\ \bibinfo {pages} {460.03}\BibitemShut {NoStop}%
\bibitem [{\citenamefont {Laureijs}\ \emph {et~al.}(2011)\citenamefont
  {Laureijs}, \citenamefont {Amiaux}, \citenamefont {Arduini}, \citenamefont
  {Augueres}, \citenamefont {Brinchmann}, \citenamefont {Cole}, \citenamefont
  {Cropper}, \citenamefont {Dabin}, \citenamefont {Duvet}, \citenamefont
  {Ealet} \emph {et~al.}}]{laureijs2011euclid}%
  \BibitemOpen
  \bibfield  {author} {\bibinfo {author} {\bibfnamefont {R.}~\bibnamefont
  {Laureijs}}, \bibinfo {author} {\bibfnamefont {J.}~\bibnamefont {Amiaux}},
  \bibinfo {author} {\bibfnamefont {S.}~\bibnamefont {Arduini}}, \bibinfo
  {author} {\bibfnamefont {J.-L.}\ \bibnamefont {Augueres}}, \bibinfo {author}
  {\bibfnamefont {J.}~\bibnamefont {Brinchmann}}, \bibinfo {author}
  {\bibfnamefont {R.}~\bibnamefont {Cole}}, \bibinfo {author} {\bibfnamefont
  {M.}~\bibnamefont {Cropper}}, \bibinfo {author} {\bibfnamefont
  {C.}~\bibnamefont {Dabin}}, \bibinfo {author} {\bibfnamefont
  {L.}~\bibnamefont {Duvet}}, \bibinfo {author} {\bibfnamefont
  {A.}~\bibnamefont {Ealet}},  \emph {et~al.},\ }\href@noop {} {\bibfield
  {journal} {\bibinfo  {journal} {arXiv e-prints}\ ,\ \bibinfo {eid}
  {arXiv:1110.3193}} (\bibinfo {year} {2011})},\ \Eprint
  {http://arxiv.org/abs/1110.3193} {arXiv:1110.3193 [astro-ph.CO]} \BibitemShut
  {NoStop}%
\bibitem [{\citenamefont {{Cautun}}\ \emph {et~al.}(2018)\citenamefont
  {{Cautun}}, \citenamefont {{Paillas}}, \citenamefont {{Cai}}, \citenamefont
  {{Bose}}, \citenamefont {{Armijo}}, \citenamefont {{Li}},\ and\ \citenamefont
  {{Padilla}}}]{cautun2018santiago}%
  \BibitemOpen
  \bibfield  {author} {\bibinfo {author} {\bibfnamefont {M.}~\bibnamefont
  {{Cautun}}}, \bibinfo {author} {\bibfnamefont {E.}~\bibnamefont {{Paillas}}},
  \bibinfo {author} {\bibfnamefont {Y.-C.}\ \bibnamefont {{Cai}}}, \bibinfo
  {author} {\bibfnamefont {S.}~\bibnamefont {{Bose}}}, \bibinfo {author}
  {\bibfnamefont {J.}~\bibnamefont {{Armijo}}}, \bibinfo {author}
  {\bibfnamefont {B.}~\bibnamefont {{Li}}}, \ and\ \bibinfo {author}
  {\bibfnamefont {N.}~\bibnamefont {{Padilla}}},\ }\href {\doibase
  10.1093/mnras/sty463} {\bibfield  {journal} {\bibinfo  {journal} {\mnras}\
  }\textbf {\bibinfo {volume} {476}},\ \bibinfo {pages} {3195} (\bibinfo {year}
  {2018})},\ \Eprint {http://arxiv.org/abs/1710.01730} {arXiv:1710.01730
  [astro-ph.CO]} \BibitemShut {NoStop}%
\bibitem [{\citenamefont {{Li}}\ \emph {et~al.}(2012)\citenamefont {{Li}},
  \citenamefont {{Zhao}}, \citenamefont {{Teyssier}},\ and\ \citenamefont
  {{Koyama}}}]{li2012roy}%
  \BibitemOpen
  \bibfield  {author} {\bibinfo {author} {\bibfnamefont {B.}~\bibnamefont
  {{Li}}}, \bibinfo {author} {\bibfnamefont {G.-B.}\ \bibnamefont {{Zhao}}},
  \bibinfo {author} {\bibfnamefont {R.}~\bibnamefont {{Teyssier}}}, \ and\
  \bibinfo {author} {\bibfnamefont {K.}~\bibnamefont {{Koyama}}},\ }\href
  {\doibase 10.1088/1475-7516/2012/01/051} {\bibfield  {journal} {\bibinfo
  {journal} {\jcap}\ }\textbf {\bibinfo {volume} {2012}},\ \bibinfo {eid} {051}
  (\bibinfo {year} {2012})},\ \Eprint {http://arxiv.org/abs/1110.1379}
  {arXiv:1110.1379 [astro-ph.CO]} \BibitemShut {NoStop}%
\bibitem [{\citenamefont {{Hinshaw}}\ \emph {et~al.}(2013)\citenamefont
  {{Hinshaw}}, \citenamefont {{Larson}}, \citenamefont {{Komatsu}},
  \citenamefont {{Spergel}}, \citenamefont {{Bennett}}, \citenamefont
  {{Dunkley}}, \citenamefont {{Nolta}}, \citenamefont {{Halpern}},
  \citenamefont {{Hill}}, \citenamefont {{Odegard}}, \citenamefont {{Page}},
  \citenamefont {{Smith}}, \citenamefont {{Weiland}}, \citenamefont {{Gold}},
  \citenamefont {{Jarosik}} \emph {et~al.}}]{hinshaw2013nine}%
  \BibitemOpen
  \bibfield  {author} {\bibinfo {author} {\bibfnamefont {G.}~\bibnamefont
  {{Hinshaw}}}, \bibinfo {author} {\bibfnamefont {D.}~\bibnamefont {{Larson}}},
  \bibinfo {author} {\bibfnamefont {E.}~\bibnamefont {{Komatsu}}}, \bibinfo
  {author} {\bibfnamefont {D.~N.}\ \bibnamefont {{Spergel}}}, \bibinfo {author}
  {\bibfnamefont {C.~L.}\ \bibnamefont {{Bennett}}}, \bibinfo {author}
  {\bibfnamefont {J.}~\bibnamefont {{Dunkley}}}, \bibinfo {author}
  {\bibfnamefont {M.~R.}\ \bibnamefont {{Nolta}}}, \bibinfo {author}
  {\bibfnamefont {M.}~\bibnamefont {{Halpern}}}, \bibinfo {author}
  {\bibfnamefont {R.~S.}\ \bibnamefont {{Hill}}}, \bibinfo {author}
  {\bibfnamefont {N.}~\bibnamefont {{Odegard}}}, \bibinfo {author}
  {\bibfnamefont {L.}~\bibnamefont {{Page}}}, \bibinfo {author} {\bibfnamefont
  {K.~M.}\ \bibnamefont {{Smith}}}, \bibinfo {author} {\bibfnamefont {J.~L.}\
  \bibnamefont {{Weiland}}}, \bibinfo {author} {\bibfnamefont {B.}~\bibnamefont
  {{Gold}}}, \bibinfo {author} {\bibnamefont {{Jarosik}}},  \emph {et~al.},\
  }\href {\doibase 10.1088/0067-0049/208/2/19} {\bibfield  {journal} {\bibinfo
  {journal} {\apjs}\ }\textbf {\bibinfo {volume} {208}},\ \bibinfo {eid} {19}
  (\bibinfo {year} {2013})},\ \Eprint {http://arxiv.org/abs/1212.5226}
  {arXiv:1212.5226 [astro-ph.CO]} \BibitemShut {NoStop}%
\bibitem [{\citenamefont {{Khoury}}\ and\ \citenamefont
  {{Weltman}}(2004)}]{khoury2004chameleon}%
  \BibitemOpen
  \bibfield  {author} {\bibinfo {author} {\bibfnamefont {J.}~\bibnamefont
  {{Khoury}}}\ and\ \bibinfo {author} {\bibfnamefont {A.}~\bibnamefont
  {{Weltman}}},\ }\href {\doibase 10.1103/PhysRevD.69.044026} {\bibfield
  {journal} {\bibinfo  {journal} {\prd}\ }\textbf {\bibinfo {volume} {69}},\
  \bibinfo {eid} {044026} (\bibinfo {year} {2004})},\ \Eprint
  {http://arxiv.org/abs/astro-ph/0309411} {arXiv:astro-ph/0309411 [astro-ph]}
  \BibitemShut {NoStop}%
\bibitem [{\citenamefont {{Brax}}\ \emph {et~al.}(2008)\citenamefont {{Brax}},
  \citenamefont {{van de Bruck}}, \citenamefont {{Davis}},\ and\ \citenamefont
  {{Shaw}}}]{Brax2008}%
  \BibitemOpen
  \bibfield  {author} {\bibinfo {author} {\bibfnamefont {P.}~\bibnamefont
  {{Brax}}}, \bibinfo {author} {\bibfnamefont {C.}~\bibnamefont {{van de
  Bruck}}}, \bibinfo {author} {\bibfnamefont {A.-C.}\ \bibnamefont {{Davis}}},
  \ and\ \bibinfo {author} {\bibfnamefont {D.~J.}\ \bibnamefont {{Shaw}}},\
  }\href {\doibase 10.1103/PhysRevD.78.104021} {\bibfield  {journal} {\bibinfo
  {journal} {\prd}\ }\textbf {\bibinfo {volume} {78}},\ \bibinfo {eid} {104021}
  (\bibinfo {year} {2008})},\ \Eprint {http://arxiv.org/abs/0806.3415}
  {arXiv:0806.3415 [astro-ph]} \BibitemShut {NoStop}%
\bibitem [{\citenamefont {{Arnold}}\ \emph {et~al.}(2019)\citenamefont
  {{Arnold}}, \citenamefont {{Fosalba}}, \citenamefont {{Springel}},
  \citenamefont {{Puchwein}},\ and\ \citenamefont
  {{Blot}}}]{arnold2019modified}%
  \BibitemOpen
  \bibfield  {author} {\bibinfo {author} {\bibfnamefont {C.}~\bibnamefont
  {{Arnold}}}, \bibinfo {author} {\bibfnamefont {P.}~\bibnamefont {{Fosalba}}},
  \bibinfo {author} {\bibfnamefont {V.}~\bibnamefont {{Springel}}}, \bibinfo
  {author} {\bibfnamefont {E.}~\bibnamefont {{Puchwein}}}, \ and\ \bibinfo
  {author} {\bibfnamefont {L.}~\bibnamefont {{Blot}}},\ }\href {\doibase
  10.1093/mnras/sty3044} {\bibfield  {journal} {\bibinfo  {journal} {\mnras}\
  }\textbf {\bibinfo {volume} {483}},\ \bibinfo {pages} {790} (\bibinfo {year}
  {2019})},\ \Eprint {http://arxiv.org/abs/1805.09824} {arXiv:1805.09824
  [astro-ph.CO]} \BibitemShut {NoStop}%
\bibitem [{\citenamefont {Alam}\ \emph {et~al.}(2021)\citenamefont {Alam},
  \citenamefont {Arnold}, \citenamefont {Aviles}, \citenamefont {Bean},
  \citenamefont {Cai}, \citenamefont {Cautun}, \citenamefont {Cervantes-Cota},
  \citenamefont {Cuesta-Lazaro}, \citenamefont {Devi}, \citenamefont
  {Eggemeier} \emph {et~al.}}]{alam2020testing}%
  \BibitemOpen
  \bibfield  {author} {\bibinfo {author} {\bibfnamefont {S.}~\bibnamefont
  {Alam}}, \bibinfo {author} {\bibfnamefont {C.}~\bibnamefont {Arnold}},
  \bibinfo {author} {\bibfnamefont {A.}~\bibnamefont {Aviles}}, \bibinfo
  {author} {\bibfnamefont {R.}~\bibnamefont {Bean}}, \bibinfo {author}
  {\bibfnamefont {Y.-C.}\ \bibnamefont {Cai}}, \bibinfo {author} {\bibfnamefont
  {M.}~\bibnamefont {Cautun}}, \bibinfo {author} {\bibfnamefont {J.~L.}\
  \bibnamefont {Cervantes-Cota}}, \bibinfo {author} {\bibfnamefont
  {C.}~\bibnamefont {Cuesta-Lazaro}}, \bibinfo {author} {\bibfnamefont {N.~C.}\
  \bibnamefont {Devi}}, \bibinfo {author} {\bibfnamefont {A.}~\bibnamefont
  {Eggemeier}},  \emph {et~al.},\ }\href {\doibase
  10.1088/1475-7516/2021/11/050} {\bibfield  {journal} {\bibinfo  {journal}
  {\jcap}\ }\textbf {\bibinfo {volume} {2021}},\ \bibinfo {eid} {050} (\bibinfo
  {year} {2021})},\ \Eprint {http://arxiv.org/abs/2011.05771} {arXiv:2011.05771
  [astro-ph.CO]} \BibitemShut {NoStop}%
\bibitem [{\citenamefont {{Babichev}}\ and\ \citenamefont
  {{Deffayet}}(2013)}]{babichev2013introduction}%
  \BibitemOpen
  \bibfield  {author} {\bibinfo {author} {\bibfnamefont {E.}~\bibnamefont
  {{Babichev}}}\ and\ \bibinfo {author} {\bibfnamefont {C.}~\bibnamefont
  {{Deffayet}}},\ }\href {\doibase 10.1088/0264-9381/30/18/184001} {\bibfield
  {journal} {\bibinfo  {journal} {Classical and Quantum Gravity}\ }\textbf
  {\bibinfo {volume} {30}},\ \bibinfo {eid} {184001} (\bibinfo {year}
  {2013})},\ \Eprint {http://arxiv.org/abs/1304.7240} {arXiv:1304.7240 [gr-qc]}
  \BibitemShut {NoStop}%
\bibitem [{\citenamefont {{Behroozi}}\ \emph {et~al.}(2013)\citenamefont
  {{Behroozi}}, \citenamefont {{Wechsler}},\ and\ \citenamefont
  {{Wu}}}]{behroozi2012rockstar}%
  \BibitemOpen
  \bibfield  {author} {\bibinfo {author} {\bibfnamefont {P.~S.}\ \bibnamefont
  {{Behroozi}}}, \bibinfo {author} {\bibfnamefont {R.~H.}\ \bibnamefont
  {{Wechsler}}}, \ and\ \bibinfo {author} {\bibfnamefont {H.-Y.}\ \bibnamefont
  {{Wu}}},\ }\href {\doibase 10.1088/0004-637X/762/2/109} {\bibfield  {journal}
  {\bibinfo  {journal} {\apj}\ }\textbf {\bibinfo {volume} {762}},\ \bibinfo
  {eid} {109} (\bibinfo {year} {2013})},\ \Eprint
  {http://arxiv.org/abs/1110.4372} {arXiv:1110.4372 [astro-ph.CO]} \BibitemShut
  {NoStop}%
\bibitem [{\citenamefont {Manera}\ \emph {et~al.}(2013)\citenamefont {Manera},
  \citenamefont {Scoccimarro}, \citenamefont {Percival}, \citenamefont
  {Samushia}, \citenamefont {McBride}, \citenamefont {Ross}, \citenamefont
  {Sheth}, \citenamefont {White}, \citenamefont {Reid}, \citenamefont
  {S{\'a}nchez} \emph {et~al.}}]{manera2013clustering}%
  \BibitemOpen
  \bibfield  {author} {\bibinfo {author} {\bibfnamefont {M.}~\bibnamefont
  {Manera}}, \bibinfo {author} {\bibfnamefont {R.}~\bibnamefont {Scoccimarro}},
  \bibinfo {author} {\bibfnamefont {W.~J.}\ \bibnamefont {Percival}}, \bibinfo
  {author} {\bibfnamefont {L.}~\bibnamefont {Samushia}}, \bibinfo {author}
  {\bibfnamefont {C.~K.}\ \bibnamefont {McBride}}, \bibinfo {author}
  {\bibfnamefont {A.~J.}\ \bibnamefont {Ross}}, \bibinfo {author}
  {\bibfnamefont {R.~K.}\ \bibnamefont {Sheth}}, \bibinfo {author}
  {\bibfnamefont {M.}~\bibnamefont {White}}, \bibinfo {author} {\bibfnamefont
  {B.~A.}\ \bibnamefont {Reid}}, \bibinfo {author} {\bibfnamefont {A.~G.}\
  \bibnamefont {S{\'a}nchez}},  \emph {et~al.},\ }\href {\doibase
  10.1093/mnras/sts084} {\bibfield  {journal} {\bibinfo  {journal} {\mnras}\
  }\textbf {\bibinfo {volume} {428}},\ \bibinfo {pages} {1036} (\bibinfo {year}
  {2013})},\ \Eprint {http://arxiv.org/abs/1203.6609} {arXiv:1203.6609
  [astro-ph.CO]} \BibitemShut {NoStop}%
\bibitem [{\citenamefont {{Blaizot}}\ \emph {et~al.}(2005)\citenamefont
  {{Blaizot}}, \citenamefont {{Wadadekar}}, \citenamefont {{Guiderdoni}},
  \citenamefont {{Colombi}}, \citenamefont {{Bertin}}, \citenamefont
  {{Bouchet}}, \citenamefont {{Devriendt}},\ and\ \citenamefont
  {{Hatton}}}]{blaizot2005momaf}%
  \BibitemOpen
  \bibfield  {author} {\bibinfo {author} {\bibfnamefont {J.}~\bibnamefont
  {{Blaizot}}}, \bibinfo {author} {\bibfnamefont {Y.}~\bibnamefont
  {{Wadadekar}}}, \bibinfo {author} {\bibfnamefont {B.}~\bibnamefont
  {{Guiderdoni}}}, \bibinfo {author} {\bibfnamefont {S.~T.}\ \bibnamefont
  {{Colombi}}}, \bibinfo {author} {\bibfnamefont {E.}~\bibnamefont {{Bertin}}},
  \bibinfo {author} {\bibfnamefont {F.~R.}\ \bibnamefont {{Bouchet}}}, \bibinfo
  {author} {\bibfnamefont {J.~E.~G.}\ \bibnamefont {{Devriendt}}}, \ and\
  \bibinfo {author} {\bibfnamefont {S.}~\bibnamefont {{Hatton}}},\ }\href
  {\doibase 10.1111/j.1365-2966.2005.09019.x} {\bibfield  {journal} {\bibinfo
  {journal} {\mnras}\ }\textbf {\bibinfo {volume} {360}},\ \bibinfo {pages}
  {159} (\bibinfo {year} {2005})},\ \Eprint
  {http://arxiv.org/abs/astro-ph/0309305} {arXiv:astro-ph/0309305 [astro-ph]}
  \BibitemShut {NoStop}%
\bibitem [{\citenamefont {{Overzier}}\ \emph {et~al.}(2013)\citenamefont
  {{Overzier}}, \citenamefont {{Lemson}}, \citenamefont {{Angulo}},
  \citenamefont {{Bertin}}, \citenamefont {{Blaizot}}, \citenamefont
  {{Henriques}}, \citenamefont {{Marleau}},\ and\ \citenamefont
  {{White}}}]{overzier2013millennium}%
  \BibitemOpen
  \bibfield  {author} {\bibinfo {author} {\bibfnamefont {R.}~\bibnamefont
  {{Overzier}}}, \bibinfo {author} {\bibfnamefont {G.}~\bibnamefont
  {{Lemson}}}, \bibinfo {author} {\bibfnamefont {R.~E.}\ \bibnamefont
  {{Angulo}}}, \bibinfo {author} {\bibfnamefont {E.}~\bibnamefont {{Bertin}}},
  \bibinfo {author} {\bibfnamefont {J.}~\bibnamefont {{Blaizot}}}, \bibinfo
  {author} {\bibfnamefont {B.~M.~B.}\ \bibnamefont {{Henriques}}}, \bibinfo
  {author} {\bibfnamefont {G.~D.}\ \bibnamefont {{Marleau}}}, \ and\ \bibinfo
  {author} {\bibfnamefont {S.~D.~M.}\ \bibnamefont {{White}}},\ }\href
  {\doibase 10.1093/mnras/sts076} {\bibfield  {journal} {\bibinfo  {journal}
  {\mnras}\ }\textbf {\bibinfo {volume} {428}},\ \bibinfo {pages} {778}
  (\bibinfo {year} {2013})},\ \Eprint {http://arxiv.org/abs/1206.6923}
  {arXiv:1206.6923 [astro-ph.CO]} \BibitemShut {NoStop}%
\bibitem [{\citenamefont {{Smith}}\ \emph {et~al.}(2017)\citenamefont
  {{Smith}}, \citenamefont {{Cole}}, \citenamefont {{Baugh}}, \citenamefont
  {{Zheng}}, \citenamefont {{Angulo}}, \citenamefont {{Norberg}},\ and\
  \citenamefont {{Zehavi}}}]{smith2017lightcone}%
  \BibitemOpen
  \bibfield  {author} {\bibinfo {author} {\bibfnamefont {A.}~\bibnamefont
  {{Smith}}}, \bibinfo {author} {\bibfnamefont {S.}~\bibnamefont {{Cole}}},
  \bibinfo {author} {\bibfnamefont {C.}~\bibnamefont {{Baugh}}}, \bibinfo
  {author} {\bibfnamefont {Z.}~\bibnamefont {{Zheng}}}, \bibinfo {author}
  {\bibfnamefont {R.}~\bibnamefont {{Angulo}}}, \bibinfo {author}
  {\bibfnamefont {P.}~\bibnamefont {{Norberg}}}, \ and\ \bibinfo {author}
  {\bibfnamefont {I.}~\bibnamefont {{Zehavi}}},\ }\href {\doibase
  10.1093/mnras/stx1432} {\bibfield  {journal} {\bibinfo  {journal} {\mnras}\
  }\textbf {\bibinfo {volume} {470}},\ \bibinfo {pages} {4646} (\bibinfo {year}
  {2017})},\ \Eprint {http://arxiv.org/abs/1701.06581} {arXiv:1701.06581
  [astro-ph.CO]} \BibitemShut {NoStop}%
\bibitem [{\citenamefont {{Coles}}\ and\ \citenamefont
  {{Jones}}(1991)}]{coles1991lognormal}%
  \BibitemOpen
  \bibfield  {author} {\bibinfo {author} {\bibfnamefont {P.}~\bibnamefont
  {{Coles}}}\ and\ \bibinfo {author} {\bibfnamefont {B.}~\bibnamefont
  {{Jones}}},\ }\href {\doibase 10.1093/mnras/248.1.1} {\bibfield  {journal}
  {\bibinfo  {journal} {\mnras}\ }\textbf {\bibinfo {volume} {248}},\ \bibinfo
  {pages} {1} (\bibinfo {year} {1991})}\BibitemShut {NoStop}%
\bibitem [{\citenamefont {{Bouchet}}\ and\ \citenamefont
  {{Hernquist}}(1992)}]{Bouchet1992}%
  \BibitemOpen
  \bibfield  {author} {\bibinfo {author} {\bibfnamefont {F.~R.}\ \bibnamefont
  {{Bouchet}}}\ and\ \bibinfo {author} {\bibfnamefont {L.}~\bibnamefont
  {{Hernquist}}},\ }\href {\doibase 10.1086/171970} {\bibfield  {journal}
  {\bibinfo  {journal} {\apj}\ }\textbf {\bibinfo {volume} {400}},\ \bibinfo
  {pages} {25} (\bibinfo {year} {1992})}\BibitemShut {NoStop}%
\bibitem [{\citenamefont {{Gaztanaga}}\ and\ \citenamefont
  {{Bernardeau}}(1998)}]{gaztanaga1997skewness}%
  \BibitemOpen
  \bibfield  {author} {\bibinfo {author} {\bibfnamefont {E.}~\bibnamefont
  {{Gaztanaga}}}\ and\ \bibinfo {author} {\bibfnamefont {F.}~\bibnamefont
  {{Bernardeau}}},\ }\href@noop {} {\bibfield  {journal} {\bibinfo  {journal}
  {\aap}\ }\textbf {\bibinfo {volume} {331}},\ \bibinfo {pages} {829} (\bibinfo
  {year} {1998})},\ \Eprint {http://arxiv.org/abs/astro-ph/9707095}
  {arXiv:astro-ph/9707095 [astro-ph]} \BibitemShut {NoStop}%
\bibitem [{\citenamefont {{Pollo}}(1997)}]{pollo1997gravitational}%
  \BibitemOpen
  \bibfield  {author} {\bibinfo {author} {\bibfnamefont {A.}~\bibnamefont
  {{Pollo}}},\ }\href@noop {} {\bibfield  {journal} {\bibinfo  {journal}
  {\actaa}\ }\textbf {\bibinfo {volume} {47}},\ \bibinfo {pages} {413}
  (\bibinfo {year} {1997})}\BibitemShut {NoStop}%
\bibitem [{\citenamefont {{Smith}}\ \emph {et~al.}(2003)\citenamefont
  {{Smith}}, \citenamefont {{Peacock}}, \citenamefont {{Jenkins}},
  \citenamefont {{White}}, \citenamefont {{Frenk}}, \citenamefont {{Pearce}},
  \citenamefont {{Thomas}}, \citenamefont {{Efstathiou}},\ and\ \citenamefont
  {{Couchman}}}]{Smith2003stable}%
  \BibitemOpen
  \bibfield  {author} {\bibinfo {author} {\bibfnamefont {R.~E.}\ \bibnamefont
  {{Smith}}}, \bibinfo {author} {\bibfnamefont {J.~A.}\ \bibnamefont
  {{Peacock}}}, \bibinfo {author} {\bibfnamefont {A.}~\bibnamefont
  {{Jenkins}}}, \bibinfo {author} {\bibfnamefont {S.~D.~M.}\ \bibnamefont
  {{White}}}, \bibinfo {author} {\bibfnamefont {C.~S.}\ \bibnamefont
  {{Frenk}}}, \bibinfo {author} {\bibfnamefont {F.~R.}\ \bibnamefont
  {{Pearce}}}, \bibinfo {author} {\bibfnamefont {P.~A.}\ \bibnamefont
  {{Thomas}}}, \bibinfo {author} {\bibfnamefont {G.}~\bibnamefont
  {{Efstathiou}}}, \ and\ \bibinfo {author} {\bibfnamefont {H.~M.~P.}\
  \bibnamefont {{Couchman}}},\ }\href {\doibase
  10.1046/j.1365-8711.2003.06503.x} {\bibfield  {journal} {\bibinfo  {journal}
  {\mnras}\ }\textbf {\bibinfo {volume} {341}},\ \bibinfo {pages} {1311}
  (\bibinfo {year} {2003})},\ \Eprint {http://arxiv.org/abs/astro-ph/0207664}
  {arXiv:astro-ph/0207664 [astro-ph]} \BibitemShut {NoStop}%
\bibitem [{\citenamefont {{Lewis}}\ \emph {et~al.}(2000)\citenamefont
  {{Lewis}}, \citenamefont {{Challinor}},\ and\ \citenamefont
  {{Lasenby}}}]{lewis2000efficient}%
  \BibitemOpen
  \bibfield  {author} {\bibinfo {author} {\bibfnamefont {A.}~\bibnamefont
  {{Lewis}}}, \bibinfo {author} {\bibfnamefont {A.}~\bibnamefont
  {{Challinor}}}, \ and\ \bibinfo {author} {\bibfnamefont {A.}~\bibnamefont
  {{Lasenby}}},\ }\href {\doibase 10.1086/309179} {\bibfield  {journal}
  {\bibinfo  {journal} {\apj}\ }\textbf {\bibinfo {volume} {538}},\ \bibinfo
  {pages} {473} (\bibinfo {year} {2000})},\ \Eprint
  {http://arxiv.org/abs/astro-ph/9911177} {arXiv:astro-ph/9911177 [astro-ph]}
  \BibitemShut {NoStop}%
\bibitem [{\citenamefont {{Li}}\ \emph {et~al.}(2013)\citenamefont {{Li}},
  \citenamefont {{Hellwing}}, \citenamefont {{Koyama}}, \citenamefont {{Zhao}},
  \citenamefont {{Jennings}},\ and\ \citenamefont {{Baugh}}}]{Li2013}%
  \BibitemOpen
  \bibfield  {author} {\bibinfo {author} {\bibfnamefont {B.}~\bibnamefont
  {{Li}}}, \bibinfo {author} {\bibfnamefont {W.~A.}\ \bibnamefont
  {{Hellwing}}}, \bibinfo {author} {\bibfnamefont {K.}~\bibnamefont
  {{Koyama}}}, \bibinfo {author} {\bibfnamefont {G.-B.}\ \bibnamefont
  {{Zhao}}}, \bibinfo {author} {\bibfnamefont {E.}~\bibnamefont {{Jennings}}},
  \ and\ \bibinfo {author} {\bibfnamefont {C.~M.}\ \bibnamefont {{Baugh}}},\
  }\href {\doibase 10.1093/mnras/sts072} {\bibfield  {journal} {\bibinfo
  {journal} {\mnras}\ }\textbf {\bibinfo {volume} {428}},\ \bibinfo {pages}
  {743} (\bibinfo {year} {2013})},\ \Eprint {http://arxiv.org/abs/1206.4317}
  {arXiv:1206.4317 [astro-ph.CO]} \BibitemShut {NoStop}%
\bibitem [{\citenamefont {{Hellwing}}\ \emph {et~al.}(2013)\citenamefont
  {{Hellwing}}, \citenamefont {{Li}}, \citenamefont {{Frenk}},\ and\
  \citenamefont {{Cole}}}]{hellwing2013hierarchical}%
  \BibitemOpen
  \bibfield  {author} {\bibinfo {author} {\bibfnamefont {W.~A.}\ \bibnamefont
  {{Hellwing}}}, \bibinfo {author} {\bibfnamefont {B.}~\bibnamefont {{Li}}},
  \bibinfo {author} {\bibfnamefont {C.~S.}\ \bibnamefont {{Frenk}}}, \ and\
  \bibinfo {author} {\bibfnamefont {S.}~\bibnamefont {{Cole}}},\ }\href
  {\doibase 10.1093/mnras/stt1430} {\bibfield  {journal} {\bibinfo  {journal}
  {\mnras}\ }\textbf {\bibinfo {volume} {435}},\ \bibinfo {pages} {2806}
  (\bibinfo {year} {2013})},\ \Eprint {http://arxiv.org/abs/1305.7486}
  {arXiv:1305.7486 [astro-ph.CO]} \BibitemShut {NoStop}%
\bibitem [{\citenamefont {{Hellwing}}\ \emph {et~al.}(2017)\citenamefont
  {{Hellwing}}, \citenamefont {{Koyama}}, \citenamefont {{Bose}},\ and\
  \citenamefont {{Zhao}}}]{hellwing2017revealing}%
  \BibitemOpen
  \bibfield  {author} {\bibinfo {author} {\bibfnamefont {W.~A.}\ \bibnamefont
  {{Hellwing}}}, \bibinfo {author} {\bibfnamefont {K.}~\bibnamefont
  {{Koyama}}}, \bibinfo {author} {\bibfnamefont {B.}~\bibnamefont {{Bose}}}, \
  and\ \bibinfo {author} {\bibfnamefont {G.-B.}\ \bibnamefont {{Zhao}}},\
  }\href {\doibase 10.1103/PhysRevD.96.023515} {\bibfield  {journal} {\bibinfo
  {journal} {\prd}\ }\textbf {\bibinfo {volume} {96}},\ \bibinfo {eid} {023515}
  (\bibinfo {year} {2017})},\ \Eprint {http://arxiv.org/abs/1703.03395}
  {arXiv:1703.03395 [astro-ph.CO]} \BibitemShut {NoStop}%
\bibitem [{\citenamefont {{Bacon}}\ and\ \citenamefont
  {{Taylor}}(2003)}]{Bacon2003}%
  \BibitemOpen
  \bibfield  {author} {\bibinfo {author} {\bibfnamefont {D.~J.}\ \bibnamefont
  {{Bacon}}}\ and\ \bibinfo {author} {\bibfnamefont {A.~N.}\ \bibnamefont
  {{Taylor}}},\ }\href {\doibase 10.1046/j.1365-8711.2003.06922.x} {\bibfield
  {journal} {\bibinfo  {journal} {\mnras}\ }\textbf {\bibinfo {volume} {344}},\
  \bibinfo {pages} {1307} (\bibinfo {year} {2003})},\ \Eprint
  {http://arxiv.org/abs/astro-ph/0212266} {arXiv:astro-ph/0212266 [astro-ph]}
  \BibitemShut {NoStop}%
\bibitem [{\citenamefont {{Jain}}\ and\ \citenamefont
  {{Taylor}}(2003)}]{Jain2003}%
  \BibitemOpen
  \bibfield  {author} {\bibinfo {author} {\bibfnamefont {B.}~\bibnamefont
  {{Jain}}}\ and\ \bibinfo {author} {\bibfnamefont {A.}~\bibnamefont
  {{Taylor}}},\ }\href {\doibase 10.1103/PhysRevLett.91.141302} {\bibfield
  {journal} {\bibinfo  {journal} {\prl}\ }\textbf {\bibinfo {volume} {91}},\
  \bibinfo {eid} {141302} (\bibinfo {year} {2003})},\ \Eprint
  {http://arxiv.org/abs/astro-ph/0306046} {arXiv:astro-ph/0306046 [astro-ph]}
  \BibitemShut {NoStop}%
\bibitem [{\citenamefont {{Massey}}\ \emph {et~al.}(2007)\citenamefont
  {{Massey}}, \citenamefont {{Rhodes}}, \citenamefont {{Ellis}}, \citenamefont
  {{Scoville}}, \citenamefont {{Leauthaud}}, \citenamefont {{Finoguenov}},
  \citenamefont {{Capak}}, \citenamefont {{Bacon}}, \citenamefont {{Aussel}},
  \citenamefont {{Kneib}}, \citenamefont {{Koekemoer}}, \citenamefont
  {{McCracken}}, \citenamefont {{Mobasher}}, \citenamefont {{Pires}},
  \citenamefont {{Refregier}} \emph {et~al.}}]{Massey2007}%
  \BibitemOpen
  \bibfield  {author} {\bibinfo {author} {\bibfnamefont {R.}~\bibnamefont
  {{Massey}}}, \bibinfo {author} {\bibfnamefont {J.}~\bibnamefont {{Rhodes}}},
  \bibinfo {author} {\bibfnamefont {R.}~\bibnamefont {{Ellis}}}, \bibinfo
  {author} {\bibfnamefont {N.}~\bibnamefont {{Scoville}}}, \bibinfo {author}
  {\bibfnamefont {A.}~\bibnamefont {{Leauthaud}}}, \bibinfo {author}
  {\bibfnamefont {A.}~\bibnamefont {{Finoguenov}}}, \bibinfo {author}
  {\bibfnamefont {P.}~\bibnamefont {{Capak}}}, \bibinfo {author} {\bibfnamefont
  {D.}~\bibnamefont {{Bacon}}}, \bibinfo {author} {\bibfnamefont
  {H.}~\bibnamefont {{Aussel}}}, \bibinfo {author} {\bibfnamefont {J.-P.}\
  \bibnamefont {{Kneib}}}, \bibinfo {author} {\bibfnamefont {A.}~\bibnamefont
  {{Koekemoer}}}, \bibinfo {author} {\bibfnamefont {H.}~\bibnamefont
  {{McCracken}}}, \bibinfo {author} {\bibfnamefont {B.}~\bibnamefont
  {{Mobasher}}}, \bibinfo {author} {\bibfnamefont {S.}~\bibnamefont {{Pires}}},
  \bibinfo {author} {\bibnamefont {{Refregier}}},  \emph {et~al.},\ }\href
  {\doibase 10.1038/nature05497} {\bibfield  {journal} {\bibinfo  {journal}
  {\nat}\ }\textbf {\bibinfo {volume} {445}},\ \bibinfo {pages} {286} (\bibinfo
  {year} {2007})},\ \Eprint {http://arxiv.org/abs/astro-ph/0701594}
  {arXiv:astro-ph/0701594 [astro-ph]} \BibitemShut {NoStop}%
\bibitem [{\citenamefont {{Hivon}}\ \emph {et~al.}(1995)\citenamefont
  {{Hivon}}, \citenamefont {{Bouchet}}, \citenamefont {{Colombi}},\ and\
  \citenamefont {{Juszkiewicz}}}]{Hivon95}%
  \BibitemOpen
  \bibfield  {author} {\bibinfo {author} {\bibfnamefont {E.}~\bibnamefont
  {{Hivon}}}, \bibinfo {author} {\bibfnamefont {F.~R.}\ \bibnamefont
  {{Bouchet}}}, \bibinfo {author} {\bibfnamefont {S.}~\bibnamefont
  {{Colombi}}}, \ and\ \bibinfo {author} {\bibfnamefont {R.}~\bibnamefont
  {{Juszkiewicz}}},\ }\href@noop {} {\bibfield  {journal} {\bibinfo  {journal}
  {\aap}\ }\textbf {\bibinfo {volume} {298}},\ \bibinfo {pages} {643} (\bibinfo
  {year} {1995})},\ \Eprint {http://arxiv.org/abs/astro-ph/9407049}
  {arXiv:astro-ph/9407049 [astro-ph]} \BibitemShut {NoStop}%
\bibitem [{\citenamefont {{Okamura}}\ \emph {et~al.}(2011)\citenamefont
  {{Okamura}}, \citenamefont {{Taruya}},\ and\ \citenamefont
  {{Matsubara}}}]{Okamura2011}%
  \BibitemOpen
  \bibfield  {author} {\bibinfo {author} {\bibfnamefont {T.}~\bibnamefont
  {{Okamura}}}, \bibinfo {author} {\bibfnamefont {A.}~\bibnamefont {{Taruya}}},
  \ and\ \bibinfo {author} {\bibfnamefont {T.}~\bibnamefont {{Matsubara}}},\
  }\href {\doibase 10.1088/1475-7516/2011/08/012} {\bibfield  {journal}
  {\bibinfo  {journal} {\jcap}\ }\textbf {\bibinfo {volume} {2011}},\ \bibinfo
  {eid} {012} (\bibinfo {year} {2011})},\ \Eprint
  {http://arxiv.org/abs/1105.1491} {arXiv:1105.1491 [astro-ph.CO]} \BibitemShut
  {NoStop}%
\bibitem [{\citenamefont {{Kitaura}}\ and\ \citenamefont
  {{Angulo}}(2012)}]{Kitaura2012}%
  \BibitemOpen
  \bibfield  {author} {\bibinfo {author} {\bibfnamefont {F.-S.}\ \bibnamefont
  {{Kitaura}}}\ and\ \bibinfo {author} {\bibfnamefont {R.~E.}\ \bibnamefont
  {{Angulo}}},\ }\href {\doibase 10.1111/j.1365-2966.2012.21614.x} {\bibfield
  {journal} {\bibinfo  {journal} {\mnras}\ }\textbf {\bibinfo {volume} {425}},\
  \bibinfo {pages} {2443} (\bibinfo {year} {2012})},\ \Eprint
  {http://arxiv.org/abs/1111.6617} {arXiv:1111.6617 [astro-ph.CO]} \BibitemShut
  {NoStop}%
\bibitem [{\citenamefont {{Fry}}\ and\ \citenamefont
  {{Gaztanaga}}(1993)}]{Fry1993}%
  \BibitemOpen
  \bibfield  {author} {\bibinfo {author} {\bibfnamefont {J.~N.}\ \bibnamefont
  {{Fry}}}\ and\ \bibinfo {author} {\bibfnamefont {E.}~\bibnamefont
  {{Gaztanaga}}},\ }\href {\doibase 10.1086/173015} {\bibfield  {journal}
  {\bibinfo  {journal} {\apj}\ }\textbf {\bibinfo {volume} {413}},\ \bibinfo
  {pages} {447} (\bibinfo {year} {1993})},\ \Eprint
  {http://arxiv.org/abs/astro-ph/9302009} {arXiv:astro-ph/9302009 [astro-ph]}
  \BibitemShut {NoStop}%
\bibitem [{\citenamefont {{Hellwing}}(2020)}]{hellwing2019skewness}%
  \BibitemOpen
  \bibfield  {author} {\bibinfo {author} {\bibfnamefont {W.~A.}\ \bibnamefont
  {{Hellwing}}},\ }\bibfield  {booktitle} {\emph {\bibinfo {booktitle} {XXXIX
  Polish Astronomical Society Meeting}},\ }\href@noop {} {\ \textbf {\bibinfo
  {volume} {10}},\ \bibinfo {pages} {315} (\bibinfo {year} {2020})},\ \Eprint
  {http://arxiv.org/abs/1912.13026} {arXiv:1912.13026 [astro-ph.CO]}
  \BibitemShut {NoStop}%
\bibitem [{\citenamefont {{Shi}}\ \emph {et~al.}(2015)\citenamefont {{Shi}},
  \citenamefont {{Li}}, \citenamefont {{Han}}, \citenamefont {{Gao}},\ and\
  \citenamefont {{Hellwing}}}]{Shi2015}%
  \BibitemOpen
  \bibfield  {author} {\bibinfo {author} {\bibfnamefont {D.}~\bibnamefont
  {{Shi}}}, \bibinfo {author} {\bibfnamefont {B.}~\bibnamefont {{Li}}},
  \bibinfo {author} {\bibfnamefont {J.}~\bibnamefont {{Han}}}, \bibinfo
  {author} {\bibfnamefont {L.}~\bibnamefont {{Gao}}}, \ and\ \bibinfo {author}
  {\bibfnamefont {W.~A.}\ \bibnamefont {{Hellwing}}},\ }\href {\doibase
  10.1093/mnras/stv1549} {\bibfield  {journal} {\bibinfo  {journal} {\mnras}\
  }\textbf {\bibinfo {volume} {452}},\ \bibinfo {pages} {3179} (\bibinfo {year}
  {2015})},\ \Eprint {http://arxiv.org/abs/1503.01109} {arXiv:1503.01109
  [astro-ph.CO]} \BibitemShut {NoStop}%
\bibitem [{\citenamefont {{Ross}}\ \emph {et~al.}(2007)\citenamefont {{Ross}},
  \citenamefont {{Brunner}},\ and\ \citenamefont {{Myers}}}]{ross2007higher}%
  \BibitemOpen
  \bibfield  {author} {\bibinfo {author} {\bibfnamefont {A.~J.}\ \bibnamefont
  {{Ross}}}, \bibinfo {author} {\bibfnamefont {R.~J.}\ \bibnamefont
  {{Brunner}}}, \ and\ \bibinfo {author} {\bibfnamefont {A.~D.}\ \bibnamefont
  {{Myers}}},\ }\href {\doibase 10.1086/519020} {\bibfield  {journal} {\bibinfo
   {journal} {\apj}\ }\textbf {\bibinfo {volume} {665}},\ \bibinfo {pages} {67}
  (\bibinfo {year} {2007})},\ \Eprint {http://arxiv.org/abs/0704.2573}
  {arXiv:0704.2573 [astro-ph]} \BibitemShut {NoStop}%
\bibitem [{\citenamefont {{Hamilton}}(1998)}]{Hamilton1998}%
  \BibitemOpen
  \bibfield  {author} {\bibinfo {author} {\bibfnamefont {A.~J.~S.}\
  \bibnamefont {{Hamilton}}},\ }in\ \href {\doibase
  10.1007/978-94-011-4960-0\_17} {\emph {\bibinfo {booktitle} {The Evolving
  Universe}}},\ \bibinfo {series} {Astrophysics and Space Science Library},
  Vol.\ \bibinfo {volume} {231},\ \bibinfo {editor} {edited by\ \bibinfo
  {editor} {\bibfnamefont {D.}~\bibnamefont {{Hamilton}}}}\ (\bibinfo {year}
  {1998})\ p.\ \bibinfo {pages} {185},\ \Eprint
  {http://arxiv.org/abs/astro-ph/9708102} {arXiv:astro-ph/9708102 [astro-ph]}
  \BibitemShut {NoStop}%
\bibitem [{\citenamefont {{Arnalte-Mur}}\ \emph {et~al.}(2017)\citenamefont
  {{Arnalte-Mur}}, \citenamefont {{Hellwing}},\ and\ \citenamefont
  {{Norberg}}}]{Arnalte-Mur2017}%
  \BibitemOpen
  \bibfield  {author} {\bibinfo {author} {\bibfnamefont {P.}~\bibnamefont
  {{Arnalte-Mur}}}, \bibinfo {author} {\bibfnamefont {W.~A.}\ \bibnamefont
  {{Hellwing}}}, \ and\ \bibinfo {author} {\bibfnamefont {P.}~\bibnamefont
  {{Norberg}}},\ }\href {\doibase 10.1093/mnras/stx196} {\bibfield  {journal}
  {\bibinfo  {journal} {\mnras}\ }\textbf {\bibinfo {volume} {467}},\ \bibinfo
  {pages} {1569} (\bibinfo {year} {2017})},\ \Eprint
  {http://arxiv.org/abs/1612.02355} {arXiv:1612.02355 [astro-ph.CO]}
  \BibitemShut {NoStop}%
\bibitem [{\citenamefont {{Hern{\'a}ndez-Aguayo}}\ \emph
  {et~al.}(2019)\citenamefont {{Hern{\'a}ndez-Aguayo}}, \citenamefont {{Hou}},
  \citenamefont {{Li}}, \citenamefont {{Baugh}},\ and\ \citenamefont
  {{S{\'a}nchez}}}]{Hernandez-Aguayo2019}%
  \BibitemOpen
  \bibfield  {author} {\bibinfo {author} {\bibfnamefont {C.}~\bibnamefont
  {{Hern{\'a}ndez-Aguayo}}}, \bibinfo {author} {\bibfnamefont {J.}~\bibnamefont
  {{Hou}}}, \bibinfo {author} {\bibfnamefont {B.}~\bibnamefont {{Li}}},
  \bibinfo {author} {\bibfnamefont {C.~M.}\ \bibnamefont {{Baugh}}}, \ and\
  \bibinfo {author} {\bibfnamefont {A.~G.}\ \bibnamefont {{S{\'a}nchez}}},\
  }\href {\doibase 10.1093/mnras/stz516} {\bibfield  {journal} {\bibinfo
  {journal} {\mnras}\ }\textbf {\bibinfo {volume} {485}},\ \bibinfo {pages}
  {2194} (\bibinfo {year} {2019})},\ \Eprint {http://arxiv.org/abs/1811.09197}
  {arXiv:1811.09197 [astro-ph.CO]} \BibitemShut {NoStop}%
\bibitem [{\citenamefont {De~Jong}\ \emph {et~al.}(2017)\citenamefont
  {De~Jong}, \citenamefont {Kleijn}, \citenamefont {Erben}, \citenamefont
  {Hildebrandt}, \citenamefont {Kuijken}, \citenamefont {Sikkema},
  \citenamefont {Brescia}, \citenamefont {Bilicki}, \citenamefont {Napolitano},
  \citenamefont {Amaro} \emph {et~al.}}]{de2017third}%
  \BibitemOpen
  \bibfield  {author} {\bibinfo {author} {\bibfnamefont {J.~T.}\ \bibnamefont
  {De~Jong}}, \bibinfo {author} {\bibfnamefont {G.~A.~V.}\ \bibnamefont
  {Kleijn}}, \bibinfo {author} {\bibfnamefont {T.}~\bibnamefont {Erben}},
  \bibinfo {author} {\bibfnamefont {H.}~\bibnamefont {Hildebrandt}}, \bibinfo
  {author} {\bibfnamefont {K.}~\bibnamefont {Kuijken}}, \bibinfo {author}
  {\bibfnamefont {G.}~\bibnamefont {Sikkema}}, \bibinfo {author} {\bibfnamefont
  {M.}~\bibnamefont {Brescia}}, \bibinfo {author} {\bibfnamefont
  {M.}~\bibnamefont {Bilicki}}, \bibinfo {author} {\bibfnamefont {N.~R.}\
  \bibnamefont {Napolitano}}, \bibinfo {author} {\bibfnamefont
  {V.}~\bibnamefont {Amaro}},  \emph {et~al.},\ }\href {\doibase
  10.1051/0004-6361/201730747} {\bibfield  {journal} {\bibinfo  {journal}
  {\aap}\ }\textbf {\bibinfo {volume} {604}},\ \bibinfo {eid} {A134} (\bibinfo
  {year} {2017})},\ \Eprint {http://arxiv.org/abs/1703.02991} {arXiv:1703.02991
  [astro-ph.GA]} \BibitemShut {NoStop}%
\bibitem [{\citenamefont {Tutusaus}\ \emph {et~al.}(2020)\citenamefont
  {Tutusaus}, \citenamefont {Martinelli}, \citenamefont {Cardone},
  \citenamefont {Camera}, \citenamefont {Yahia-Cherif}, \citenamefont {Casas},
  \citenamefont {Blanchard}, \citenamefont {Kilbinger}, \citenamefont {Lacasa},
  \citenamefont {Sakr} \emph {et~al.}}]{tutusaus2020euclid}%
  \BibitemOpen
  \bibfield  {author} {\bibinfo {author} {\bibfnamefont {I.}~\bibnamefont
  {Tutusaus}}, \bibinfo {author} {\bibfnamefont {M.}~\bibnamefont
  {Martinelli}}, \bibinfo {author} {\bibfnamefont {V.~F.}\ \bibnamefont
  {Cardone}}, \bibinfo {author} {\bibfnamefont {S.}~\bibnamefont {Camera}},
  \bibinfo {author} {\bibfnamefont {S.}~\bibnamefont {Yahia-Cherif}}, \bibinfo
  {author} {\bibfnamefont {S.}~\bibnamefont {Casas}}, \bibinfo {author}
  {\bibfnamefont {A.}~\bibnamefont {Blanchard}}, \bibinfo {author}
  {\bibfnamefont {M.}~\bibnamefont {Kilbinger}}, \bibinfo {author}
  {\bibfnamefont {F.}~\bibnamefont {Lacasa}}, \bibinfo {author} {\bibfnamefont
  {Z.}~\bibnamefont {Sakr}},  \emph {et~al.},\ }\href {\doibase
  10.1051/0004-6361/202038313} {\bibfield  {journal} {\bibinfo  {journal}
  {\aap}\ }\textbf {\bibinfo {volume} {643}},\ \bibinfo {eid} {A70} (\bibinfo
  {year} {2020})},\ \Eprint {http://arxiv.org/abs/2005.00055} {arXiv:2005.00055
  [astro-ph.CO]} \BibitemShut {NoStop}%
\bibitem [{\citenamefont {Abbott}\ \emph {et~al.}(2021)\citenamefont {Abbott},
  \citenamefont {Adamow}, \citenamefont {Aguena}, \citenamefont {Allam},
  \citenamefont {Amon}, \citenamefont {Annis}, \citenamefont {Avila},
  \citenamefont {Bacon}, \citenamefont {Banerji}, \citenamefont {Bechtol} \emph
  {et~al.}}]{abbott2021dark}%
  \BibitemOpen
  \bibfield  {author} {\bibinfo {author} {\bibfnamefont {T.}~\bibnamefont
  {Abbott}}, \bibinfo {author} {\bibfnamefont {M.}~\bibnamefont {Adamow}},
  \bibinfo {author} {\bibfnamefont {M.}~\bibnamefont {Aguena}}, \bibinfo
  {author} {\bibfnamefont {S.}~\bibnamefont {Allam}}, \bibinfo {author}
  {\bibfnamefont {A.}~\bibnamefont {Amon}}, \bibinfo {author} {\bibfnamefont
  {J.}~\bibnamefont {Annis}}, \bibinfo {author} {\bibfnamefont
  {S.}~\bibnamefont {Avila}}, \bibinfo {author} {\bibfnamefont
  {D.}~\bibnamefont {Bacon}}, \bibinfo {author} {\bibfnamefont
  {M.}~\bibnamefont {Banerji}}, \bibinfo {author} {\bibfnamefont
  {K.}~\bibnamefont {Bechtol}},  \emph {et~al.},\ }\href {\doibase
  10.3847/1538-4365/ac00b3} {\bibfield  {journal} {\bibinfo  {journal} {\apjs}\
  }\textbf {\bibinfo {volume} {255}},\ \bibinfo {eid} {20} (\bibinfo {year}
  {2021})},\ \Eprint {http://arxiv.org/abs/2101.05765} {arXiv:2101.05765
  [astro-ph.IM]} \BibitemShut {NoStop}%
\bibitem [{\citenamefont {{Walmsley}}\ \emph {et~al.}(2022)\citenamefont
  {{Walmsley}}, \citenamefont {{Lintott}}, \citenamefont {{G{\'e}ron}},
  \citenamefont {{Kruk}}, \citenamefont {{Krawczyk}}, \citenamefont
  {{Willett}}, \citenamefont {{Bamford}}, \citenamefont {{Kelvin}},
  \citenamefont {{Fortson}}, \citenamefont {{Gal}}, \citenamefont {{Keel}},
  \citenamefont {{Masters}}, \citenamefont {{Mehta}}, \citenamefont
  {{Simmons}}, \citenamefont {{Smethurst}} \emph
  {et~al.}}]{walmsley2022galaxy}%
  \BibitemOpen
  \bibfield  {author} {\bibinfo {author} {\bibfnamefont {M.}~\bibnamefont
  {{Walmsley}}}, \bibinfo {author} {\bibfnamefont {C.}~\bibnamefont
  {{Lintott}}}, \bibinfo {author} {\bibfnamefont {T.}~\bibnamefont
  {{G{\'e}ron}}}, \bibinfo {author} {\bibfnamefont {S.}~\bibnamefont {{Kruk}}},
  \bibinfo {author} {\bibfnamefont {C.}~\bibnamefont {{Krawczyk}}}, \bibinfo
  {author} {\bibfnamefont {K.~W.}\ \bibnamefont {{Willett}}}, \bibinfo {author}
  {\bibfnamefont {S.}~\bibnamefont {{Bamford}}}, \bibinfo {author}
  {\bibfnamefont {L.~S.}\ \bibnamefont {{Kelvin}}}, \bibinfo {author}
  {\bibfnamefont {L.}~\bibnamefont {{Fortson}}}, \bibinfo {author}
  {\bibfnamefont {Y.}~\bibnamefont {{Gal}}}, \bibinfo {author} {\bibfnamefont
  {W.}~\bibnamefont {{Keel}}}, \bibinfo {author} {\bibfnamefont {K.~L.}\
  \bibnamefont {{Masters}}}, \bibinfo {author} {\bibfnamefont {V.}~\bibnamefont
  {{Mehta}}}, \bibinfo {author} {\bibfnamefont {B.~D.}\ \bibnamefont
  {{Simmons}}}, \bibinfo {author} {\bibnamefont {{Smethurst}}},  \emph
  {et~al.},\ }\href {\doibase 10.1093/mnras/stab2093} {\bibfield  {journal}
  {\bibinfo  {journal} {\mnras}\ }\textbf {\bibinfo {volume} {509}},\ \bibinfo
  {pages} {3966} (\bibinfo {year} {2022})},\ \Eprint
  {http://arxiv.org/abs/2102.08414} {arXiv:2102.08414 [astro-ph.GA]}
  \BibitemShut {NoStop}%
\bibitem [{\citenamefont {Porredon}\ \emph {et~al.}(2021)\citenamefont
  {Porredon}, \citenamefont {Crocce}, \citenamefont {Fosalba}, \citenamefont
  {Elvin-Poole}, \citenamefont {Rosell}, \citenamefont {Cawthon}, \citenamefont
  {Eifler}, \citenamefont {Fang}, \citenamefont {Ferrero}, \citenamefont
  {Krause} \emph {et~al.}}]{porredon2021dark}%
  \BibitemOpen
  \bibfield  {author} {\bibinfo {author} {\bibfnamefont {A.}~\bibnamefont
  {Porredon}}, \bibinfo {author} {\bibfnamefont {M.}~\bibnamefont {Crocce}},
  \bibinfo {author} {\bibfnamefont {P.}~\bibnamefont {Fosalba}}, \bibinfo
  {author} {\bibfnamefont {J.}~\bibnamefont {Elvin-Poole}}, \bibinfo {author}
  {\bibfnamefont {A.~C.}\ \bibnamefont {Rosell}}, \bibinfo {author}
  {\bibfnamefont {R.}~\bibnamefont {Cawthon}}, \bibinfo {author} {\bibfnamefont
  {T.}~\bibnamefont {Eifler}}, \bibinfo {author} {\bibfnamefont
  {X.}~\bibnamefont {Fang}}, \bibinfo {author} {\bibfnamefont {I.}~\bibnamefont
  {Ferrero}}, \bibinfo {author} {\bibfnamefont {E.}~\bibnamefont {Krause}},
  \emph {et~al.},\ }\href {\doibase 10.1103/PhysRevD.103.043503} {\bibfield
  {journal} {\bibinfo  {journal} {\prd}\ }\textbf {\bibinfo {volume} {103}},\
  \bibinfo {eid} {043503} (\bibinfo {year} {2021})},\ \Eprint
  {http://arxiv.org/abs/2011.03411} {arXiv:2011.03411 [astro-ph.CO]}
  \BibitemShut {NoStop}%
\end{thebibliography}%

\end{document}